\documentclass{emulateapj}
\usepackage{float,epsfig}
\usepackage{lscape}
\usepackage{natbib}
\newcommand{\LA}{\mbox{\raisebox{-0.6ex}{$\stackrel{\textstyle<}{\sim}$}}}
\newcommand{\cxo}{{\sl Chandra}}

\newcommand{\ngc}{{NGC~2403}}
\newcommand{\msun}{M$_{\odot}$}
\newcommand{\ergl}{erg~s$^{-1}$}

\newcommand{\hi}{H{\sc i}}
\newcommand{\ergcms}{ergs~cm$^{-2}$~s$^{-1}$}
\newcommand{\ha}{H$\alpha$}
\newcommand{\hii}{H{\sc ii}}
\newcommand{\nh}{$N_{\rm H}$}

\newcommand{\sst}{{\sl Spitzer}}
\newcommand{\galex}{{\sl GALEX}}
\newcommand{\etal}{et al.}
\newcommand{\um}{$\mu$m}
\slugcomment{Submitted to Astronomical Journal}

\begin{document}

\title{An X-ray View of Star Formation 
 in the Central 3~kpc of NGC~2403}

\author{
Mihoko~Yukita\altaffilmark{1},
Douglas~A.~Swartz\altaffilmark{2},
Allyn~F.~Tennant\altaffilmark{3}, and
Roberto~Soria\altaffilmark{4} 
}
\altaffiltext{1}{University of Alabama in Huntsville, Dept. of Physics,
    Huntsville, AL, USA}
\altaffiltext{2}{Universities Space Research Association,
    NASA Marshall Space Flight Center, VP62, Huntsville, AL, USA}
\altaffiltext{3}{Space Science Office,
    NASA Marshall Space Flight Center, VP62, Huntsville, AL, USA}
\altaffiltext{4}{Mullard Space Science Laboratory,
    University College London, Holmbury St. Mary, Surrey RH5 6NT, UK}

\begin{abstract}

Archival \cxo\ observations are used to 
study the X-ray emission associated with star formation 
in the central region of the nearby SAB(s)cd galaxy \ngc.
The distribution of 
 X-ray emission is compared 
 to the morphology visible at other wavelengths using  
 complementary \sst, \galex, and ground-based \ha\ imagery.
In general, the brightest extended X-ray emission is 
 associated with \hii\ regions and to other star-forming structures but
 is more pervasive; existing also in regions devoid of strong \ha\ 
 and UV emission.
This X-ray emission has the spectral properties of diffuse hot gas
 ($kT\sim0.2$~keV) whose likely origin is in 
 gas shock-heated by stellar winds and supernovae with
 $\lesssim$20\% coming from faint unresolved X-ray point sources.
This hot gas 
 may be slowly-cooling extra-planar remnants of past outflow events,
or a disk component
 that either lingers 
 after local star formation activity has ended 
or that has vented from active star-forming regions into a porous interstellar medium.

\end{abstract}

\section{Introduction}
Whether merger-induced collapse \citep{barnes92,cole00},
 bar-driven inflow \citep{kormendy04},
 or dynamical friction \citep{noguchi00} 
builds structure at the center of a particular galaxy,
it is likely that star formation and (if present) central black hole growth
will be strongly regulated by feedback from massive stars (and
AGN activity). 

Since much of the current central massive object growth and star formation is occurring in small, low-density, disk-dominated spirals rather than
the massive but fuel-starved ellipticals \citep{heckman04},
nearby late-type galaxies are the ideal laboratories to view the growth of
galactic structure in the current epoch.
Here, we investigate star formation, feedback, and the growth of the central 
 region of \ngc.
We focus on the unique perspective enabled by X-rays in viewing the
 violent and inherently high-energy phenomena associated with 
 these dynamical processes and compare X-ray behavior to that 
 exhibited at other wavelengths. 

\ngc\ is a SAB(s)cd galaxy in the M81 group of galaxies
 ($D=3.2$~Mpc, 1\arcmin$=$1~kpc, \citeauthor{madore91} \citeyear{madore91}).
It is the most massive, $\sim$10$^{10}$~\msun, galaxy in the
 second largest of three M81 subgroups.
There are 7 dSph and dIrr  
 satellites of \ngc\ known within this subgroup \citep{karachentsev02}
ranging in size from $\sim$3$\times$10$^7$ to $\sim$3$\times$10$^9$~\msun\
\citep{karachentsev04}.

As with many disk-dominated late-type spirals, 
 \ngc\ lacks a central bulge \citep{kent87} but does host a luminous compact
 nuclear star cluster  
 first identified in IR images \citep{davidge02}.
The mass and luminosity of this cluster  
 are comparable to those found in many late-type spirals \citep{boker02}
but it is older and less compact than typical for late-type galaxies
 \citep{yukita07}. 
The dominant age of stars within the nuclear star cluster is $\sim$1.4~Gyr 
\citep{yukita07}. There are numerous younger
asymptotic giant branch, red, and early-type supergiants within the inner disk 
with stellar ages of 100~Myr \citep{davidge02}.
Similarly, the \ha\ surface brightness near the center is lower than
in surrounding regions while the youngest ($\sim$2--10~Myr) 
and most massive \hii\ regions
\citep{drissen99} are 0.7 to 1.6~kpc from the center.  
\citet{davidge02} speculate that this age gradient may be due to
the growth of a superbubble in
the central region that has  
quenched star-formation in the central region while triggering
activity further out through compression of surrounding gas during bubble 
expansion.
In this picture, nuclear star formation may be an episodic phenomenon with 
roughly a few 100~Myr interval between major star-forming events.

Such a star formation cycle requires a replenishing source of cold gas.
\citet{sheth05}, using molecular gas maps, showed that bulges can be built
 by bar-driven gas inflow 
but that the process requires of order a Hubble time in galaxies like \ngc.
Other studies have shown that the molecular and atomic gas in the central
regions of \ngc\ amount to only a small fraction of the dynamical mass
\citep[e.g.,][]{thornley95}. 
Importantly, the central total gas
surface density \citep[e.g.,][]{thornley95, martin01}, at least on kiloparsec spatial scales, is below the
critical value for star formation \citep{kennicutt89} 
yet the global star-formation rate in \ngc\ is a moderate 1.2~\msun~yr$^{-1}$.

\begin{figure*}
\begin{center}
\includegraphics[angle=-90,width=0.9\columnwidth]{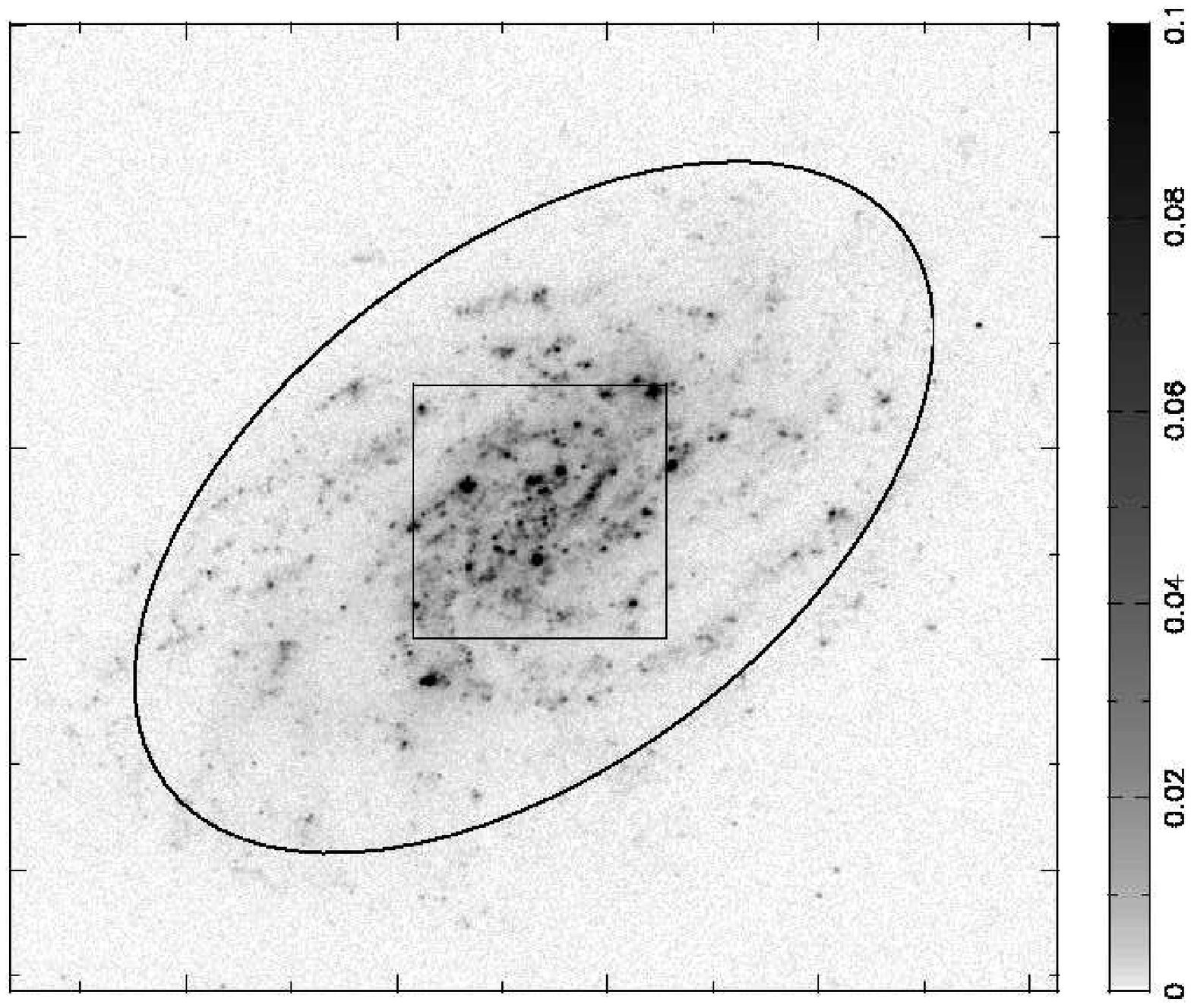}
\hspace{20pt}
\includegraphics[angle=-90,width=0.9\columnwidth]{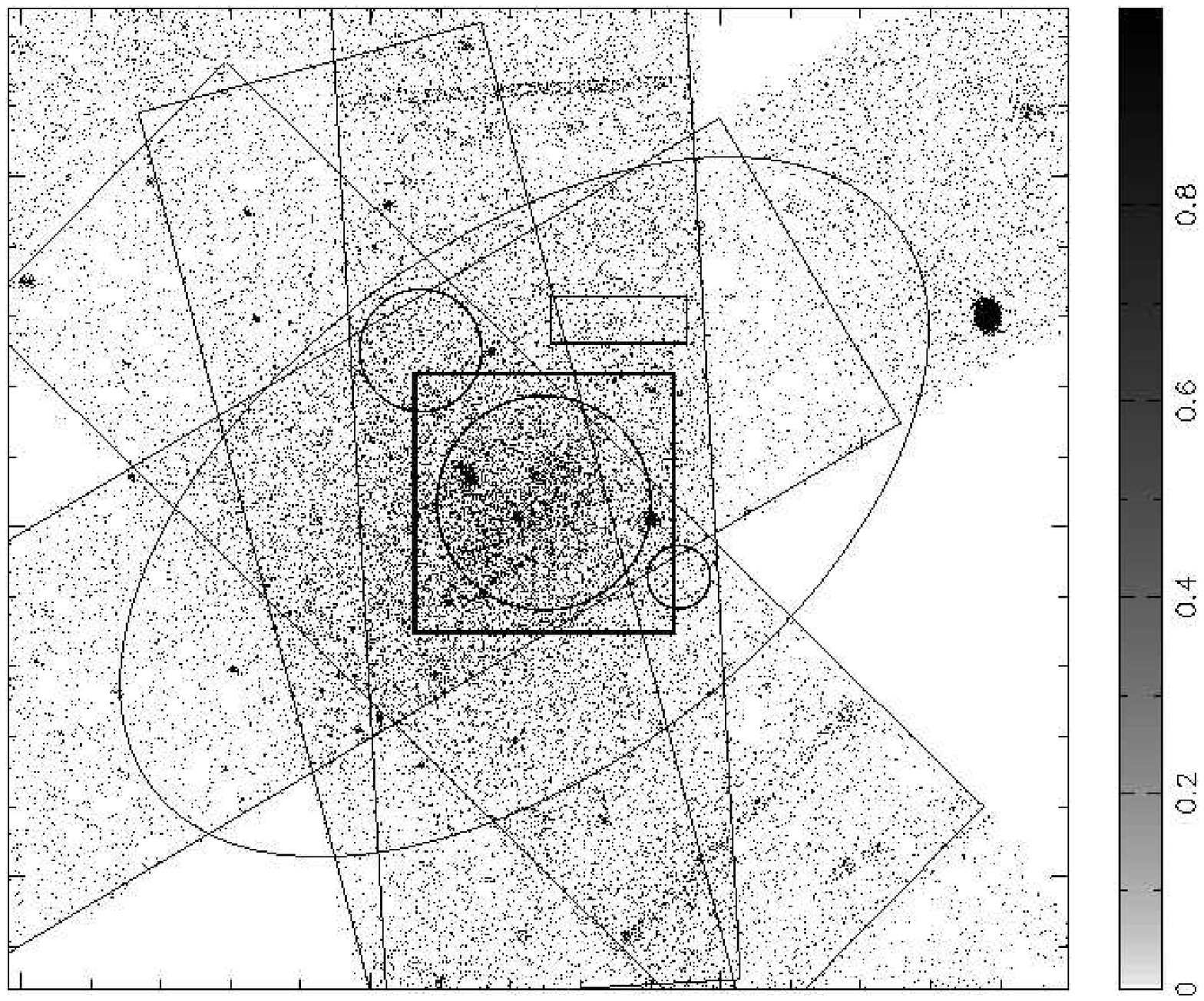}
\vspace{10pt}
\figcaption{Left: The \galex\ FUV image of \ngc. The 21\farcm9 $\times$12\farcm3 ellipes makes 
$D_{25}$ isophote  of NGC 2403. The central 6\arcmin$\times$6\arcmin\ region is shown as a square.
The gray scale indicates intensity level
in the unit of ct s$^{-1}$ pixel$^{-2}$. One pixel corresponds to 1\farcs5. 
Right: The same region in the merged X-ray data.  
The outer boundaries of the ACIS S2-S3-S4 combination are
 shown as thin-lined rectangles for each of the 4 
 overlaid observations.
The central 6\arcmin$\times$6\arcmin\ region is shown as a thick-lined square.  
The large circle in the center is the 2\farcm5 radius circle used for the X-ray residual emission spectral analysis (see text).  
The two smaller circles and the small rectangle identify the background 
 regions used for analysis of this residual emission. The gray scale indicates intensity levels
in units of ct pixel$^{-2}$. One pixel corresponds to 0\farcs492. 
\label{f:cxo_chip}}
\end{center}
\end{figure*}

More pertinent to the current study are
several independent investigations of the dynamics of the \hi\ gas in \ngc\
\citep{sicking97,schaap00,fraternali02a}.
The \hi\ rotation curve shows a flat gravitational potential
 typical of late-type spirals
but there is also slower-rotating neutral hydrogen extending up to 15~kpc 
above
 the disk.
Many properties of this anomalous \hi\ gas can be
 explained as
 galactic fountains \citep{shapiro76,bregman80,spitzer90}
but an additional,
external, source of cold gas such as infalling clouds
 or even small satellite galaxies might also be required
 \citep{fraternali06,fraternali08,struck09}.
High-resolution VLA observations of \ngc\  \citep{fraternali02a}
 show very anomalous kinematic features on small scales that may be
 evidence of such clouds.
Direct evidence of fountain activity is less conclusive.
Measurements of \ha\ emission line widths at several locations across
 the galaxy \citep{fraternali04} found only a few small-scale
 fountain-like features.
To match the observed \hi\ inflow with hot gas outflow
 would require many thousands of  these small-scale fountains.
Analysis of the spatial distribution of diffuse
 hot X-ray-emitting gas in \ngc\ \citep{fraternali02b} 
concluded that as little as a few
 percent or as much as the majority of the inflowing \hi\ can be
 matched by outflowing diffuse hot gas, depending on assumptions made.
Thus, it is not yet clear whether the buildup of the central regions of \ngc\ 
is
 being fueled cyclically through galactic fountains, or by
accreting gas from the intergalactic medium, or through mergers of
 small-scale gaseous satellites.

If galactic fountains are at work and are the source of the majority of the 
observed infalling \hi\ gas, then there
should be an imprint of this process in the X-ray emission from the disk
correlated with the source of the fountains; namely localized star-forming 
regions that heat the gas through massive stars to the point of breakout
from the disk.


We confine our study of \ngc\ to the inner 6\arcmin$\times$6\arcmin\ nuclear 
region (corresponding to 5.6$\times$5.6~kpc; 
 Figure~\ref{f:cxo_chip} left panel).
This region is small enough to be fully imaged within a single observation
 by most instruments used in this study
  yet includes a large portion of the actively star-forming
 regions of \ngc\ including 5 of the 6 giant \hii\ regions identified by
\citet{drissen99}.

We begin (\S~\ref{s:X}) with an independent analysis of the X-ray data;
examining X-ray-detected discrete sources and the underlying residual X-ray emission. This analysis extends the previous work of 
 \citet{schlegel03} and of \citet{fraternali02b} by including 
3 subsequent \cxo\ observations which allow us to better quantify
the spectral and temporal behavior of the X-ray emission. 
Our results are consistent with both these previous investigations.
We expand our analysis 
to include optically-identified supernova remnants (SNRs) and \hii\ regions.
Although individually X-ray-faint, we are able to 
characterize their bulk (average) X-ray temperatures and other X-ray
properties using stacking analysis.

We then turn to analysis of individual massive 
star-forming regions (\S~\ref{s:blobs}) in an attempt to
determine 
the X-ray emission properties of these regions as a function of their
 age, mass, and extinction properties.
We use a combination of (ground-based) \ha, mid-IR (\sst), and UV (\galex)
images to define these regions and to determine their basic physical characteristics.
We then use our knowledge of the differences in 
X-ray properties between \hii\ regions
and SNRs assembled in \S~\ref{s:X} to interpret the properties of the hot gas.

\section{\cxo\ X-ray Observations and Preliminary Analysis} \label{s:X}

In this section, archival \cxo\ images are examined to derive the X-ray 
 properties of four different source populations in \ngc; 
 namely, bright X-ray-detected sources
 (mainly luminous X-ray binaries and background AGNs), 
 optically-identified SNRs, 
 optically-identified \hii\ regions, and the underlying unresolved
 X-ray emission.
This allows us to parameterize these different types of sources in terms of
their X-ray temperatures, luminosity distributions, and emission measures. 
This information
will be used in  \S~\ref{s:blobs} to investigate individual massive star-forming 
regions defined at other wavelengths.

\ngc\ was observed in full-frame mode with \cxo\ ACIS-S on four occasions
for a total of $\sim$180~ks (Table~\ref{tb:cxolog}).  
We obtained level 1 event lists for all four observations from the \cxo\ data archive\footnote{http://cda.harvard.edu/chaser/} and reprocessed them 
using the CIAO (version 3.3.0.1) tool {\tt acis\_process\_events} 
and calibration database CALDB 3.2.1. 
Reprocessing removed pixel randomization,  applied CTI- and time-dependent gain corrections, and removed events with bad grades or bad status bits as well as bad and hot pixels.  We created lightcurves for each observation using a 1~ks binning to check for periods of high background.  
We excluded intervals with total count rates 
$>$3$\sigma$ above the mean rate for each observation.  The final Good Time Intervals for the observations are listed in Table~\ref{tb:cxolog}.

\begin{table}
\begin{center}
\caption{\cxo\ Observation log \label{tb:cxolog}}
\begin{tabular}{rrrr}
\tableline \tableline
\multicolumn{1}{c}{Date}    &  ObsID &  Instruments \& Mode &  GTI    \\ 

2001/04/17   &2014  &ACIS-S  TE        &  36.1 ks\\
2004/08/13   &4628  &ACIS-S  TE        &  47.1 ks \\
2004/10/03   &4629  &ACIS-S  TE        &  45.1 ks \\
2004/12/22   &4630  &ACIS-S  TE        &  50.6 ks\\
\tableline
\end{tabular}

\end{center}
\end{table}

SN~2004dj was used to define a common registration among the data sets. 
For this purpose, a circular Gaussian was fit to the image of the supernova to obtain an accurate centroid
 then the coordinates were adjusted to agree with the known position of SN~2004dj 
(J2000 R.A.$=$7$^h$37$^m$17.044$^s$, Decl.$=$$+$65\arcdeg35\arcmin57.84\arcsec, 
 \citeauthor{argo04} \citeyear{argo04}).  
To register pre-supernova data, we bootstrapped using a bright point source near the aimpoint that is common to both pre- and post-supernova data.  
The four co-aligned \cxo\ data sets were combined using the FTOOL utility {\tt fmerge}  to form a fifth merged dataset. 
The right panel of 
Figure~\ref{f:cxo_chip} displays the merged X-ray image.
The longest cumulative exposure is near the center of the 
 galaxy where we have selected a
 6\arcmin$\times$6\arcmin\ region for analysis as indicated.
Also shown are the galaxy's $D_{25}$ ellipse, corresponding to
 the 25~mag-sec$^{-2}$ contour in $B$.
The 6\arcmin$\times$6\arcmin\ region contains many point-like 
sources, extended star forming regions, and an extended region of
 X-ray emission near the galactic center.
With the exception of the bright point-like X-ray sources,
 we are primarily concerned with the soft X-ray emission components of \ngc.
For the soft emission, we fit spectra in the 0.4$-$2.0~keV range but quote results
 such as luminosities in the more familiar 0.5$-$2.0~keV band. 
For point sources, the fitting range is extended to higher X-ray energies.

\begin{deluxetable}{rrrrc} 
\tablecolumns{5} 
\tablewidth{0pc} 
\tablecaption{X-ray Point sources \label{tb:x58srcs}} 
\tablehead{
\colhead{R.A.} & \colhead{Decl}   & \colhead{Count Rate $^{a}$}    & \colhead{$L_{\rm X}$ $^{b}$} &  \colhead{Variability} \\
\colhead{J2000} & \colhead{J2000}   & \colhead{10$^{-4}$ cts s}    & \colhead{10$^{36}$\ergl} &  \colhead{} 
}
\startdata 
 07 36 24.5 &   +65 37 13.1 &     1.2   $\pm$   0.7  &   0.5  $\pm$ 0.3  &     \\         
 07 36 24.5  &  +65 37 34.9 &  $-$0.3  	$\pm$   0.4  & $-$0.1 $\pm$ 0.1  &     \\         
 07 36 25.2  &  +65 38 40.6 &     7.9  	$\pm$   1.7  &   3.1  $\pm$ 0.7  & S    \\         
 07 36 25.6  &  +65 35 39.7 &    1158   $\pm$   12   & 456.6  $\pm$ 4.7  & L S  \\         
 07 36 25.9  &  +65 37 57.0 &     0.1  	$\pm$   0.6  &   0.0  $\pm$ 0.2  &     \\         
 07 36 26.7  &  +65 36 20.9 &     0.8  	$\pm$   0.4  &   0.3  $\pm$ 0.2  &     \\         
 07 36 32.6  &  +65 34 52.6 &     0.9  	$\pm$   0.3  &   0.4  $\pm$ 0.1  &     \\         
 07 36 32.5  &  +65 38 20.0 &     0.1  	$\pm$   0.6  &   0.0  $\pm$ 0.2  &     \\         
 07 36 32.5  &  +65 39 00.7 &    29.7  	$\pm$   3.0  &  11.7  $\pm$ 1.2  & S    \\         
 07 36 33.0  &  +65 34 57.8 &     3.9  	$\pm$   0.6  &   1.6  $\pm$ 0.2  & L S  \\         
 07 36 34.0  &  +65 35 40.7 &     0.7  	$\pm$   0.3  &   0.3  $\pm$ 0.1  &     \\         
 07 36 34.1  &  +65 38 54.7 &    54.4  	$\pm$   3.9  &  21.4  $\pm$ 1.6  & L   \\         
 07 36 35.6  &  +65 36 08.4 &     3.1  	$\pm$   0.5  &   1.2  $\pm$ 0.2  &     \\         
 07 36 37.9  &  +65 38 15.7 &     2.0  	$\pm$   0.8  &   0.8  $\pm$ 0.3  &     \\         
 07 36 37.9  &  +65 37 37.2 &     1.2  	$\pm$   0.4  &   0.5  $\pm$ 0.2  &     \\         
 07 36 42.0  &  +65 36 51.8 &    52.6  	$\pm$   2.0  &  20.7  $\pm$ 0.8  &     \\         
 07 36 43.0  &  +65 37 09.7 &     0.7  	$\pm$   0.3  &   0.3  $\pm$ 0.1  &     \\         
 07 36 45.2  &  +65 35 57.3 &     3.1  	$\pm$   0.5  &   1.2  $\pm$ 0.2  &     \\         
 07 36 45.8  &  +65 36 40.5 &     2.4  	$\pm$   0.5  &   0.9  $\pm$ 0.2  &     \\         
 07 36 46.1  &  +65 36 13.6 &  $-$0.0  	$\pm$   0.1  & $-$0.0  $\pm$ 0.1  & L   \\         
 07 36 46.8  &  +65 35 58.1 &     1.0  	$\pm$   0.3  &   0.4  $\pm$ 0.1  &     \\         
 07 36 47.5  &  +65 36 19.0 &     7.3  	$\pm$   0.8  &   2.9  $\pm$ 0.3  & L S  \\         
 07 36 47.6  &  +65 36 22.9 &     5.5  	$\pm$   0.7  &   2.1  $\pm$ 0.3  &     \\         
 07 36 50.1  &  +65 36 03.7 &    20.9  	$\pm$   1.3  &   8.2  $\pm$ 0.5  & L   \\         
 07 36 52.0  &  +65 36 41.0 &     2.5  	$\pm$   0.5  &   1.0  $\pm$ 0.2  &     \\         
 07 36 52.5  &  +65 36 46.0 &     0.9  	$\pm$   0.3  &   0.4  $\pm$ 0.1  &     \\         
 07 36 53.8  &  +65 33 32.1 &     0.9  	$\pm$   0.2  &   0.4  $\pm$ 0.1  &     \\         
 07 36 53.9  &  +65 35 35.2 &     0.5  	$\pm$   0.2  &   0.2  $\pm$ 0.1  &     \\         
 07 36 55.4  &  +65 36 08.3 &     2.5  	$\pm$   0.4  &   1.0  $\pm$ 0.1  &     \\         
 07 36 55.6  &  +65 35 40.9 &   436.8  	$\pm$   5.0  & 172.3  $\pm$ 2.0  & L S  \\         
 07 36 56.1  &  +65 37 15.6 &     2.4  	$\pm$   0.5  &   0.9  $\pm$ 0.2  &     \\         
 07 36 57.2  &  +65 36 03.7 &     2.8  	$\pm$   0.4  &   1.1  $\pm$ 0.2  &     \\         
 07 37 00.7  &  +65 36 06.2 &     1.1  	$\pm$   0.3  &   0.4  $\pm$ 0.1  &     \\         
 07 37 00.8  &  +65 34 17.9 &    14.2  	$\pm$   0.9  &   5.6  $\pm$ 0.4  & L S  \\         
 07 37 01.8  &  +65 38 27.2 &  $-$0.1  	$\pm$   0.2  & $-$0.0  $\pm$ 0.1  &     \\         
 07 37 02.2  &  +65 34 45.8 &     0.4  	$\pm$   0.2  &   0.2  $\pm$ 0.1  &     \\         
 07 37 02.4  &  +65 36 01.5 &     1.2  	$\pm$   0.3  &   0.5  $\pm$ 0.1  &     \\         
 07 37 02.6  &  +65 37 10.8 &     2.4  	$\pm$   0.4  &   0.9  $\pm$ 0.1  &     \\         
 07 37 02.9  &  +65 34 37.9 &     1.4  	$\pm$   0.3  &   0.6  $\pm$ 0.1  &     \\         
 07 37 06.4  &  +65 34 51.9 &     0.7  	$\pm$   0.2  &   0.3  $\pm$ 0.1  &     \\         
 07 37 06.9  &  +65 36 21.7 &     0.7  	$\pm$   0.2  &   0.3  $\pm$ 0.1  &     \\         
 07 37 07.0  &  +65 36 14.1 &     1.1  	$\pm$   0.3  &   0.5  $\pm$ 0.1  &     \\         
 07 37 07.1  &  +65 35 56.6 &     2.1  	$\pm$   0.4  &   0.8  $\pm$ 0.1  &     \\         
 07 37 07.4  &  +65 34 55.9 &    21.1  	$\pm$   1.1  &   8.3  $\pm$ 0.4  & L S  \\         
 07 37 09.2  &  +65 35 44.2 &    23.8  	$\pm$   1.2  &   9.4  $\pm$ 0.5  & L   \\         
 07 37 09.4  &  +65 35 01.4 &     0.3  	$\pm$   0.2  &   0.1  $\pm$ 0.1  &     \\         
 07 37 09.6  &  +65 33 05.8 &     3.0  	$\pm$   0.4  &   1.2  $\pm$ 0.2  &     \\         
 07 37 09.9  &  +65 35 46.9 &     0.7  	$\pm$   0.2  &   0.3  $\pm$ 0.1  &     \\         
 07 37 10.2  &  +65 33 11.2 &     3.5  	$\pm$   0.5  &   1.4  $\pm$ 0.2  &     \\         
 07 37 11.3  &  +65 38 14.2 &     1.0  	$\pm$   0.3  &   0.4  $\pm$ 0.1  &     \\         
 07 37 11.6  &  +65 33 45.9 &    51.7  	$\pm$   1.7  &  20.4  $\pm$ 0.7  &  S   \\         
 07 37 13.1  &  +65 35 58.2 &     3.0  	$\pm$   0.4  &   1.2  $\pm$ 0.2  &     \\         
 07 37 14.9  &  +65 34 29.0 &    23.7  	$\pm$   1.2  &   9.4  $\pm$ 0.5  & S    \\         
 07 37 16.1  &  +65 33 28.9 &     1.9  	$\pm$   0.3  &   0.8  $\pm$ 0.1  &     \\         
 07 37 17.1  &  +65 35 57.9 &    42.4  	$\pm$   1.6  &  16.7  $\pm$ 0.6  & L S  \\         
 07 37 17.6  &  +65 36 23.4 &     1.3  	$\pm$   0.3  &   0.5  $\pm$ 0.1  &     \\         
 07 37 17.9  &  +65 37 26.5 &     1.2  	$\pm$   0.3  &   0.5  $\pm$ 0.1  &     \\         
 07 37 18.0  &  +65 35 09.4 &     1.0  	$\pm$   0.3  &   0.4  $\pm$ 0.1  &     \\         
\enddata
\tablecomments{Units of right ascension are hours, minutes, and seconds, and units of declination are degrees, arcmintes, and arcseconds.} 
\tablenotetext{a}{Count rates for ASIC-S3, taking the 0.5 $-$ 2.0 keV energy band.}
\tablenotetext{b}{$L_{\rm X}$ for the 0.5 $-$ 2.0 keV energy band. $L_{\rm X}$ is computed using
the bremsstrahlung model obtained from Table~\ref{tb:xptf} (see text).}
\end{deluxetable}

\subsection{X-ray Source and Region Definitions}

The source-finding tool described by \citet{tennant06} was applied to all five
datasets to search for discrete X-ray sources. The search was limited to the central 6\arcmin$\times$6\arcmin\
region and to events within the full \cxo\ energy range 0.3$-$8.0~keV.   
Fifty-eight sources were detected in the merged dataset
with a S/N above 2.8 and with a minimum of 5 $\sigma$ above background uncertainty (corresponding to a detection limit of about 8$-$10  counts for a typical on-axis source). 
This tally included all sources detected in the individual observations 
(provided they were within the FOV of the 
individual images) and all those previously reported by \citet{schlegel03}  
 within our 
6\arcmin$\times$6\arcmin\ field.  The X-ray point sources and their properties 
are listed in Table~\ref{tb:x58srcs}.

Since we are primarily interested in star-formation and the X-ray properties
of recent and current star-forming events in this study, we also examined the 
X-ray emission from known SNRs and \hii\ regions. 
For this purpose, we first masked out the X-ray-detected point-like 
 sources in the field of the merged image using a mask radius equal to 
 3 times the half-width of a Gaussian approximation to the 
 instrumental point spread function (PSF).
Larger radii were used, as necessary, for a few bright sources with 
 strong emission in their PSF wings.
Spectral information was then extracted from events
falling within the area defined by optically identified SNRs from [SII]/\ha\ ratio by 
\citet{matonick97} using their listed locations and
source radii. One arcsecond ( $\sim$ 16 pc ) radii are used 
 for the SNRs whose sizes are unlisted.   Local backgrounds were also extracted from surrounding annuli.
There are 29 cataloged SNRs within our 6\arcmin$\times$6\arcmin\ field.
There are also three candidate radio SNRs in our field
\citep{pannuti07}.
One (denoted $\mu$ by \citeauthor{pannuti07} 2007)
 is already included in the list of optically-identified SNRs.
Another (denoted TH2 by \citeauthor{pannuti07} 2007) is a strong
 X-ray source, $L_{\rm X} = 2\times 10^{37}$~\ergl, as 
 previously noted by \citet{pannuti07}.
However, no radio spectral indices are available for any of 
 these candidate radio SNRs.
Therefore, we opt to omit TH2 from our list of candidate SNRs
 as it is anomolously bright if it is, in fact, a SNR.
It remains classified as a discrete X-ray source in our tabulation. 

A similar procedure was attempted using the \hii\ regions defined by 
\citet{sivan90} based on photographic images but
many of these cataloged objects did not correspond to \hii\ regions visible
in a recent continuum-subtracted \ha\ CCD image. 
This is due mostly to crude modeling of the \hii\ structure which resulted
 in poor estimates of source positions and sizes.
Therefore, we constructed 
our own catalog of \hii\ regions defined as circular approximations
to the areas enclosed by  surface brightness 
contours corresponding to a level of 20$\times$ the background 
in our continuum-subtracted \ha\ image.
There are 58 \hii\ regions defined in this way
within our 6\arcmin$\times$6\arcmin\ field
including 5 of the 6 giant \hii\ regions identified previously
by \citet{drissen99}.

There are four point-like sources detected by our source-finding trial
 that spatially coincide with SNRs. 
Two more detected sources lie
within a bright extended \hii\ region.
Upon further examination, all these sources were determined 
to be steady thermal sources 
 and were therefore assigned to the SNR (\hii) 
category rather than to the general category of point sources for purposes
of our analysis. Conversely, 
the PSF wings of bright point sources overlap with the positions
of two of the cataloged SNRs. These two
SNRs were therefore excluded from our analysis because of this contamination. 
The three historical supernovae, SN~2004dj (a strong X-ray source), 
SN~2002kg and SN~1954J (both fainter than our detection limit) 
and the X-ray-bright candidate radio SNR
were also not considered further in this study.

Finally, we also examined the properties of the X-ray emission remaining 
after the X-ray point sources and the optically-identified SNRs and \hii\
regions are excluded. 
This residual component is likely 
comprised of unresolved (faint) XRBs and other low-luminosity objects related
to the stellar content of \ngc, extended diffuse hot gas within the disk and halo of \ngc, and unresolved background (primarily AGN) and foreground (local diffuse 
Galactic) emission. 
Inspection of the residual emission image shows a clear excess 
 above the background in the central regions of \ngc.

We examine the X-ray properties of these four source types 
in the following subsections.

\subsection{Discrete X-ray-Detected Sources} \label{s:pts}

\begin{table}
\begin{center}
\caption{Point sources fitting results \label{tb:xptf}}
\begin{tabular}{lrr}
\tableline \tableline 
Fit Parameters & Bremsstrahlung & Powerlaw \\
\tableline 
                               	
 $N_{\rm H}$ (10$^{21}$cm$^{-2}$)                &  1.9$^{+0.1}_{-0.1}$   & 2.5 $^{+0.2}_{-0.2}$\\
 $kT$ (keV)                                &  4.7$^{+0.6}_{-0.5}$   & \\
   $\Gamma$                                &                        & 1.87$^{+0.06}_{-0.06}$  \\ 
 Normalization (10$^{-4}$)                 &  5.9$^{+0.2}_{-0.2}$   & 5.4$^{+0.3}_{-0.3}$ \\
 $L_{\rm X}$(0.5-2.0 keV)/10$^{38}$ \ergl\ &  7.7$^{+0.3}_{-0.4}$   & 7.7$^{+0.5}_{-0.5}$\\
  $L_{\rm X}$(2.0-10 keV)/10$^{38}$ \ergl\ &  16.7$^{+0.8}_{-1.1}$  & 20.0$^{+0.8}_{-0.8}$\\
 $\chi^{2}$/dof                            &   185/197              & 211/197 \\
\tableline
\end{tabular}
\end{center}
\end{table}

\begin{figure}
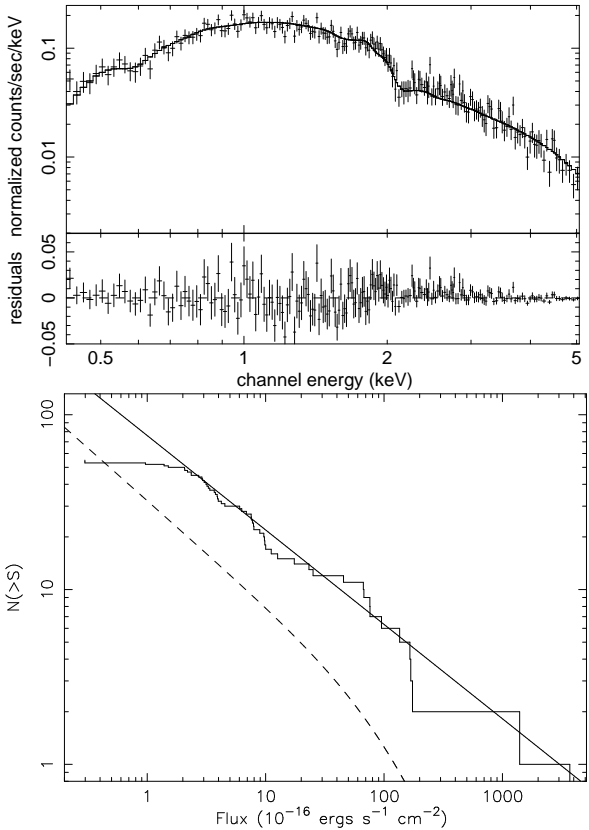

\begin{center}
\includegraphics[angle=-90,width=0.9\columnwidth]{f2a.eps}
\includegraphics[angle=-90,width=0.9\columnwidth]{f2b.eps}
\vspace{10pt}
\figcaption{Top: The co-added (stacked) 
 spectrum of all 58 discrete sources detected
 in the \cxo\ X-ray data.
The co-added spectrum was constructed using only the first observation.
Also shown are the best-fit model (an absorbed bremsstahlung) and the fit 
 residuals.  
Bottom: The X-ray luminosity function on the energy range of 0.5$-$2.0 keV 
 (solid line) for the 58 sources. 
The best-fit model  was used to convert count rates to flux  
 (3.1$\times$10$^{-4}$~ct~s$^{-1}$ corresponds to  10$^{-15}$\ergcms).  
The single powerlaw model fitted to the luminosity function for 
 sources brighter than 2$\times$10$^{-16}$~\ergcms\ $\sim$2$\times$10$^{35}$\ergl\
 is shown as a solid curve and gives a slope of $-$0.54. 
The dashed curve represents the expected contribution to the luminosity
 function from cosmic X-ray background sources as given by the 
 Moretti \etal\ (2003) analytic expression. 
\label{f:X_pts}}
\end{center}
\end{figure}

\citet{schlegel03} have already presented the properties of 
the discrete source population detected in the first \cxo\ observation of \ngc.
Here, our interest is only in the average spectral shape of the discrete sources and 
their luminosity distribution. 
The average spectrum can be  represented by  
co-adding the source spectra from only the first observation.
This provides sufficient 
source counts 
for characterizing the spectrum 
yet avoids complications that arise when combining observations made at 
different times during the \cxo\ mission.  
The spectra of the individual point sources (and corresponding backgrounds)
were co-added and weighted ancillary response 
(ARF) and response matrix (RMF) files were
generated using CIAO tool {\tt acisspec}.
The resulting spectrum was grouped to ensure at least 20 counts in each 
spectral bin.
The source spectrum was modelled
 using the XSPEC spectral-fitting utility (version 11.3.2t).

We applied both absorbed powerlaw and bremsstrahlung models to the co-added 
discrete source spectrum.  
The fitting results are shown in Table~\ref{tb:xptf}, and the spectrum is shown in 
 Figure~\ref{f:X_pts}.  
Although we are mainly interested in the energy range 0.5$-$2.0 keV, 
 a more precise fit can be made by extending the spectral fits to at least 
5~keV.
Over this range, the best-fitting model is the thermal bremsstrahlung model.
 This model is used  to estimate the luminosities of the 
individual sources. 
For this purpose, separate ARF and RMF files
 for each  source and for each observation were created and used in the fit.
The hydrogen column density and bremsstrahlung temperature were held
 fixed at the values from Table~\ref{tb:xptf} during the fitting; 
 only the model normalization was allowed to vary. 
The resulting 0.5$-$2.0~keV luminosities, averaged over the four observations,
 are listed in Table~\ref{tb:x58srcs}.

The flux estimates were used to check for variability of sources between
and during individual \cxo\ observations. 
This check helps to differentiate 
 between diffuse emission associated with star-formation that has a relatively 
 soft X-ray spectrum and no variability and emission from XRBs that is 
 spectrally hard and often variable over time.

If the flux from a source during one observation deviated from its average
 over the other available observations by $>$3$\sigma$, then it was flagged as 
long-term variable and designated by the letter ``L'' in column 5 of Table~\ref{tb:x58srcs}.
(Because of differences in spacecraft roll and target aimpoint,
 not all sources were within the field of view during all four observations.)  
Short-term variability during individual observations was checked by binning 
the source light curve (event arrival time) for each observation using 2~ks 
temporal bins. The shape of the light curve was compared to a constant using 
the $\chi^2$ statistic to check for variability.   Short-term variability is 
also given in column 5 of Table~\ref{tb:x58srcs} by the letter ``S''.

The cumulative X-ray luminosity function (XLF) for the 58 sources detected within the
6\arcmin$\times$6\arcmin\ central region of \ngc\ is shown in 
Figure~\ref{f:X_pts}.
Luminosities for each source were computed from their average flux
over up to four observations, depending on availability, on the energy range 0.5$-$2.0~keV.  
Using only the 50 sources brighter than 
a flux limit flux of 2.0$\times$10$^{-16}$~\ergcms\ in the 0.5$-$2.0~keV energy range 
(corresponding to a luminosity at the distance of \ngc\ 
of 2.4$\times$10$^{35}$~\ergl), 
a single power law gives a best-fitting slope of $-0.54\pm0.02$, 
roughly consistent with the value of $-0.59\pm0.02$ reported by \citet{schlegel03}. 
For reference, a flux of 10$^{-15}$~\ergcms\ in the 0.5$-$2.0~keV band 
corresponds to a flux of 2.1$\times$10$^{-15}$~\ergcms\ in the 2.0--10.0~keV band 
assuming our best-fit co-added discrete source spectral shape.

An estimate of the contribution from unrelated background AGN to the detected
source population in the field can be 
made by comparing the observed XLF to the 
analytical fits to the deep field cosmic X-ray background 
log$N$--log$S$ \citep{moretti03}.
This fit is shown in Figure~\ref{f:X_pts}. An estimated 21 $\pm$ 1 of the 50 sources, or 
42$\pm$5\%, are likely background AGNs. 

\subsection{Supernova Remnants} \label{s:snr}

\begin{figure}
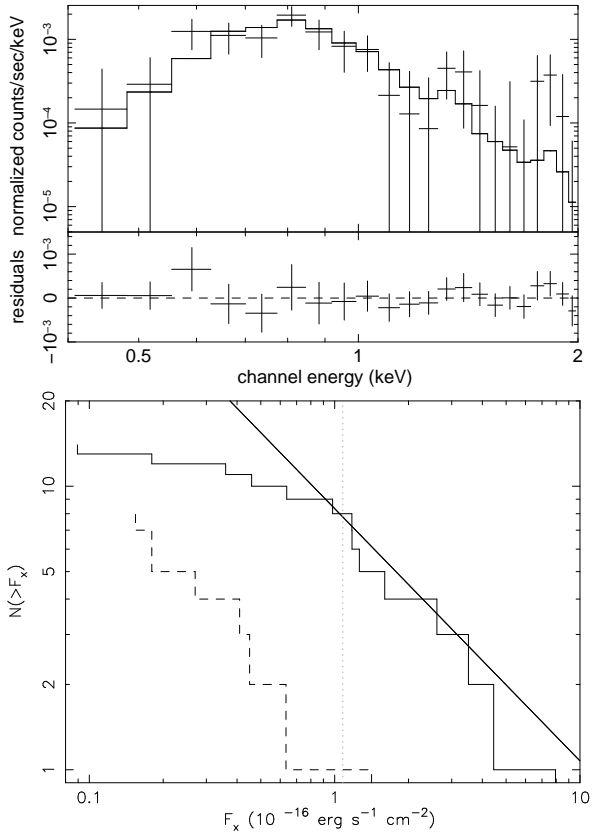

\begin{center}
\includegraphics[angle=-90,width=0.9\columnwidth]{f3a.eps}
\includegraphics[angle=-90,width=0.9\columnwidth]{f3b.eps}
\vspace{10pt}
\figcaption{Top: The co-added (stacked) spectrum of the
 24 optically-identified SNRs imaged on S3.  
The best-fitting model 
 (an absorbed {\tt apec} model with hydrogen column density, $N_{\rm H}$,
 allowed to be a free parameter) is shown along with the fit residual.
The spectrum has been re-binned to have a width of 73 eV for display purposes.
Bottom: The luminosity function of the background-subtracted SNRs (solid line) 
and of the pseudo-random positions (dashed line, see text).
The single powerlaw model fitted to the luminosity function of 
 sources brighter than 1.1$\times$10$^{-16}$~\ergcms\ $\sim$2$\times$10$^{35}$\ergl\
 is shown as a curve and gives a slope of $-$0.89. 
\label{f:X_snr}}
\end{center}
\end{figure}

Most of the individual SNRs (and \hii\ regions discussed in the next subsection)
are too faint to be detected as discrete X-ray sources. The possibility exists
that many of the   
X-ray events detected within the spatial regions defined by the 
individual optical SNRs are unrelated ``background'' events. 
Figure~\ref{f:X_snr} shows the XLF for the SNRs. Also shown is a pseudo-random
sampling of the background defined using the same locations and sizes of the 
cataloged SNRs but with their RA coordinates reflected about the center of \ngc. 
(For both populations, only those regions with $>$1 background-subtracted net counts
are shown.)
This second XLF indicates that only the (X-ray) 
brightest 4 or 5 cataloged SNRs are 
truly sources of X-ray emission. 
Fitting a single power law model to the XLF of the eight SNRs with $>$4 source 
counts in the 0.5$-$2.0 keV energy range (corresponding to 1.1$\times$10$^{-16}$~\ergcms)
 gives a best-fitting slope of $-$0.89 $\pm$ 0.14.

There are four optically-identified SNRs detected in X-rays 
 via the source-finding tool.   
The X-ray-brightest SNR is also the brightest in \ha\ line emission. 
However, there is no overall correlation between X-ray and \ha\ luminosity
 in the SNR sample.  
This is consistent with the analysis of \citet{pannuti07}.  
Using the two SNRs whose 
 diameters are listed in \citet{matonick97}, 
we also find the mean diameter of X-ray-detected SNRs is smaller (40~pc) 
 than the mean diameter of X-ray non-detections (70~pc)
 again comfirming the \citet{pannuti07} result.

A co-added spectrum of the cataloged SNRs was created 
 by merging the first three \cxo\ images (the fourth observation does not
 adequately cover the central 6\arcmin$\times$6\arcmin\ region of interest). 
Twenty-four SNRs (of 27 total) are imaged on S3 in the first three observations. 
The same averaging 
method used for the discrete sources (\S~\ref{s:pts}) was used here
to build the co-added spectrum.
Weighted ARF and RMF files generated from the second observation were used 
for the spectral fitting. 
We 
checked that this did not adversely affect our fitting results by repeating the 
spectral analysis using ARFs and RMFs generated from the first and third
observations. 

\begin{table}
\begin{center}
\caption{SNRs X-ray fitting results \label{tb:xsnrf}}
\begin{tabular}{lrrr}
\tableline \tableline 
 Fit Parameters & {\tt apec} & {\tt apec} & {\tt apec} \\ 
\tableline 
$N_{\rm H}$ (10$^{21}$cm$^{-2}$)         & 0.4                   & 1.4             & 2.8$^{+3.3}_{-2.3}$  \\
 $T_e$ (MK)                         & 4.8$^{+2.6}_{-1.0}$   & 4.2$^{+1.6}_{-0.7}$  & 3.5$^{+2.1}_{-1.4}$  \\
 Normalization$^a$                  & 0.1$^{+0.5}_{-0.1}$   & 0.3$^{+0.1}_{-0.1}$  & 0.5$^{+2.3}_{-0.5}$  \\
 $L_{\rm X}$/10$^{37}$ \ergl\       & 0.3$^{+0.1}_{-0.1}$   & 0.3$^{+0.1}_{-0.1}$  & 0.3$^{+0.5}_{-0.2}$  \\ 
 $L_{\rm X}^{int}$/10$^{37}$ \ergl\ & 0.3                   & 0.6                  & 0.8                  \\ 
  C-stat/bins                       & 107.4/108             & 105.9/108            & 105.3/108            \\
\tableline
\multicolumn{4}{c}{Derived Parameters$^b$}  \\
\tableline
 $V$    [$f$] ($10^{63}$ cm$^{3}$)       &0.13  & 0.13   & 0.13  \\
 $n_e$ [$f^{-1/2}$] (cm$^{-3}$)          &0.04 & 0.05  & 0.07 \\
 $P/k$ [$f^{-1/2}$] (10$^5$ K cm$^{-3}$) &3.5  & 4.1   & 5.1  \\
 $M_x$ [$f^{+1/2}$] (10$^5$ \msun)       &0.05 & 0.07  & 0.11 \\
 $E_{th}$ [$f^{+1/2}$] (10$^{53}$ erg)   & 0.09 & 0.1   & 0.1  \\
 $t_c$ [$f^{+1/2}$] (Myr)                & 48   & 30    & 16   \\
\tableline
\multicolumn{4}{l}{$^a$ $K=(10^{-9}/4\pi D^2) \int n_e n_p dV$} \\
\multicolumn{4}{l}{$^b$volume filling factor scaling in [ ] in Column 1.}\\
\end{tabular}
\end{center}
\end{table}

We used a surrounding annulus to define a local background for the individual SNRs,
 after masking out the detected X-ray point sources. 
The total number of X-ray events in the co-added spectrum between 0.4 and 2.0 keV is 153. 
This reduces to $94.8 \pm 13.1$ after background subtraction. 
The XSPEC model {\tt apec}, representing 
 thermal emission from an optically-thin collisionally-ionized plasma,
 combined with a model {\tt phabs} for an intervening absorption column
 were applied to the spectrum.
The parameters of the {\tt apec} model are the plasma temperature,
 elemental abundances, and emission integral.
Elemental abundances were fixed to the solar values
 as given in \citet{anders89}.
Three model fits were attempted as detailed in Table~\ref{tb:xsnrf}. 
These differed in the values assigned to the absorbing columns. 
In two of these models, the  absorbing column was held fixed at 
 either the Galactic value along the line of sight, 
 $N_{\rm H}^g=4\times10^{20}$~cm$^{-2}$, implying no local absorption
 within NGC~2403, or $N_{\rm H}=14\times10^{20}$~cm$^{-2}$ which is 
 equivalent to the mean value of the optical extinction derived 
 for active star-forming regions (\S~\ref{s:blobs}).
The third model allows the absorbing column to be a free parameter 
 in the fit. 
In this and in models of the \hii\ regions and of the residual X-ray
 emission to be discussed below, allowing the absorbing column,
 plasma temperature, and normalization (emission integral) to vary
 in the fitting process often leads to large uncertainties in some of
 these parameter values.
The reason for this is simply that the plasma temperature is quite 
 low, of order a few $10^6$~K (a few 0.1~keV), so that most of the
 emission is at low energies where the sensitivity to absorption
 is most acute. 
This introduces a degeneracy in which an equally acceptable 
 fit is possible by increasing the column density, \nh, while
 simultaneously increasing the model normalization, usually with
a moderate decrease in the fitted temperature.

The X-ray luminosities of the individual SNRs were computed by 
adopting the spectral model that best fits the stacked spectrum
 (Table~\ref{tb:xsnrf}, column 4). This model was applied to the
distribution of events of each individual SNR using the best-fit 
values of $T_{e}$ and \nh,  
and allowing only the model normalization to vary in the final 
fitting. This is equivalent to scaling by the number of counts 
in the individual spectra.
The X-ray count rates, corresponding luminosities, and \ha\ luminosities 
 are listed in Table~\ref{tb:snr}.

\begin{table*}
\begin{center}
\caption{SNRs \label{tb:snr}}
\begin{tabular}{rrrrr}

\tableline \tableline
  RA &  Dec &  Count Rate $^{a}$& $L_{\rm X}$ $^{b}$                     & $L_{H\alpha}$   \\
 (J2000) & (J2000) & 10$^{-5}$ cts s$^{-1}$  &  10$^{35}$\ergl& 10$^{37}$\ergl \\
\tableline   
 07 36 42.8 & +65 34 52.7 & $-$0.8 $\pm$   1.0  & $-$0.3 $\pm$   0.4  & 0.3  \\     
 07 36 45.7 & +65 36 35.7 &    1.1 $\pm$   1.5  &    0.4 $\pm$   0.6  & 1.7 \\      
 07 36 45.7 & +65 36 40.7 &   25.8 $\pm$   4.8  &   10.0 $\pm$   1.8  &17.1  \\     
 07 36 49.0 & +65 34 31.0 &    3.3 $\pm$   3.8  &    1.3 $\pm$   1.5  & 2.4  \\     
 07 36 52.0 & +65 33 41.0 & $-$0.5 $\pm$   0.9  & $-$0.2 $\pm$   0.3  & 0.5  \\     
 07 36 52.7 & +65 35 50.5 & $-$0.8 $\pm$   1.0  & $-$0.3 $\pm$   0.4  & 4.8 \\      
 07 36 53.3 & +65 36 00.3 &    3.8 $\pm$   2.0  &    1.5 $\pm$   0.8  & 1.0  \\     
 07 36 53.5 & +65 33 42.0 &    2.4 $\pm$   2.1  &    0.9 $\pm$   0.8  & 5.5  \\     
 07 36 53.6 & +65 35 11.6 & $-$0.5 $\pm$   0.9  & $-$0.2 $\pm$   0.3  & 2.9  \\     
 07 36 56.2 & +65 34 06.0 &    5.1 $\pm$   2.7  &    2.0 $\pm$   1.1  & 1.8  \\     
 07 36 57.0 & +65 36 04.8 & $-$0.5 $\pm$   0.9  & $-$0.2 $\pm$   0.3  & 0.3  \\     
 07 37 01.6 & +65 34 13.0 &    7.8 $\pm$   3.4  &    3.0 $\pm$   1.3  & 2.9  \\     
 07 37 02.0 & +65 34 37.0 &    0.0 $\pm$   0.9  &    0.0 $\pm$   0.4  & 5.5  \\     
 07 37 02.0 & +65 33 42.0 &    0.3 $\pm$   0.9  &    0.1 $\pm$   0.3  & 0.4  \\     
 07 37 02.2 & +65 37 20.7 &    1.0 $\pm$   1.2  &    0.4 $\pm$   0.5  & 2.3  \\     
 07 37 02.2 & +65 36 02.0 &    3.7 $\pm$   1.8  &    1.5 $\pm$   0.7  & 2.9  \\     
 07 37 03.0 & +65 33 45.0 &    0.0 $\pm$   0.9  &    0.0 $\pm$   0.4  & 0.3  \\     
 07 37 03.0 & +65 34 38.0 &   10.1 $\pm$   2.9  &    3.9 $\pm$   1.1  & 1.0  \\     
 07 37 06.0 & +65 36 03.2 & $-$2.4 $\pm$   1.8  & $-$0.9 $\pm$   0.7  & 2.0  \\     
 07 37 06.1 & +65 36 10.3 & $-$0.8 $\pm$   1.7  & $-$0.3 $\pm$   0.7  & 2.2  \\     
 07 37 07.1 & +65 37 10.6 & $-$1.3 $\pm$   1.0  & $-$0.5 $\pm$   0.4  & 1.2  \\     
 07 37 10.6 & +65 33 10.0 &    2.8 $\pm$   1.9  &    1.1 $\pm$   0.7  & 0.3  \\     
 07 37 12.5 & +65 33 46.0 &    0.5 $\pm$   0.8  &    0.2 $\pm$   0.3  & 2.0  \\     
 07 37 16.1 & +65 33 29.0 &   13.2 $\pm$   3.3  &    5.1 $\pm$   1.3  & 0.8  \\     
\tableline   
\multicolumn{5}{l}{Units of right ascension are hours, minutes, and seconds, and units}\\
\multicolumn{5}{l}{of declination are degrees, arcmintes, and arcseconds.}\\
\multicolumn{5}{l}{$^{a}$ Count rates for ACIS-S3, taking the 0.5 $-$ 2.0 keV energy band.}\\
\multicolumn{5}{l}{$^{b}$ Luminosity in the 0.5 $-$ 2.0 keV energy band. $L_{\rm X}$ is computed }\\
\multicolumn{5}{l}{using the one temperature model with variable $N_{\rm H}$ model as listed }\\
\multicolumn{5}{l}{in Table~\ref{tb:xsnrf} (see text)}\\
\end{tabular}
\end{center}                                                                                          
\end{table*}

A number of other physical parameters can be derived from the fitted
 values of the plasma temperature and emission integral provided some
 estimate of the volume occupied by the hot gas can be made. 
The total volume of the 24 SNRs  
is $1.3\times10^{62}$~cm$^3$
 assuming a spherical geometry for each remnant and adopting
 the radii reported in \citeauthor{matonick97} (1997; the same radii
 used here to define the X-ray spectral source regions).
The actual volume occupied by the hot gas is some fraction, $f$, of this
 total volume. 
Assuming the ion and electron number densities are equal, 
which is roughly true for a hydrogen-dominated hot plasma, then the 
emission integral $\int n_e n_H dV \sim n_e^2 f V$ so that the electron 
density $n_e \propto (K /f V)^{1/2}$ where $K$ is the spectral model normalization parameter which is itself proportional to the emission integral.

From $n_e$ and the flux-weighted mean temperature $T_e$, 
we can estimate the hot gas pressure
$P/k = 2n_e T_e$, mass $M_X=n_e \mu m_p V$ 
 where $m_p$ is the proton mass and $\mu=1.4$ is the mass per proton,
thermal energy $E_{th}=3n_e VkT_e$ and the cooling time $t_c=E_{th}/L_{\rm bol}$
of the X-ray emitting plasma
as listed in Table~\ref{tb:xsnrf}. 
Here, $L_{\rm bol}$ is estimated by integrating the spectral model 
 from 0.02~keV to 20~keV within XSPEC. For plasma temperature typical of SNRs
(and of other thermal sources analyzed in this paper), $L_{bol}$  is $\sim$ 5 times 
$L_{\rm X}$, where $L_{\rm X}$ is the observed X-ray luminosity quoted on the standard
0.5$-$2.0~keV range. 
These derived quantities are all functions of the 
 volume filling factor, $f$ as indicated in the table.

The total X-ray luminosity from the 24 SNRs is only 3$\times$10$^{36}$~\ergl.
Inspection of Figure~\ref{f:X_snr} shows that most of this emission comes
 from the few brightest sources with the most luminous being 
 $\sim$1$\times$10$^{36}$~\ergl.
The thermal energy content of the SNRs, $\sim$10$^{52}f^{1/2}$~erg,
 corresponds to up to $\sim$40\% (in the limit $f=1$) 
 of the initial kinetic energy released by the supernova explosions,
 assuming 10$^{51}$~erg per event.
The total mass of hot gas is substantial. 
It corresponds to an average mass per SNR of 
 $M_X\sim 300 \pm 100 f^{1/2}$~\msun\ (uncertainties given by the range
of values deduced from the three X-ray spectral models).
This can be compared to the amount of material swept up by
the SNRs, $\sim$ 5000 $n_H$ \msun\ per SNR, assuming a uniform
density $n_H$~cm$^{-3}$, over the SNR volume.


\subsection{\hii\ Regions} \label{s:hii}

\begin{figure}
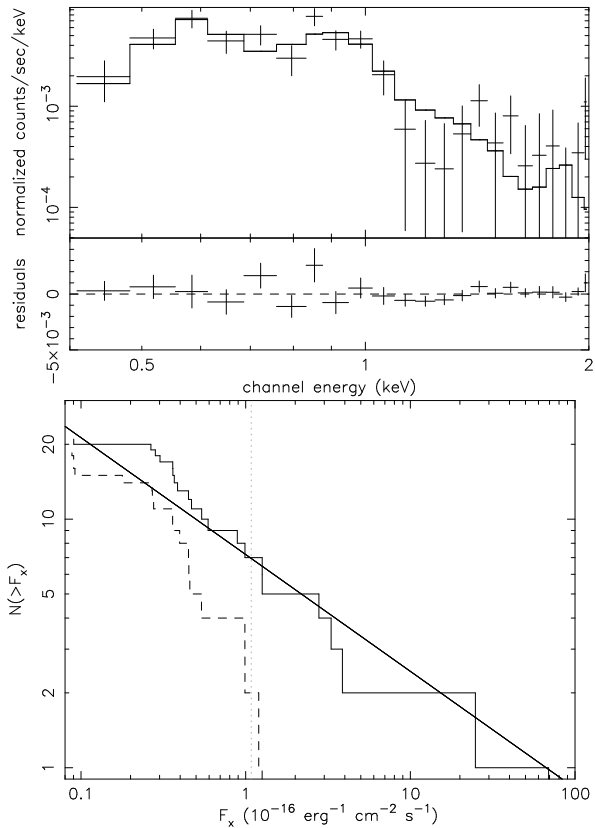

\begin{center}
\includegraphics[angle=-90,width=0.9\columnwidth]{f4a.eps}
\includegraphics[angle=-90,width=0.9\columnwidth]{f4b.eps}
\vspace{10pt}
\figcaption{Top: The stacked spectrum of the identified \hii\ regions.  
The best-fitting model (an absorbed two-temperature {\tt apec} model 
 with hydrogen column density, $N_{\rm H}$, as a free parameter) is 
shown along with the fit residuals. 
The spectrum has been re-binned to have a width of 73 eV 
 for display purposes. 
Bottom: The luminosity function of the background-subtracted \hii\ regions
 (solid line) and of the pseudo-random positions (dashed line, see text).
The single powerlaw model fitted to the luminosity function of 
 sources brighter than 1.1$\times$10$^{-16}$~\ergcms\ $\sim$2$\times$10$^{35}$\ergl\
 is shown as a curve and gives a slope of $-$0.47. 
\label{f:X_hii}}
\end{center}
\end{figure}

We performed an analysis of the 48 \hii\ regions, imaged on S3 in the first three 
observations, analogous to that of the SNRs.
Figure~\ref{f:X_hii} shows the XLF for the \hii\ regions and the pseudo-random
sampling defined using RA coordinates reflected about the center of \ngc. 
Again, only the most X-ray-luminous \hii\ regions are significant detections.
Similar to the SNRs, we also fitted a single power law model to the XLF
of the 7 brightest \hii\ regions, ($>$4 counts and corresponding to 
1.1$\times$10$^{-16}$~\ergcms).  
The best-fitting slope is $-0.47 \pm 0.13$, somewhat shallower
 than for the SNRs.

The co-added X-ray spectrum is shown in Figure~\ref{f:X_hii}.  Similar to the SNR spectrum, we also applied thermal models to the \hii\ region spectrum.   
There are sufficient source counts 
 (858 cts, 422$\pm$32 after the background subtraction) 
 in the \hii\ region spectrum to
 allow 2-temperature model fits in addition to the 
 1-temperature models that were applied to the SNR spectrum. 
The fitting results and derived values are listed in Table~\ref{tb:xhiif}.
We assumed pressure equilibrium between the two temperature components and
 identical filling factors to derive the values of $n_e$, $E_{th}$, and $M_{x}$.  

Note that the best-fitting 1-temperature model with
 variable column density results in a very high model normalization
 which leads to unreasonable values of the derived parameters.
However, this does not seem to be the issue with the 2-temperature model.
This model (column 6 in Table~\ref{tb:xhiif}) results in 
 $N_{\rm H} \sim N_{\rm H}^g$ (though with large uncertainty)
 and hence with values of the remaining
 parameters similar to those of the fixed \nh\ model.

The X-ray luminosities of the individual \hii\ regions were computed in 
the same manner as for the SNRs;  using the best-fitting 
two temperature spectral model with variable $N_{\rm H}$ in this case.  
The X-ray count rates, corresponding luminosities, and \ha\ luminosities are listed in Table~\ref{tb:hii}.

There are twice as many \hii\ regions in our sample as there are SNRs
in our SNR sample. 
The X-ray temperatures of the \hii\ regions are about 40\% lower 
 compared to the SNRs.
The total X-ray luminosity is $\sim$1.5$\times$10$^{37}$~\ergl\ for the 
 \hii\ regions which is only a factor of 3 higher, on average, for 
 individual \hii\ regions compared to the SNRs.
Similarly, the average hot gas mass is $\sim$1000~\msun,
per \hii\ region, based on the best-fitting 2-temperature spectral model.
This is about a factor of 3 larger than the hot gas mass in individual SNRs.

\begin{table*}
\begin{center}
\caption{\hii\ regions X-ray fitting results \label{tb:xhiif}}
\begin{tabular}{lrrrrr}

\tableline \tableline 
 Fit Parameters & {\tt apec} &  {\tt apec}  &  {\tt apec} & {\tt apec+apec} & {\tt apec+apec} \\ 
\tableline 

$N_{\rm H}$ (10$^{21}$cm$^{-2}$)      & 0.4                 & 1.4                    & 7.8$^{+1.5}_{-0.8}$   & 0.4                   & 0.4$^{+2.3}_{-0.0}$  \\
$T_{e_{1}}$ (MK)                & 3.5$^{+0.4}_{-0.3}$ & 3.0$^{+0.6}_{-0.2}$    & 1.1$^{+0.4}_{-0.2}$   & 1.8$^{+0.6}_{-0.5}$   & 1.8$^{+0.6}_{-0.5}$  \\
Normalization K$_{1}$ $^a$      & 0.7$^{+0.1}_{-0.1}$ & 1.5$^{+0.3}_{-0.4}$    & 2153$^{+33590}_{-1991}$  & 0.8$^{+0.1}_{-0.3}$ & 0.8$^{+4.7}_{-0.2}$ \\
$T_{e_{2}}$ (MK)                &                     &                        &                       & 8.8$^{+2.0}_{-1.2}$   & 8.8$^{+1.0}_{-1.4}$  \\
Normalization K$_{2}$           &                     &                        &                       & 0.3$^{+0.1}_{-0.1}$   & 0.3$^{+0.2}_{-0.1}$  \\  
$L_{\rm X}$/10$^{37}$ \ergl\    & 1.2$^{+0.2}_{-0.2}$ & 1.2$^{+0.2}_{-0.3}$    &  1.3$^{+0.5}_{-1.0}$  & 1.4$^{+0.2}_{-0.4}$   & 1.4$^{+0.1}_{-0.2}$  \\
$L_{\rm X}^{int}$/10$^{37}$ \ergl\ & 1.5              & 2.7                    & 365                   &  1.7                  & 1.7                  \\ 
  C-stat/bin                    &  154.0/108          & 151.9/108              &   129.4/108           & 120.3/108             & 120.3/108            \\
\tableline
 \multicolumn{6}{l}{Derived Parameters$^b$}  \\ 
\tableline
 $V$    [$f$] ($10^{63}$ cm$^{3}$)       & 2.2   & 2.2   & 2.2  & 2.2       & 2.2        \\
 $n_e$ [$f^{-1/2}$] (cm$^{-3}$)          & 0.02  & 0.03  & 1.1  & 0.07,0.01 & 0.07,0.01  \\
 $P/k$ [$f^{-1/2}$] (10$^5$ K cm$^{-3}$) & 1.4   & 1.9   &  40  & 2.5       & 2.5        \\
 $M_x$ [$f^{+1/2}$] (10$^5$ \msun)       &  0.5  & 0.8   &  28  & 0.5       & 0.5        \\
 $E_{th}$ [$f^{+1/2}$] (10$^{53}$ erg)   &  0.6  & 0.9   &  18  & 1.1       & 1.1        \\
 $t_c$ [$f^{+1/2}$] (Myr)                &  55   & 32   &  0.1 &  32        & 32        \\
\tableline
 \multicolumn{6}{l}{$^a$ $K=(10^{-9}/4\pi D^2) \int n_e n_p dV$} \\
\multicolumn{6}{l}{$^b$volume filling factor scaling in [ ] in Column 1.}\\
\end{tabular}
\end{center}
\end{table*}

\begin{deluxetable*}{rrrrr} 
\tablecolumns{5} 
\tablewidth{0pc} 
\tablecaption{\hii\ regions \label{tb:hii}} 
\tablehead{
\colhead{R.A.} & \colhead{Decl}   & \colhead{Count Rate $^{a}$}    & \colhead{$L_{{\rm x}}$ $^{b}$} &  \colhead{$L_{H\alpha}$} \\
\colhead{J2000} & \colhead{J2000}   & \colhead{10$^{-4}$ cts s}    & \colhead{10$^{35}$\ergl} &  \colhead{10$^{37}$\ergl} \\
}
\startdata 
 07 36 35.7  & +65 35  09.0   &    3.8  $\pm$  2.9  &    1.6 $\pm$   1.3 &  16.4   \\   
 07 36 38.0  & +65 36  09.5   &    1.6  $\pm$  1.5  &    0.7 $\pm$   0.6 &   2.3   \\   
 07 36 38.4  & +65 36  23.7   &    3.0  $\pm$  2.3  &    1.3 $\pm$   1.0 &   5.7   \\   
 07 36 39.6  & +65 37  11.5   &    0.3  $\pm$  0.9  &    0.1 $\pm$   0.4 &   2.2   \\   
 07 36 41.7  & +65 38  06.7   & $-$0.2  $\pm$  1.4  & $-$0.1 $\pm$   0.6 &  31.6   \\   
 07 36 44.6  & +65 35  07.7   &    3.8  $\pm$  2.4  &    1.6 $\pm$   1.0 &   6.1   \\   
 07 36 45.6  & +65 36  59.9   &    9.8  $\pm$  7.3  &    4.2 $\pm$   3.1 & 147.7   \\   
 07 36 45.7  & +65 36  37.4   &    0.8  $\pm$  1.6  &    0.4 $\pm$   0.7 &   0.9   \\   
 07 36 45.9  & +65 36  40.6   & $-$0.3  $\pm$  0.8  & $-$0.1 $\pm$   0.3 &   0.1   \\   
 07 36 45.9  & +65 37  14.6   & $-$0.8  $\pm$  1.9  & $-$0.3 $\pm$   0.8 &   3.5   \\   
 07 36 46.7  & +65 36  36.9   &    1.4  $\pm$  2.9  &    0.6 $\pm$   1.2 &   5.4   \\   
 07 36 46.9  & +65 37  05.3   & $-$1.7  $\pm$  2.5  & $-$0.7 $\pm$   1.1 &   3.6   \\   
 07 36 47.1  & +65 35  42.5   & $-$1.9  $\pm$  1.8  & $-$0.8 $\pm$   0.8 &   6.6   \\   
 07 36 47.2  & +65 36  44.8   & $-$0.8  $\pm$  1.7  & $-$0.3 $\pm$   0.7 &   2.3   \\   
 07 36 47.4  & +65 37  00.1   & $-$0.4  $\pm$  1.9  & $-$0.2 $\pm$   0.8 &   1.2   \\   
 07 36 47.6  & +65 35  53.5   &    1.1  $\pm$  1.8  &    0.5 $\pm$   0.8 &   2.5   \\   
 07 36 47.9  & +65 34  34.8   &    0.0  $\pm$  0.9  &    0.0 $\pm$   0.4 &   2.2   \\   
 07 36 48.1  & +65 33  26.3   &    0.0  $\pm$  1.6  &    0.0 $\pm$   0.7 &   7.0   \\   
 07 36 49.2  & +65 36  33.5   & $-$1.6  $\pm$  2.2  & $-$0.7 $\pm$   0.9 &   8.5   \\   
 07 36 49.3  & +65 36  52.9   & $-$3.8  $\pm$  1.4  & $-$1.6 $\pm$   0.6 &  14.3   \\   
 07 36 50.0  & +65 35  11.2   & $-$0.5  $\pm$  1.4  & $-$0.2 $\pm$   0.6 &   4.8   \\   
 07 36 50.4  & +65 35  28.6   & $-$1.6  $\pm$  1.7  & $-$0.7 $\pm$   0.7 &   2.8   \\   
 07 36 50.9  & +65 34  57.7   & $-$2.4  $\pm$  1.2  & $-$1.0 $\pm$   0.5 &   6.1   \\   
 07 36 52.3  & +65 36  47.3   &   74.1  $\pm$ 10.3  &   31.9 $\pm$   4.4 & 140.7   \\   
 07 36 52.9  & +65 36  20.7   & $-$2.1  $\pm$  1.2  & $-$0.9 $\pm$   0.5 &   3.5   \\   
 07 36 53.2  & +65 36  34.7   & $-$0.8  $\pm$  1.0  & $-$0.3 $\pm$   0.4 &   0.8   \\   
 07 36 53.2  & +65 34  14.4   & $-$1.1  $\pm$  1.0  & $-$0.5 $\pm$   0.4 &   3.8   \\   
 07 36 53.4  & +65 35  28.4   & $-$0.3  $\pm$  1.3  & $-$0.1 $\pm$   0.6 &   1.2   \\   
 07 36 54.5  & +65 35  47.8   & $-$1.6  $\pm$  1.7  & $-$0.7 $\pm$   0.7 &   4.1   \\   
 07 36 55.6  & +65 34  11.0   & $-$0.3  $\pm$  1.9  & $-$0.1 $\pm$   0.8 &   1.9   \\   
 07 36 57.8  & +65 37  24.5   &    2.7  $\pm$  4.4  &    1.1 $\pm$   1.9 &  67.1   \\   
 07 36 59.4  & +65 35  09.1   &    0.8  $\pm$  1.9  &    0.3 $\pm$   0.8 &   1.3   \\   
 07 36 60.0  & +65 35  06.8   & $-$2.4  $\pm$  1.4  & $-$1.1 $\pm$   0.6 &   2.1   \\   
 07 37 00.3  & +65 35  11.0   &    1.1  $\pm$  2.6  &    0.5 $\pm$   1.1 &   5.8   \\   
 07 37 00.8  & +65 35  25.2   &   11.5  $\pm$  4.3  &    5.0 $\pm$   1.9 &  15.7   \\   
 07 37 03.7  & +65 36  19.9   & $-$0.8  $\pm$  0.9  & $-$0.3 $\pm$   0.4 &   2.9   \\   
 07 37 05.8  & +65 35  52.3   &    1.1  $\pm$  1.5  &    0.5 $\pm$   0.7 &   3.9   \\   
 07 37 07.0  & +65 36  41.1   &  206.1  $\pm$ 16.6  &   88.8 $\pm$   7.2 & 448.2   \\   
 07 37 10.6  & +65 35  42.2   & $-$0.2  $\pm$  1.9  & $-$0.1 $\pm$   0.8 &   6.3   \\   
 07 37 11.1  & +65 34  36.7   & $-$0.8  $\pm$  1.0  & $-$0.3 $\pm$   0.4 &   3.0   \\   
 07 37 11.6  & +65 33  56.2   &    1.8  $\pm$  1.5  &    0.8 $\pm$   0.6 &   1.5   \\   
 07 37 11.7  & +65 33  59.5   &    0.9  $\pm$  1.6  &    0.4 $\pm$   0.7 &   2.4   \\   
 07 37 11.8  & +65 33  49.0   & $-$1.6  $\pm$  2.0  & $-$0.7 $\pm$   0.8 &   5.3   \\   
 07 37 14.0  & +65 36  10.0   & $-$1.9  $\pm$  1.1  & $-$0.8 $\pm$   0.5 &   1.7   \\   
 07 37 16.4  & +65 34  03.5   &    1.3  $\pm$  2.4  &    0.6 $\pm$   1.0 &   5.3   \\   
 07 37 18.6  & +65 35  41.7   &    1.2  $\pm$  2.5  &    0.5 $\pm$   1.1 &   8.2   \\   
 07 37 18.7  & +65 33  50.6   &    8.3  $\pm$  3.7  &    3.6 $\pm$   1.6 &  36.2   \\   
\enddata
\tablecomments{Units of right ascension are hours, minutes, and seconds, and units of declination are degrees, arcmintes, and arcseconds.} 
\tablenotetext{a}{Count rates for ASIC-S3, taking the 0.5 $-$ 2.0 keV energy band.}
\tablenotetext{b}{Luminosity in the 0.5 $-$ 2.0 keV energy band.  $L_{\rm X}$ is computed using
the two temperature model with variable $N_{\rm H}$ model listed in Table~\ref{tb:xhiif}. }
\end{deluxetable*}

\subsection{Residual Emission} \label{s:residual}

\begin{figure*}
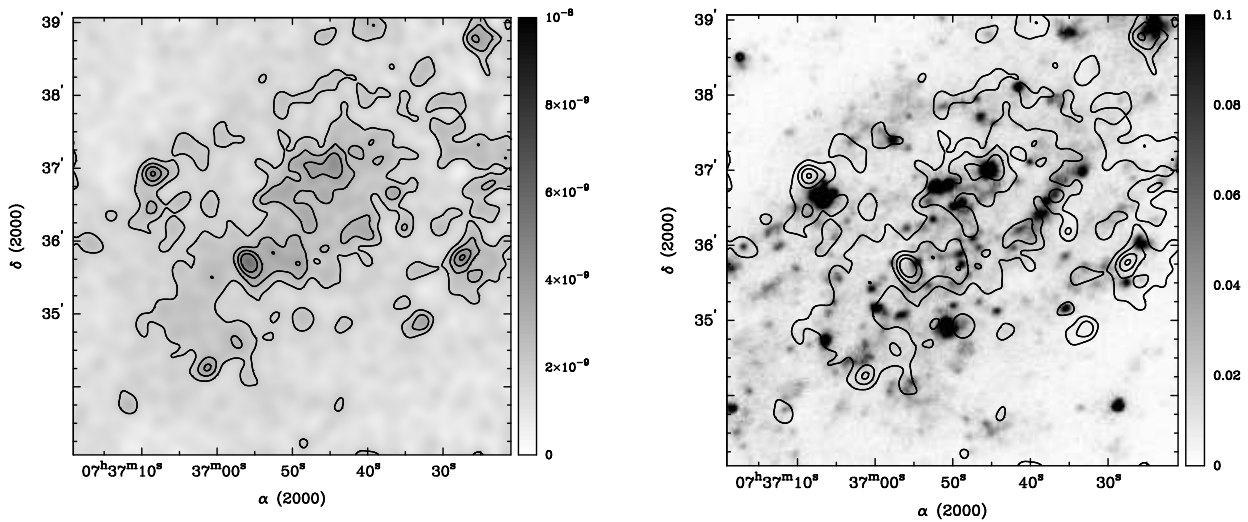

\begin{center}
\includegraphics[angle=-90,width=0.9\columnwidth]{f5b.eps}
\hspace{20pt}
\includegraphics[angle=-90,width=0.9\columnwidth]{f5c.eps}
\vspace{10pt}
\figcaption{The residual X-ray emission from \ngc.
Left:  The soft X-ray image (0.4$-$2.0 keV) of the central 6\arcmin$\times$6\arcmin\
 region. All four observations have been co-added and exposure-corrected
 and all discrete sources, optically-identified SNRs and \hii\ regions have been removed.
The image has been aggressively smoothed with the CIAO tool {\tt aconvolve}.  
The contours indicate 10, 20, 30, and 40$\sigma$ above the background level. The gray scale indicates intensity levels
in units of photon~cm$^{-2}$~s$^{-1}$~pixel$^{-1}$. One pixel corresponds to 0\farcs492.
Right: The X-ray residual emission contours overlaid on the
\galex\ FUV image. 
The gray scale of the image indicates intensity levels
in units of ct~s$^{-1}$~pixel$^{-2}$. One pixel corresponds to 1\farcs5.
\label{f:X_sm}}
\end{center}
\end{figure*}

\begin{table*}
\begin{center}
\caption{Residual emission X-ray fitting results \label{tb:xrf}}
\begin{tabular}{lrrrrr}
%
\tableline \tableline 
 Fit Parameters & {\tt apec} & {\tt apec}  & {\tt apec} & {\tt apec+apec} & {\tt apec+apec} \\ 
\tableline 
%
$N_{\rm H}$ (10$^{21}$cm$^{-2}$) & 0.4                  &1.4                  & 5.6$^{+1.0}_{-1.7}$   & 0.4                   & 5.2$^{+3.6}_{-1.0}$   \\
$T_{e_{1}}$ (MK)           & 3.1$^{+0.3}_{-0.2}$  &2.8$^{+0.2}_{-0.1}$  & 1.7$^{+0.2}_{-0.3}$   & 2.3$^{+0.4}_{-0.3}$   & 1.7$^{+0.1}_{-1.7}$   \\
Normalization K$_{1}$ $^a$ & 9.0$^{+0.9}_{-1.0}$  &17.2$^{+1.9}_{-1.8}$ & 820$^{+1716}_{-459}$  & 7.7$^{+1.3}_{-1.2}$   & 642$^{+2639}_{-432}$  \\
$T_{e_{2}}$ (MK)           &                      &                     &                       & 8.6$^{+1.3}_{-1.2}$   & 45.5$^{+697}_{-27.0}$ \\
Normalization K$_{2}$      &                      &                     &                       & 2.5$^{+0.6}_{-0.7}$   & 9.2$^{+13.8}_{-5.4}$  \\    
$L_{\rm X}$/10$^{38}$ \ergl\ & 1.3$^{+0.2}_{-0.2}$&1.3$^{+0.2}_{-0.2}$  &  1.5$^{+0.2}_{-1.1}$  & 1.5$^{+0.3}_{-0.3}$   & 1.7$^{+0.4}_{-0.9}$    \\ 
$L_{\rm X}^{int}$/10$^{38}$ \ergl\ & 1.6          &3.1                  &  70                   & 1.9                   & 56.9                   \\ 
  C-stat/bin               &  140.7/108           & 133.9/108           &   106.6/108           & 105.9/108             & 97.1/108               \\
\tableline
 \multicolumn{6}{l}{Derived Parameters$^b$}  \\ 
\tableline
 $V_{sphere} ^c$ [$f$] ($10^{63}$ cm$^{3}$)      & 1535 &  1535 &1535 &1535       & 1535       \\
 $n_e$ [$f^{-1/2}$] (cm$^{-3}$)         & 0.003  &  0.004  & 0.026 &  0.006,0.002 &  0.08,0.003  \\
 $P/k$ [$f^{-1/2}$] (10$^5$ K cm$^{-3}$)& 0.2    & 0.2     & 0.8   & 0.3          &  2.5         \\
 $M_x$ [$f^{+1/2}$] (10$^5$ \msun)      & 47     & 66      & 460   & 43           &  167         \\
 $E_{th}$ [$f^{+1/2}$] (10$^{53}$ erg)  & 51    &  67     &  270   & 90           &  830         \\
 $t_c$ [$f^{+1/2}$] (Myr)               & 311   &  203      & 7.3  & 342          &  30         \\
\tableline
 $V_{disk} ^d$    [$f$] ($10^{63}$ cm$^{3}$)    & 97 &  97 &97 &97       & 97       \\
 $n_e$ [$f^{-1/2}$] (cm$^{-3}$)         & 0.01  &  0.01  & 0.10  &  0.023,0.006 &  0.31,0.01  \\
 $P/k$ [$f^{-1/2}$] (10$^5$ K cm$^{-3}$)& 0.7   & 0.8    & 3.3   & 1.1          &  10         \\
 $M_x$ [$f^{+1/2}$] (10$^5$ \msun)      & 11   & 17     & 114   & 11           &  42          \\
 $E_{th}$ [$f^{+1/2}$] (10$^{53}$ erg)  & 12   & 17    &  66    & 23         & 211          \\
 $t_c$ [$f^{+1/2}$] (Myr)               & 78   & 51      & 1.8   & 86         &  7.5         \\
\tableline
\multicolumn{6}{l}{For two tempareture models, the average $n_e$ is estimated as $M_x / (\mu m_p V)$ and used in text.} \\
 \multicolumn{6}{l}{$^a$ $K=(10^{-9}/4\pi D^2) \int n_e n_p dV$} \\
 \multicolumn{6}{l}{$^b$Volume filling factor scaling in [ ] in Column 1.}\\
 \multicolumn{6}{l}{$^c$A 2.3~kpc radius spherical emission volume.}\\
 \multicolumn{6}{l}{$^d$A disk with the thickness 200pc and 2.3~kpc radius.}\\
\end{tabular}
\end{center}
\end{table*}

The left panel of
Figure~\ref{f:X_sm} displays a smoothed image of the soft X-ray emission
 from the underlying unresolved component.
This residual emission is
defined as the net emission left after masking out all detected point sources
(including SN~2004dj and TH2), \hii\ regions and SNRs. 
For spectral fitting, an X-ray map of this component was created 
using only the first \cxo\ observation. This is a compromise in that a deeper
image could be made using an image merged from two or more individual 
observations but the area of the region of overlap sampled by these multiple
observations is too small to define both a sizable 
source region and a surrounding 
background region that are both wholly contained within the overlap region
(see Figure~\ref{f:cxo_chip}).
The first \cxo\ observation placed the center of \ngc\ near the S3 aimpoint
and thus is the most useful for our purposes.
The source region is defined as a disk of radius 2\farcm5 (2.3~kpc) centered on \ngc.
The background is composed of three non-contiguous regions of low surface
brightness as shown in Figure~\ref{f:cxo_chip}. To construct the smoothed image
depicted in Figure~\ref{f:X_sm}, all excluded regions were filled with a local background level using the Poisson method of the CIAO tool 
{\tt dmfilth}\footnote{See http://cxc.harvard.edu/ciao/threads/diffuse\_emission/}, then divided by an 
exposure map evaluated at an energy of 0.5 keV. 
This exposure-corrected image was then smoothed 
using the CIAO tool {\tt aconvolve} using a Gaussian function with a width of 10 pixels ($\sim$ 5\arcsec).

There are several nearly point-like regions
that remain in the residual emission image. 
These may be faint point sources that
were hidden in the PSF wings of brighter nearby sources
or extended non-spherical features in \hii\ regions not covered by our
(circular) masks.
We examined the five brightest of 
these regions and find they contribute only 4\%
of the counts in the residual emission spectrum.
Therefore, these regions were left in as part of the residual emission.

In general, the residual 
emission fills a broad region of the central part of \ngc\
extending from
northwest of the nucleus (located at the center of the image)
to southeast of the nucleus. This region is not a particularly
strong source of emission at UV, mid-IR, nor \ha\ 
wavelengths. 
For example, a grayscale UV image is shown in the
 right panel of Figure~\ref{f:X_sm} with the X-ray contours overlaid.
This image suggests that much of the residual X-ray emission 
 is centrally located relative to the UV-bright zones that contain stars
 \LA~100 Myrs old. 

Several  models were fitted to the spectrum of this
 residual X-ray emission.  
Area-weighted ARFs and RMFs were created  for this purpose
using the CIAO script {\tt specextract}.  
The spectral fitting results are listed in Table~\ref{tb:xrf}.

\begin{figure}
\begin{center}
\includegraphics[angle=-90,width=\columnwidth]{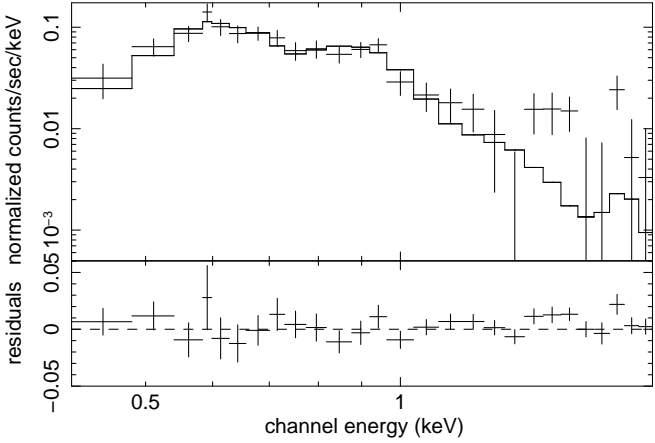}
\vspace{10pt}
\figcaption{
The X-ray spectrum of the residual emission from the central 
 6\arcmin$\times$6\arcmin\ region of \ngc.
Data from only the first observation are included. 
The best-fitting model, a two-temperature {\tt apec} model with 
 hydrogen column density fixed to the galactic value, and the fit residuals 
 are shown. 
The spectrum has been re-binned to have a width of 73 eV 
 for display purposes. 
\label{f:X_res}}
\end{center}
\end{figure}

The three thermal models with the absorption column density frozen during the 
 fitting all give low values for the model normalization,
 electron density, pressure, hot gas mass, and thermal energy.
The best-fitting of these models is the 2-temperature model with 
 temperatures of 2.3 and 8.6 MK with most of the flux in the cooler
 component (Table~\ref{tb:xrf}, column 5), shown in Figure~\ref{f:X_res}.
Fixing the  absorption column density to the Galactic value implies
 that the residual emission lies above the disk of \ngc\ where there
 is little or no overlying cold gas. 
This is a reasonable assumption if the X-ray-emitting gas was ejected
 from the galaxy through galactic-scale winds or smaller fountains
 or chimneys.
Fixing the absorption column density to the equivalent 
 mean value of the optical extinction toward 
  the active star-forming regions in \ngc, 
 $N_{\rm H}=1.4\times 10^{21}$~cm$^{-2}$,
 corresponds to a modest layer of overlying cold gas.

The models with the absorption column density free to vary during
 the fit result in statistically improved fits with
 absorbing columns higher than found for any of the other types of X-ray 
 sources considered in the previous subsections.
Such high columns imply that the hot
 gas is located behind a layer of neutral gas such as would be
 the case if the hot gas were confined to the disk. 
However, the observed H{\sc i} column densities through the 
central disk of \ngc\ range from only
$\sim$10$^{21}$~cm$^{-2}$ \citep{thornley95}
to $\sim$3$\times$10$^{21}$~cm$^{-2}$ \citep{schaap00,fraternali02a}  
which is considerably less than implied by these X-ray model fits,
suggesting the models are unphysical.
Furthermore, 
inspection of the images of this hot gas component and comparison to images
 at other wavelengths suggest that the hot gas is not emitted from
 regions with high levels of overlying cold gas.

Therefore, the models with the absorption column fixed are the
most realistic models.  
They imply $(4-7)\times 10^6$~\msun\ of hot gas is present in the
 central regions of \ngc, assuming a 2.3~kpc radius spherical emission volume.
This hot gas is rather tenuous, $n_e \sim 0.003$ to 0.004~cm$^{-3}$,
with, formally, a long cooling time of 200 $-$ 340~Myr. We note that the actual
X-ray emitting region is slightly smaller than the 2.3~kpc radius spectrum extracting
region, so that the assumed volume may be larger than the volume containing the hot gas.
If we assume instead that the hot gas is confined to 
 a disk with thickness 200~pc and 2.3~kpc radius, then the inferred density
increases by about a factor of 4 in compensation for the
smaller volume for a given emission integral.  This reduces
the cooling time to 50 $-$ 90~Myrs.  The true emission volume is probably somewhere
between our sphere and disk estimates.  Table~\ref{tb:xrf} contains derived properties
of the residual gas for both these geometries.

An estimate of the contribution to this residual X-ray emission from unresolved 
sources fainter than, but otherwise similar to, the discrete source population 
(\S\ref{s:pts}) can be made by
extrapolating the fitted point-source 
 luminosity function slope (Figure~\ref{f:X_pts})
to lower luminosities.
For this, we scale the XLF to include
only those sources within the 2\farcm5 residual emission region. 
Using the same absorbed bremsstrahlung model as before (\S~\ref{s:pts})
 results in an estimated unresolved point source contribution of 
 2$\times$10$^{37}$~\ergl\ in the 0.5 to 2.0~keV energy range 
 or about a 10--15\% contribution to the diffuse emission.
As an alternative estimate, we added a bremsstrahlung model component to 
 the hot gas spectral model to estimate the contribution 
from unresolved sources. 
Adding a bremsstrahlung component does not improve statistical significance
 of the fit but can contribute up to $\sim$20\% of 
the flux in the 0.5$-$2.0~keV band. 
The contributions from SNRs and \hii\ regions 
 to the residual X-ray emission can also be estimated from their XLFs
(Figures~\ref{f:X_snr}, and \ref{f:X_hii}).  These estimates are 
9$\times$10$^{36}$ \ergl\ for the SNRs 
and 1$\times$10$^{36}$ for the \hii\ regions. The 
estimated contribution from SNRs is relatively high because the XLF
slope is steep for this source population. 
In any event, faint SNRs and \hii\ regions can contribute only a 
small fraction to the total X-ray luminosity of the residual emission
 which is 1.6$\times$10$^{38}$~\ergl\ or higher.

\subsection{Discussion of Preliminary Analysis of X-ray Observations}

The bulk X-ray properties of the SNRs, \hii\ regions, and diffuse (residual)
 emission are summarized in Tables~\ref{tb:xsnrf}, \ref{tb:xhiif}, and \ref{tb:xrf}, respectively. 
They show that the hot gas densities in SNRs and \hii\ regions are 
comparable but that the SNR temperature is somewhat higher resulting in 
higher pressures, $P \propto n T_{\rm X}$, by about a factor of 2. 
This is not suprising since the SNR, which 
must have ages of order 10$^4$~yr or less \citep{matonick97},
are  expected to be further from pressure equilibrium with their surroundings
than the fully-developed \hii\ regions which have ages of
10$^6$ to $>$10$^7$~yr.

Another difference between the SNRs and \hii\ regions is the amount of 
hot gas estimated from the X-ray data. 
There is, on average, about 300$\pm$100~\msun\ of hot gas per SNR compared to
about 1000~\msun\ per \hii\ region. 
Of course, \hii\ regions are composed of many stars and SNs whereas the 
SNRs identified by \citet{matonick97} are morphologically consistent with
single events. 
Evidently, SNe occurring within the relatively tenuous and hot interior
 of \hii\ regions are much less efficient at producing hot gas than
 are isolated SNe.
The ratio of the hot gas mass to the swept up mass for SNRs is 
therefore $\sim$0.06$-$0.10
and for the \hii\ regions it is only $\sim$0.03 for 
an assumed ambient density of 1~H~cm$^{-3}$. 
This indicates, again, that
 shock heating and thermal evaporation from the swept-up shells
of young isolated SNRs is more efficient than in \hii\ regions.

The diffuse (residual) emission has a density and pressure lower 
 by about a factor of 10 (assuming the spherical volume) compared to SNRs and \hii\ regions
but a temperature that is comparable to both these sources.  If we assume a disk geometry,
then the density is closer to the density in SNRs and \hii\ regions.
There is another parameter to consider. The volume filling factor, $f$, 
may be lower for this residual emission 
compared to the SNRs and \hii\ regions because, by definition,
this residual shares the galaxy disk with the much colder neutral and
molecular gas and dust. 
Assuming $f\sim 1$ for the \hii\ regions and pressure equilibrium
suggests a filling factor $f\sim(0.2-0.3)$ for the residual emission with
correspondingly higher density and shorter cooling time.
Alternatively, we may assume pressure equilibrium between the hot
residual gas component and the colder gas and dust.
This cold gas resides primarily
in clouds of typical temperatures of $T\sim100$~K and 
densities of 60$-$100~cm$^{-3}$.
This corresponds to a typical pressure $P/k\sim10^4$~cm$^{-3}$~K which would
imply, at the observed temperature of $(2-3)\times10^6$~K,
 a hot gas density of $\sim$0.003$-$0.005~cm$^{-3}$ consistent 
with the derived value (Table~\ref{tb:xrf}) if $f\sim1$.
We note \citep{fraternali02a} that the mean H{\sc i} density in
the central regions of \ngc\ is about 0.4~cm$^{-3}$.
Thus its filling factor must be only 
 around 5\% for a cloud density of 80~cm$^{-3}$. 
A similar argument can be made for the H$_2$ content \citep{sheth05}.
Thus the filling factors for the cold gas may be quite small so that the bulk
of the volume in the central regions of \ngc\ may be filled with 
hot gas at high filling factor ($f\sim1$).
This, in turn, would imply a considerable overpressure in the \hii\ regions
 which is not wholly unreasonable.

In any case, the mass of hot gas in this residual component,
$M_{\rm X}\lesssim7\times10^6$~\msun, is a small fraction of the mass
contained in \hi\ (5$\times$10$^8$~\msun) and comparable to the 
 molecular (8$\times$10$^6$~\msun) gas mass.

We cannot tell with certainty where this X-ray emitting residual gas is 
located. 
There is not a strong spatial correlation between this gas 
 and star-formation indicators including \ha\ and UV emission.
Combining this fact with an estimated cooling time 
 of up to 340~Myr (depending on the geometry assumed and the filling factor)
 allows for several possibilities.
The residual gas may have
 been created {\sl in situ} during star-formation activity that ended
 up to 100~Myr ago (in regions currently lacking strong UV emission) 
 or it may have been produced more recently but
 has escaped from its place of origin either into the halo or into
 low-density regions of the disk.

At this point, we only have crude estimates of the ages of the 
 various star-forming regions that are present in the central regions of \ngc.
The ages of \hii\ regions are $\lesssim$10~Myr, 
 equivalent to the lifetimes of the least-massive LyC-producing stars. 
The ages of UV-emitting regions can be as old as 100~Myr.
What is needed is a better estimate of the age of individual 
star-forming regions. From this, we can better estimate 
how X-ray properties evolve through time
and so better constrain the origin of the residual X-ray emission~--~whether
 it traces recent star-formation activity or is a relic
of past activity. This is the subject of the next section.

\section{Individual Star-Forming Regions and their Analysis}\label{s:blobs}

We now turn to the analysis of individual young star-forming regions
 that have well-defined ages and masses.
We choose young regions because they are more likely
 sources of X-ray emission from hot gas than are older clusters;
 and have more likely retained their identity as a coherent (though not
 necessarily bound) and coeval system under the prevailing tidal forces.
Theoretically, X-ray emission from stellar winds and hot gas associated 
 with supernovae (SNe) occurs only within the first $\sim$40~Myr in the lifetime of a 
 star cluster corresponding to the longest-lived stars that become core
 collapse SNe, roughly the 8$-$10~\msun\ stars.
Star clusters in this age range emit strongly in UV light and 
 (during the first $\sim$10~Myr) in \ha\ line emission.
However, these star-formation indicators are sensitive to extinction
 by overlying dust.
Dust heated by UV and optical radiation emits this energy 
 in the mid-IR and longer wavelength bands.
Therefore, we define our regions as those bright in both 
 UV radiation from massive young
 stars and in mid-IR radiation from reprocessing by dust. 
In this way, we can better estimate the extinction
 along the line of sight using the mid-IR flux and thus correct the UV 
 emission to obtain a better estimate of the age and mass of the star clusters.

Star-formation tends to propagate spatially 
 by compressing surrounding 
 colder regions of the interstellar medium through the action of massive star
 winds and supernovae. 
Thus, small isolated regions 
 are more likely to have a single characteristic age than are large
 extended regions.
Therefore, we apply our source-finding algorithm to identify isolated 
 star-forming regions quantitatively rather than to rely on a pre-determined
 object size (such as a fixed aperture or spatial grid, for instance, which is
 known to sample a range of stellar ages; cf., \citeauthor{calzetti05} 2005).
The drawback to our selection method is that the flux from 
 individual regions may be weak at wavelengths other than the UV and mid-IR 
 we use to select them.

\subsection{Complementary Observations}

\begin{figure}
\begin{center}
\includegraphics[angle=-90,width=0.8\columnwidth]{f7a.eps}
\includegraphics[angle=-90,width=0.8\columnwidth]{f7b.eps}
\includegraphics[angle=-90,width=0.8\columnwidth]{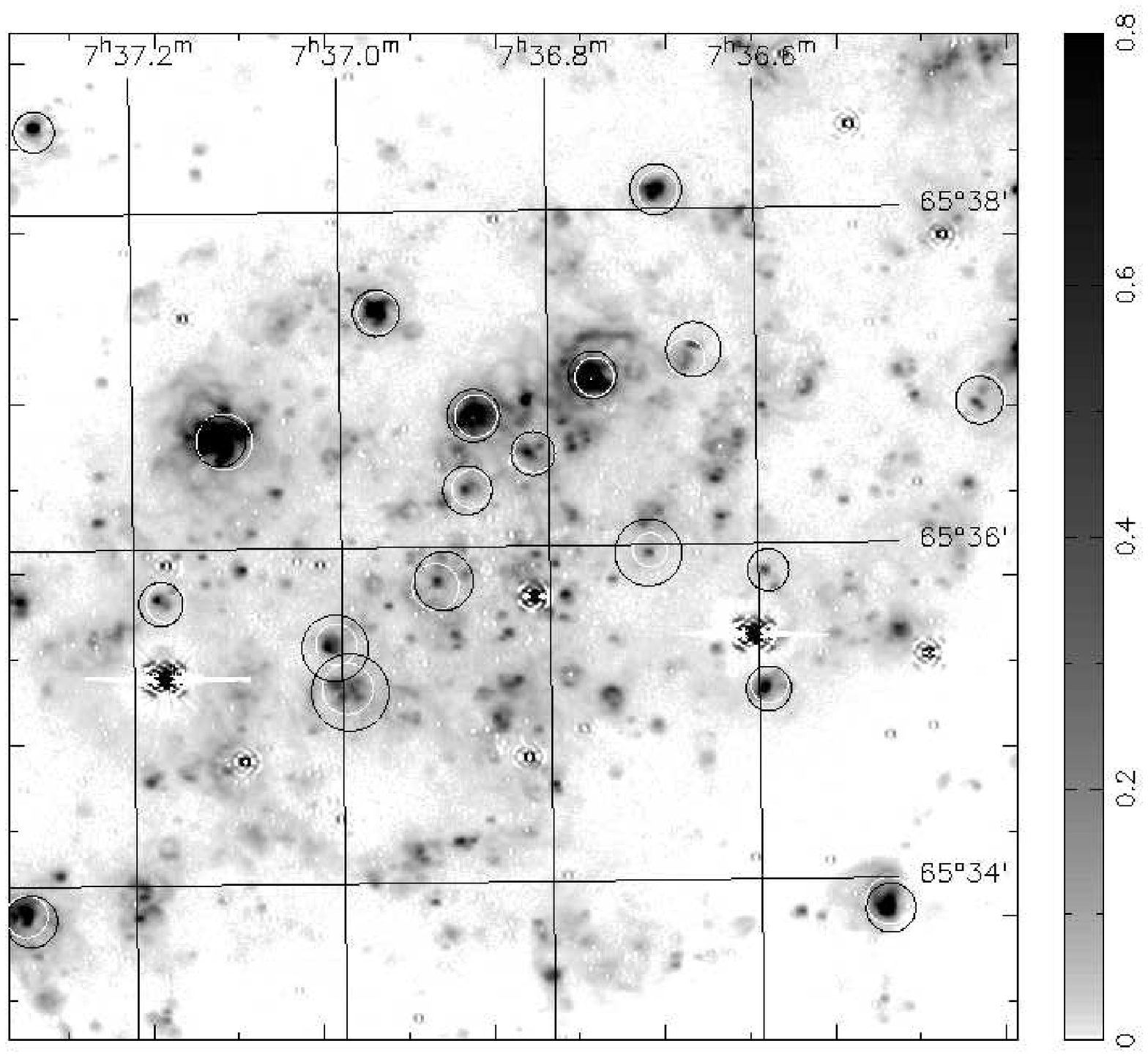}
\vspace{10pt}
\figcaption{Identification of star-forming regions in \ngc.
Top: The central 6\arcmin$\times$6\arcmin\ region of \ngc\ at
FUV (\galex\ $\lambda$1529 \AA). 
The positions of the 58 sources detected in the  FUV band are encircled in white.
The positions of the 47 sources detected at 24\,\um\ are encircled in black.  
The sizes of the circles indicate photometric apertures
 defined as the 3$\sigma$ width of a circular Gaussian model fitted to the 
 respective surface brightness distribution. The gray scale is the same as 
the right panel of Figure~\ref{f:X_sm}.
Middle: The same region at
 24\,\um\ (\sst). Symbols same as for the FUV image. The gray scale indicate intensity levels
in units  MJy~str$^{-1}$ for 24\,\um\ image.  One pixel corresponds to 1\farcs5/pixel.
Bottom:  The 19 sources common to both FUV and 24\,\um\ images overlaid on the 
 continuum-subtracted \ha\ image.  
Sixteen of the 19 sources were also selected as \hii\ regions 
 as described in \S~\ref{s:hii}.   
The gray scale indicate intensity level
in the units ct s$^{-1}$ pixels$^{-2}$, which corresponds to 2.1 
$\times 10^{-6}$ Jy pixels$^{-2}$.  The image has a pixel scale of 0\farcs304.
\label{f:allblobs}}
\end{center}
\end{figure}

We use \galex\ UV and \sst\ mid-IR measurements to identify young star-forming
 regions. 
We also use \ha, a traditional star-formation tracer, in our analysis
 to confirm a young age for the regions.


\galex\ observed \ngc\ on December 5, 2003 (Tilenum 5087) as part of the 
 Nearby Galaxy Survey \citep{gildepaz07}.  
Corrected intensity maps at FUV (1529 \AA\ central wavelength) and 
 NUV (2321 \AA) were obtained from the \galex\ 
 archive\footnote{http://galex.stsci.edu/GR2/}.  
The images have a pixel scale of 1.5\arcsec\ and a spatial resolution of about 
 5\arcsec.  
We applied the standard flux calibrations of 
 1.40$\times$10$^{-15}$ \ergcms\ cts$^{-1}$ \AA$^{-1}$ to the FUV 
 and 2.06 $\times$ 10$^{-16}$ \ergcms cts$^{-1}$ \AA$^{-1}$ to the NUV images. 


\ngc\ was observed with \sst\ IRAC and MIPS as a part of the 
 Spitzer Infrared Nearby Galaxies Survey (SINGS) legacy 
 program \citep{kennicutt03}.  
MIPS images at 24\,$\mu$m were taken on 2004 October 13 and 16 
 (key 5549568 and 5549824, respectively). 
The IRAC 3.6\,$\mu$m, and 4.5\,$\mu$m images were 
 taken on 2004 October 8 (key 5505792) and 12  (key 5505536). 
Final mosaiced images provided by the SINGS program were used here.
Details of their data reduction and calibration are reported in the 
SINGS data delivery paper\footnote{http://data.spitzer.caltech.edu/popular/sings/20070410\_enhanced\_v1/Documents/sings\_fifth\_delivery\_v2.pdf}.  
The 24\,\um\ image has an $\sim$5\arcsec\ resolution 
 which is comparable to \galex.  
The final mosaiced image has a pixel scale of 1.5\arcsec.  
Both the IRAC images have a spatial resolution of $\sim$2\arcsec\ and a 
 pixel scale of 0.75\arcsec.

As part of the SINGS project,
\ngc\ was observed with the Kitt Peak National Observatory 2.1m telescope in 
\ha\ bands on 2001 November 8.  
We obtained a continuum-subtracted \ha\ image, made by subtracting an
 $R$-band image from a narrow-filter image at \ha\
 (D. Calzetti, private communication). 
The pixel scale of the \ha\ image is 0.3\arcsec\ and the resolution is 
 about 1\arcsec.
The \ha\ flux is estimated from this image by assuming the [NII]/\ha\ ratio and 
 final \ha\ correction factor as given in \citet{prescott07}.

As with the X-ray data, SN~2004dj was used to define a common registration among the data sets. 
Unsaturated, foreground stars were used to  register pre-supernova ground-based, and \galex\ images.

\subsection{Star-forming Region Definitions}

The same source finding tool applied to the X-ray images 
 was used to identify candidate star-forming regions in the FUV and 24\,\um\ images. 
This tool is optimized for detecting point sources but can be
 applied to moderately-extended sources by increasing the characteristic size of 
 the model PSF parameter.
We used a circular Gaussian PSF in the application because
 most of the star-forming regions appeared centrally-peaked at
 the moderate resolution of the 24\,\um\ and FUV images.  
We examined their radial profiles {\it post facto} using the higher
 resolution \ha\ images to confirm that they are basically extended
 and either centrally-peaked or centrally-cratered so that the Gaussian model is
 adequate for estimating both the emission centroids and the 
 spatial extent.

We detected 58 bright sources in the central 6\arcmin$\times$6\arcmin\
 FUV image (Figure~\ref{f:allblobs}, top panel) 
 defined as those with a $S/N$ above 10 and with a minimum of
 15 source counts per unit uncertainty in the background.  
Similarly, 47 bright sources are detected in the 24\,\um\ image 
 (Figure~\ref{f:allblobs}, middle panel) with a $S/N$ above 50 and 
 with 100 or more electrons per unit uncertainty in the background. 
There are 19 sources common to both the FUV and 24\,\um\ images.  These
19 regions are shown in Figure~\ref{f:allblobs} (bottom panel) overlaid on
the \ha\ image.
(Sixteen of these regions were selected as \hii\ regions based on
their \ha\ brightness in \S\ref{s:hii}.)

This constitutes our star-forming region sample.
The sample is only representative of the star-forming regions in \ngc;
a larger sample could be made by including regions of lower $S/N$.
Note (Figure~\ref{f:allblobs}) that there are sources brighter than
these 19 in one or the other of the two selection wavebands, but not in both.
Examining all available images, none of the sample regions have properties 
of foreground stars or of background AGN.  

\begin{figure*}
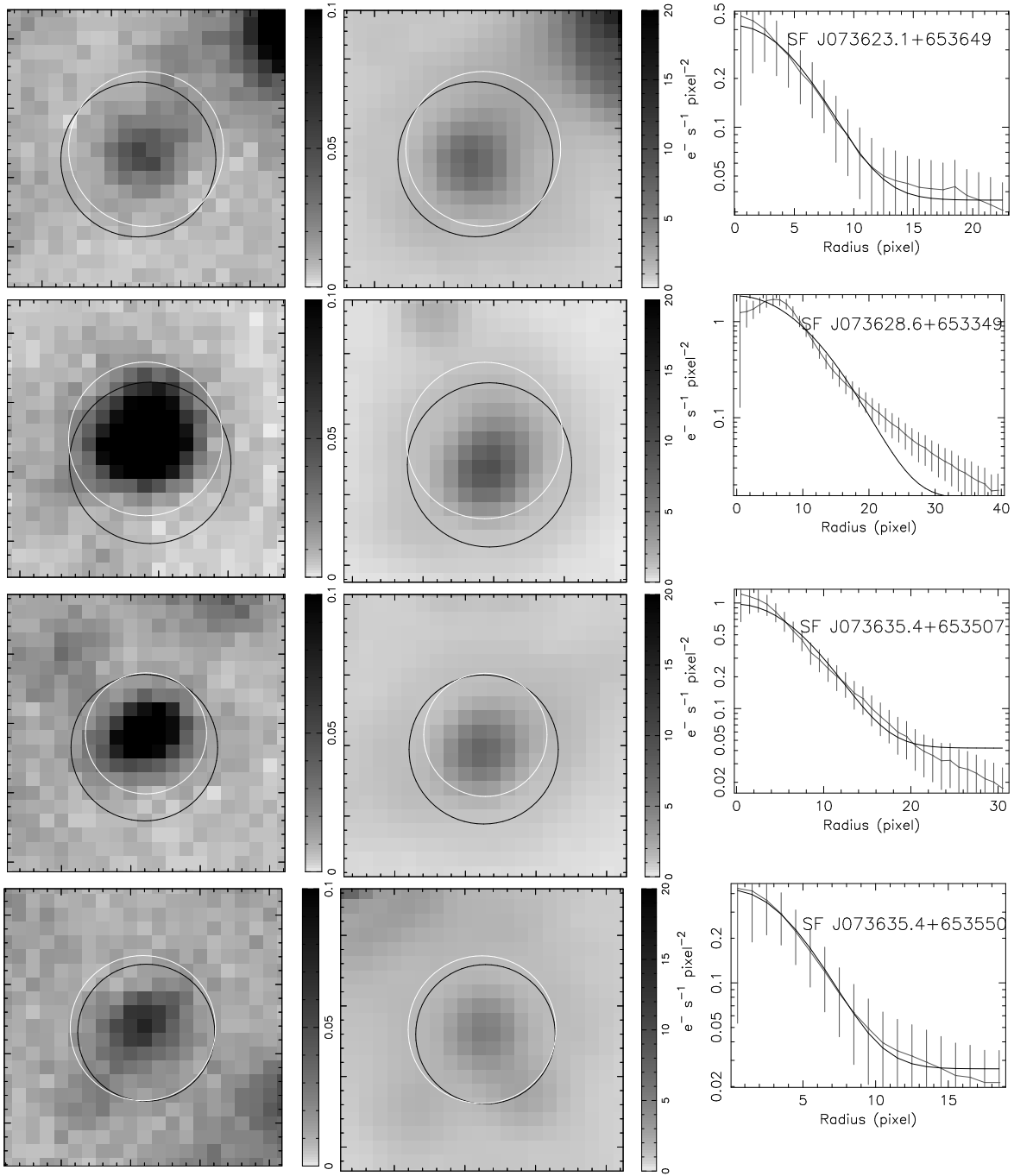

\begin{center}
\includegraphics[angle=-90,width=0.59\columnwidth]{f8al.eps}
\includegraphics[angle=-90,width=0.59\columnwidth]{f8am.eps}
\includegraphics[angle=-90,width=0.59\columnwidth]{f8ar.eps}

\includegraphics[angle=-90,width=0.59\columnwidth]{f8bl.eps}
\includegraphics[angle=-90,width=0.59\columnwidth]{f8bm.eps}
\includegraphics[angle=-90,width=0.59\columnwidth]{f8br.eps}

\includegraphics[angle=-90,width=0.59\columnwidth]{f8cl.eps} 
\includegraphics[angle=-90,width=0.59\columnwidth]{f8cm.eps} 
\includegraphics[angle=-90,width=0.59\columnwidth]{f8cr.eps} 

\includegraphics[angle=-90,width=0.59\columnwidth]{f8dl.eps}
\includegraphics[angle=-90,width=0.59\columnwidth]{f8dm.eps}
\includegraphics[angle=-90,width=0.59\columnwidth]{f8dr.eps}
\vspace{10pt}
\figcaption{
30\arcsec$\times$30\arcsec\ close-up views of the 19 star-forming regions. 
FUV images are on the left and 24\,\um\ images are in the middle.
Circle locations, sizes, colors, and gray scales are the same as in Figure~\ref{f:allblobs}.
On the right are higher-resolution \ha\ radial profiles 
 shown with best-fitting Gaussian model curves.  
Each radial profile extends to 5 Gaussian widths.  One pixel corresponds to 0.\arcsec304.
(First four are shown here.  The rest of the 15 regions are shown at the end of the  
 paper.)
\label{f:19blobs}}
\end{center}
\end{figure*}

The observed luminosities and angular sizes of the 19 regions are listed in Table~\ref{tb:19obsdata}.  We performed aperture photometry on the individual sources with corresponding aperture sizes for each region and waveband.  Although these 19 sources are extended, in order to define sizes of aperture, we apply the circular Gaussian model. 
The radii in each region and each band are determined as 3$\sigma$  circular Gaussian widths, which contains 99\% of flux of the sources.
For most regions, a surrounding annulus that extends to twice the source's aperture radius was used for background.
In more crowded regions, a non-contiguous nearby region was used instead.
Luminosities are computed using definitions similar to those used by the 
 SINGS team \citep[e.g.,][]{calzetti05};
 specifically, $L_{band}=\lambda L_{\lambda,band}$ for broad bands and  
  $L_{band}=\delta\lambda L_{\lambda,band}$ for narrow bands. 
Luminosities were corrected for the Galactic extinction along the line of sight, 
$E_{B-V} = 0.040$~mag \citep{schlegel98}, with the \citet{cardelli89} dust model. The aperture correction values listed on the \sst\ web page were applied to the 24\,\um\ data.  
Thumbnail images in the FUV and 24\,\um\ bands are shown in Figure~\ref{f:19blobs}.
The circles shown in these images define the adopted radii in each band.
Note that they are rarely concentric.
Also shown in Figure~\ref{f:19blobs} are the \ha\ radial profiles for each region.

The sizes of the regions in the FUV band are systematically smaller
 than in the \ha\ and 24\,\um\ bands.
The natural interpretation is that the FUV originates from young star clusters
whereas the source of \ha\ and 24\,\um\ is re-radiation from
surrounding warm ionized gas and cold dust regions, respectively, that lie at
the outskirts of wind-blown bubbles created by the stars.
This is also the reason why the regions are not concentric in the different 
 wavebands.
The \ha\ radial profiles, Figure~\ref{f:19blobs}, which often show a crater-like
morphology is also indicative of this shell-like structure.

\subsection{Age, Mass, \& Extinction Determinations} \label{s:seds}

\begin{deluxetable*}{rrrrrrrrrrr} 
\tabletypesize{\footnotesize}
\tablecolumns{11} 
\tablewidth{0pc} 
\tablecaption{Luminosities and sizes of 19 star-forming regions \label{tb:19obsdata}} 
\tablehead{
\colhead{SRC} & \colhead{$L_{FUV}$} & \colhead{R$_{FUV}$} & \colhead{$L_{NUV}$} &  \colhead{$L_{3.6\mu m}$} & \colhead{$L_{4.5\mu m}$} & \colhead{ $L_{24\mu m}$} & \colhead{R$_{24\mu m}$} & \colhead{$L_{H\alpha}$} & \colhead{R$_{H\alpha}$} & \colhead{$L_{\rm X}^{obs}$}\\
\colhead{} & \colhead{10$^{39}$\ergl}   & \colhead{arcsec}  & \colhead{10$^{39}$\ergl} &  \colhead{10$^{38}$\ergl}& \colhead{10$^{38}$\ergl}& \colhead{10$^{39}$\ergl}& \colhead{arcsec} &  \colhead{10$^{38}$\ergl} & \colhead{arcsec} & \colhead{10$^{35}$\ergl}
}
\startdata 
SF J073623.1+653649 &    4.8 $\pm$  0.3 &  8.4 &     2.9 $\pm$  0.1 &   4.1 $\pm$ 0.2  &    2.7 $\pm$  0.1 &  4.1 $\pm$   0.1 &  8.4 &  0.8  &  8.5 &   9.7 $\pm$   4.4 \\
SF J073628.6+653349 &   30.2 $\pm$  1.5 &  8.3 &    20.4 $\pm$  0.6 &   5.6 $\pm$ 0.2  &    4.8 $\pm$  0.1 &  4.9 $\pm$   0.1 &  8.7 &  7.4  &  8.8 &   8.9 $\pm$   4.4 \\
SF J073635.4+653507 &   10.0 $\pm$  0.5 &  6.5 &     5.6 $\pm$  0.2 &   5.7 $\pm$ 0.2  &    3.3 $\pm$  0.1 &  2.3 $\pm$   0.1 &  7.9 &  3.0  &  8.0 &   8.5 $\pm$   4.2 \\
SF J073635.4+653550 &    4.9 $\pm$  0.3 &  7.8 &     2.8 $\pm$  0.1 &   3.2 $\pm$ 0.2  &    2.0 $\pm$  0.1 &  1.9 $\pm$   0.0 &  7.4 &  0.7  &  7.5 &   1.2 $\pm$   2.9 \\
SF J073639.5+653709 &    5.4 $\pm$  0.3 &  6.7 &     3.7 $\pm$  0.1 &   2.9 $\pm$ 0.2  &    1.8 $\pm$  0.1 &  3.1 $\pm$   0.1 &  9.7 &  1.2  &  9.8 &   7.7 $\pm$   4.6 \\
SF J073641.6+653806 &   18.6 $\pm$  1.0 &  7.0 &    10.6 $\pm$  0.3 &   6.3 $\pm$ 0.2  &    4.7 $\pm$  0.2 &  5.0 $\pm$   0.1 &  9.1 &  4.5  &  9.2 &   7.1 $\pm$   4.3 \\
SF J073642.3+653557 &    4.3 $\pm$  0.3 &  6.0 &     3.3 $\pm$  0.1 &   3.7 $\pm$ 0.3  &    2.3 $\pm$  0.2 &  4.7 $\pm$   0.1 & 11.8 &  1.0  & 11.9 &  21.6 $\pm$   6.5 \\
SF J073645.4+653700 &  148.9 $\pm$  7.5 &  6.8 &   106.5 $\pm$  3.2 &  30.8 $\pm$ 0.6  &   22.7 $\pm$  0.4 & 43.9 $\pm$   0.9 &  8.5 & 16.7  &  8.6 &  24.8 $\pm$   6.3 \\
SF J073648.9+653633 &   23.1 $\pm$  1.2 &  7.3 &    19.8 $\pm$  0.6 &   8.3 $\pm$ 0.4  &    7.2 $\pm$  0.2 &  5.2 $\pm$   0.1 &  7.7 &  1.5  &  7.8 &   9.0 $\pm$   4.2 \\
SF J073652.3+653646 &   69.6 $\pm$  3.5 &  7.5 &    53.3 $\pm$  1.6 &  19.7 $\pm$ 0.5  &   13.9 $\pm$  0.3 & 36.4 $\pm$   0.7 &  9.3 & 15.2  &  9.4 &  63.0 $\pm$   9.2 \\
SF J073652.7+653620 &    7.5 $\pm$  0.4 &  6.7 &     5.9 $\pm$  0.2 &   2.6 $\pm$ 0.4  &    0.8 $\pm$  0.1 &  3.7 $\pm$   0.1 &  8.7 &  1.1  &  8.8 &   7.9 $\pm$   4.3 \\
SF J073654.1+653547 &   13.1 $\pm$  0.7 &  8.2 &    12.0 $\pm$  0.4 &   3.7 $\pm$ 0.4  &    2.5 $\pm$  0.2 & 10.9 $\pm$   0.2 & 10.4 &  2.0  & 10.5 &  24.1 $\pm$   6.5 \\
SF J073657.8+653723 &   16.3 $\pm$  0.8 &  7.3 &    11.2 $\pm$  0.3 &  21.3 $\pm$ 0.5  &   14.5 $\pm$  0.3 & 18.2 $\pm$   0.4 &  8.2 &  7.8  &  8.4 &  11.8 $\pm$   5.0 \\
SF J073659.6+653508 &   22.6 $\pm$  1.2 &  9.0 &    16.3 $\pm$  0.5 &  14.0 $\pm$ 0.5  &    9.8 $\pm$  0.3 &  9.5 $\pm$   0.2 & 13.6 &  4.0  & 13.8 &  26.7 $\pm$   7.8 \\
SF J073700.4+653524 &    9.0 $\pm$  0.5 &  7.3 &     7.3 $\pm$  0.2 &   2.1 $\pm$ 0.2  &    2.0 $\pm$  0.1 &  4.7 $\pm$   0.1 & 11.7 &  3.0  & 11.9 &  21.9 $\pm$   6.5 \\
SF J073706.6+653638 &  150.9 $\pm$  7.6 &  9.9 &   107.5 $\pm$  3.2 &  81.9 $\pm$ 1.4  &   61.2 $\pm$  1.0 & 168.1 $\pm$  3.4 &  9.0 & 35.7  &  9.1 &  98.0 $\pm$  11.2 \\
SF J073710.4+653540 &    3.9 $\pm$  0.3 &  6.9 &     5.2 $\pm$  0.2 &   3.5 $\pm$ 0.2  &    2.5 $\pm$  0.1 &  4.3 $\pm$   0.1 &  7.8 &  1.2  &  7.8 &   0.9 $\pm$   2.9 \\
SF J073717.4+653829 &   18.7 $\pm$  1.0 &  6.8 &    12.3 $\pm$  0.4 &   2.1 $\pm$ 0.1  &    2.0 $\pm$  0.1 &  4.2 $\pm$   0.1 &  7.2 &  3.2  &  7.3 &   0.5 $\pm$   3.0 \\
SF J073718.0+653347 &   16.9 $\pm$  0.9 &  7.4 &    11.1 $\pm$  0.3 &   5.4 $\pm$ 0.2  &    9.4 $\pm$  0.2 & 13.1 $\pm$   0.3 &  9.1 &  6.7  &  9.2 &   5.2 $\pm$   4.4 \\
\enddata
\tablecomments{FUV, NUV, and \ha\ luminosities are Galactic exinction corrected. }
\end{deluxetable*} 

\begin{deluxetable*}{rrrrrrrrrrr} 
\tabletypesize{\footnotesize}
\tablecolumns{9} 
\tablewidth{0pc} 
\tablecaption{Derived values of 19 star-forming regions \label{tb:19derived}} 
\tablehead{
\colhead{SRC} & \colhead{Mass} & \colhead{Age} & \colhead{$A_{\rm v}$} &  \colhead{ $N_{\rm H}$} & \colhead{$L_{\rm w}$} & \colhead{$L_{FUV}^{int}$} & \colhead{$L_{{\rm H\alpha}}^{int}$}  & \colhead{$L_{\rm X}^{int}$} \\
\colhead{} & \colhead{10$^{5}$ \msun}   & \colhead{Myr}  & \colhead{} &  \colhead{10$^{21}$ cm$^{-1}$}& \colhead{10$^{38}$\ergl}& \colhead{10$^{39}$\ergl}& \colhead{10$^{38}$\ergl} &  \colhead{10$^{35}$\ergl} 
}
\startdata
SF J073623.1+653649 &  0.4  & 9  & 0.67 & 1.7  &  5.2  &    9.8 $\pm$   0.3 &  1.3   &  23.5 $\pm$  10.7    \\
SF J073628.6+653349 &  0.7  & 8  & 0.22 & 0.8  &  9.9  &    33.8 $\pm$  0.3 &  8.6   &  13.7 $\pm$   6.8    \\
SF J073635.4+653507 &  0.8  & 16 & 0.32 & 1.0  &  7.5  &    12.8 $\pm$  0.2 &  3.7   &  14.6 $\pm$   7.2    \\
SF J073635.4+653550 &  0.3  & 11 & 0.44 & 1.2  &  3.5  &    7.4 $\pm$   0.2 &  0.9   &   2.2 $\pm$   5.6    \\
SF J073639.5+653709 &  0.3  & 9  & 0.5  & 1.4  &  3.9  &    8.8 $\pm$   0.2 &  1.7   &  16.1 $\pm$   9.7    \\
SF J073641.6+653806 &  0.7  & 9  & 0.35 & 1.1  &  9.0  &    24.8 $\pm$  0.3 &  5.7   &  12.7 $\pm$   7.8    \\
SF J073642.3+653557 &  0.4  & 9  & 0.65 & 1.6  &  5.2  &    08.5 $\pm$  0.3 &  1.6   &  50.2 $\pm$  15.1    \\
SF J073645.4+653700 &  2.4  & 1  & 0.3  & 1.0  &  26.7 &   185.6 $\pm$  0.8 &  20.6  &  49.2 $\pm$  12.5    \\
SF J073648.9+653633 &  0.9  & 9  & 0.25 & 0.9  &  11.6 &    26.9 $\pm$  0.3 &  1.8   &  14.6 $\pm$   6.9    \\
SF J073652.3+653646 &  1.5  & 1  & 0.44 & 1.2  &  16.7 &   104.7 $\pm$  0.6 &  20.8  & 142.5 $\pm$  20.8    \\
SF J073652.7+653620 &  0.3  & 8  & 0.44 & 1.2  &  4.2   &    11.3 $\pm$ 0.3 &  1.4   &  15.0 $\pm$   8.2    \\
SF J073654.1+653547 &  0.3  & 2  & 0.54 & 1.4  &  4.7   &    22.5 $\pm$ 0.4 &  2.9   &  61.8 $\pm$  16.7    \\
SF J073657.8+653723 &  17.5 & 60 & 0.67 & 1.7  &  0.0  &    33.2 $\pm$  0.5 &  12.7  &  27.5 $\pm$  11.5     \\
SF J073659.6+653508 &  10   & 50 & 0.4  & 1.2  &  0.0  &    32.2 $\pm$  0.4 &  5.4   &  49.5 $\pm$  14.4    \\
SF J073700.4+653524 &  0.3  & 7  & 0.46 & 1.3  &  4.7   &    13.8 $\pm$ 0.3 &  4.2   &  43.8 $\pm$  13.0    \\
SF J073706.6+653638 &  6.4  & 1  & 0.71 & 1.7  &  71.2 &   325.2  $\pm$ 1.3 &  59.3  & 302.5  $\pm$ 34.5    \\
SF J073710.4+653540 &  0.4  & 9  & 0.57 & 1.5  &  5.2  &    6.9 $\pm$   0.3 &  1.8   &   2.1 $\pm$   6.5    \\
SF J073717.4+653829 &  0.2  & 2  & 0.27 & 0.9  &  3.1  &    22.4 $\pm$  0.3 &  3.9   &   0.8 $\pm$   5.7    \\
SF J073718.0+653347 &  0.4  & 1  & 0.56 & 1.5  &  4.5  &    29.7 $\pm$  0.4 &  9.9   &  14.1 $\pm$  11.9    \\
\enddata
\tablecomments{ ~Listed $N_{\rm H}$ values include a Galactic component of $N_{\rm H}$ $=$ 0.4 $\times$ 10$^{21}$ cm$^{-2}$.}
\end{deluxetable*}

The five wavelength bands FUV, NUV 
 (both from \galex), 3.6\,$\mu$m, 4.5\,$\mu$m, and 24\,$\mu$m
 (all from \sst) were used to define the spectral energy distributions 
 (SEDs) of light from the individual regions. 
These were compared to  theoretical spectra of instantaneous 
 starbursts based on the models of \citet{leitherer99} as improved by 
 \citet{vazquez05} by convolving these spectra with the spectral 
 response functions of the appropriate filters.
For computing the theoretical spectra, 
 we assumed the \citet{kroupa01} initial mass function (IMF), 
namely a power-law IMF on the range 
 0.1$-$100 \msun\ with a break at 0.5~\msun\ and slopes of 1.3 below and 2.3
 above the break. 
A solar metallicity was assumed;
 consistent with estimates from the literature \citep[e.g.,][]{martin96,garnett97}. 
We accounted approximately for the wavelength-dependent attenuation by dust by 
 using the starburst dust model of \citet{calzetti01}. 
We modeled the re-emission of this radiation in the mid-IR by assuming 
 energy conservation and a blackbody emission profile at a temperature of 75~K.
This temperature was estimated by fitting a blackbody curve to the \sst\ MIPS 24, 70, and 160\,$\mu$m measurements from the brightest of our star-forming regions.   
(This region is fully resolved and isolated even in the 160$\mu$m image
 so that a distinct local background could be identified.)
Varying the blackbody temperature by $\sim$30\% has little impact on
 our results. 
However, much higher or lower temperatures result in unrealistic estimates
 of the age and mass of several of the star-forming regions.

By using five-band SEDs we are thus able to sample the UV region which 
 is most sensitive to age in young star clusters yet is most affected by 
 dust attenuation, the near-IR at 3.6 and 4.5\,$\mu$m that is least affected 
 by dust and hence is sensitive primarily to cluster mass,
 and the mid-IR 24\,$\mu$m which lies within the dust emission band 
 that extends from several microns out to much longer wavelengths. 
The 24\,$\mu$m waveband is longward of the PAH emission 
 features which greatly complicate the dust emission spectrum 
  \citep[e.g.,][]{perez06} 
 but shortward of the lower spatial resolution long wavelength \sst\ bands.

We used a least-squares fit of the model SEDs to the observed 
 luminosities using equal weights for each spectral band. 
Operationally, 
 the model extinction and cluster mass parameters
 were allowed to vary, at a fixed cluster age, until a 
 minimum $\chi^2$ was found.
We then repeated this process for discrete cluster ages 
 ranging from 1 to 200~Myr (in steps of 1~Myr for ages $\le$20~Myr
 and steps of 10~Myr otherwise)
 to obtain the overall best-fitting age, mass, and 
 extinction combination.
That is, we do not interpolate the model spectra
 to intermediate ages.
 
The results are summarized in Table~\ref{tb:19derived}. 
In addition to the best-fitting
 mass, age, and extinction values based on the SEDs fitting,
 Table~\ref{tb:19derived} also includes the corresponding hydrogen column density,
 $N_{\rm H}=(A_V/R_V)5.8\times10^{21}$~cm$^{-2}$, assuming an
 optical parameter $R_V=3.1$, 
 and the corresponding observed \ha, and FUV luminosities after
 correcting for extinction.

Among the most massive young clusters in our sample 
 are the giant \hii\ regions NGC2403-I and NGC2403-II 
 studied by \citet{drissen99}.
\citeauthor{drissen99} estimate the ages of these clusters as between
 2 and 6~Myr based on their detection of Wolf-Rayet stars in these regions.
A crude estimate of the initial cluster mass can be made from
 the number of WR stars detected. 
This estimate is very sensitive to the upper mass cutoff assumed of the
 initial mass function and, of course, to the true age of the cluster.
Interestingly, the age and mass values obtained from the studies of 
 \citeauthor{drissen99} are consistent with our findings and 
 independently validates our SED fitting method.
 \citet{garnett99} estimated line-of-sight extinction to several
 of our sample star-forming regions from Balmer line ratios.  
Our estimates are in agreement with these results within $A_V$ of 0.3 mag.  
For example, we derive $A_V=0.7$~mag
 for the brightest \hii\ region, compared to $A_V=0.6$~mag from Balmer line ratios
 \citep{garnett99} and to $A_V=0.9$~mag from extinction toward blue giants  
 in the region \citep{drissen99}.

The best-fitting ages of the regions fall into three distinct
 ranges, $<$3~Myr, from 7 to 16~Myr, and $>$50~Myr which we
 designate young, intermediate, and old, respectively.  
We believe the two regions designated as old 
 are likely younger than their best-fitting ages.
Inspection of the distribution of the $\chi^2$ fit statistic shows, 
 for these two old regions,
 that there is a second local minimum at 13~Myr which is probably 
 a more realistic age estimate.
Note that they have moderately high FUV luminosities, by selection, 
 which, for 
 their fitted age, forces their best-fitting mass to be very high 
 relative to the other groups. There is not a similar trend in the 
 fitted extinction values.

There are distinct differences among the members of the 
 remaining two age groups.
Members of the young group are typically much more luminous,
 per unit mass, at UV, \ha, and 24\,\um\ compared to the 
 intermediate age group members.
Based on the star cluster model, the UV luminosity is expected to
 rise slightly during the first 4~Myr then drop rapidly as the most massive
 stars become supernovae.
Similarly, the \ha\ luminosity, which comes from recombination of 
 circumstellar gas photoionized by massive stars, also drops quickly
as these massive stars disappear.  
The observed trends are consistent with this scenario.

Dust emission, accounting for most of the 24\,\um\ luminosity, also 
 represents re-radiation of starlight. 
It is most sensitive to UV but also responds to longer wavelength light.
One would expect, therefore, that the 24\,\um\ luminosity per unit mass
 would decrease more slowly than would the UV and \ha\ luminosities.
However, as the cluster evolves and its wind-driven bubble expands,
 surrounding dust clouds will evaporate and   
 the solid angle subtended by remaining dust clouds will decrease. 
This could account for the rapid drop with age observed in the 24\,\um\ 
 luminosity.

These trends do not apply to the near-IR 3.6 and 4.5\,\um\ luminosities.
Light at these wavelengths comes directly from stars of practically
all masses (ages) and is not strongly affected by dust. Thus, the 
near-IR luminosity per unit mass is nearly independent of the age of 
the underlying star cluster; as observed.

One puzzling result is that the young star forming regions in our
 sample are much more massive than the intermediate-age regions.
The average mass of the young regions is 1.9$\times$10$^5$~\msun\ 
 and only 0.5$\times$10$^5$~\msun\ for the intermediate-age regions.
Notably, there is one region that is 3 times as massive as any other.
Excluding this region, the average mass of the young group is still
 twice that of the intermediate group.
This may be an indication of a selection bias. 
If more massive star clusters more efficiently destroy or 
disperse surrounding dust clouds, then our selection criterion
 that requires regions be bright in both UV and 24\,\um\ would
tend to select against more massive clusters. 
Naturally, the destruction of dust takes time so this bias may
 only work against massive intermediate-age star-forming regions.
Comparison of the two panels of Figure~\ref{f:allblobs} shows that
 there are several regions bright in UV that lack a strong IR counterpart.
In particular, there is a very 
 UV-bright region just below the center of the field
at $L_{\rm FUV}^{\rm int} \sim 7\times 10^{40}$~\ergl.
This cluster was not selected because it lacks strong
dust emission. It is likely of intermediate age as it
also is weak in \ha\ emission. 
Assuming an age of 10-15~Myr suggests a mass in the 
range of about 2$\times$10$^5$ to 4$\times$10$^5$
based on the values of $L_{\rm FUV}^{\rm int}$ for
the clusters listed in
Table~\ref{tb:19derived}. This makes this cluster more massive than
all but the two most massive young clusters.

Crude estimates of the star-formation rate in the central regions of 
\ngc\ can be made from the values listed in Table~\ref{tb:19derived}.
The total mass in stars in the young age group clusters is 
 1.1$\times$10$^6$~\msun. These stars were formed in the recent 
2~Myr interval for a mean star-formation rate of 0.5~\msun~yr$^{-1}$.
Adding the contribution from the intermediate-age group gives 
1.7$\times$10$^6$~\msun\ of stars formed over a 16~Myr period or
 a 0.1~\msun~yr$^{-1}$ average rate.
These rates can be compared to rates deduced from \ha\ luminosities.
For the young age group clusters this rate is only
 0.1~\msun~yr$^{-1}$ and is only slightly higher when the intermediate-age
 clusters are included (see Table~\ref{tb:19derived}).
The two methods give quite different estimates. It is 
unlikely that the escape fraction of ionizing radiation is large
 enough to compensate for the discrepancy.
Finally, the values can be compared to
 the galaxy-wide star-formation rate of 1.3~\msun~yr$^{-1}$
 based on the \ha\ luminosity and assuming a nominal 
 $A({\rm H}\alpha)=1$~mag reported by \citet{kennicutt03}.
This suggests that 10\% or up to  38\% of the current 
 galaxy-wide star formation
 is occurring within these 19 star clusters. This is not an 
unreasonable estimate since the brightest few clusters in the sample
 clearly dominate the relevant emission from \ngc.

\begin{table*}
\begin{center}
\caption{Young regions X-ray fitting parameters \label{tb:xyoung}}
\begin{tabular}{lrrrrr}

%
\tableline \tableline 
 Fit Parameters & {\tt apec} &  {\tt apec}  & {\tt apec} & {\tt apec+apec} & {\tt apec+apec} \\ 
\tableline 
%
$N_{\rm H}$ (10$^{21}$cm$^{-2}$)   & 0.4                 & 1.4                 & 4.7$^{+1.7}_{-1.1}$   & 0.4                   & 2.9$^{+2.0}_{-2.0}$   \\
$T_{e_{1}}$ (MK)             & 3.4$^{+0.3}_{-0.3}$ & 3.0$^{+0.5}_{-0.2}$ & 2.1$^{+0.3}_{-0.5}$   & 2.8$^{+0.4}_{-0.3}$   & 1.6$^{+0.4}_{-0.6}$   \\
Normalization K$_{1}$ $^a$   & 1.2$^{+0.2}_{-0.2}$ & 2.3$^{+0.4}_{-0.5}$ & 28$^{+1600}_{-28}$    & 2.3$^{+0.4}_{-0.5}$   & 12$^{+111}_{-11}$     \\
$T_{e_{2}}$ (MK)             &                     &                     &                       & 58$^{+690}_{-41}$     & 6.3$^{+1.0}_{-1.4}$   \\
Normalization K$_{2}$        &                     &                     &                       & 1.1$^{+1.5}_{-0.7}$   & 1.0$^{+2.5}_{-0.7}$   \\
$L_{\rm X}$/10$^{37}$ \ergl\ & 2.0$^{+0.2}_{-0.2}$ & 2.0$^{+0.3}_{-0.3}$ &  2.0$^{+0.7}_{-1.2}$  & 2.2$^{+0.3}_{-0.3}$   & 2.1$^{+0.7}_{-0.8}$   \\
$L_{\rm X}^{int}$/10$^{37}$ \ergl\  & 2.3          & 4.2                 & 43.7                  &  2.7                  & 12.1                  \\
  C-stat/bin                 &  167/108            & 166/108           &   154/108             & 145/108               & 144/108             \\
\tableline
 \multicolumn{5}{l}{Derived Parameters$^b$}  \\ 
\tableline
 $V$    [$f$] ($10^{63}$ cm$^{3}$)      & 2.0   & 2.0   & 2.0  & 2.0       & 2.0        \\
 $n_e$ [$f^{-1/2}$] (cm$^{-3}$)         & 0.03  & 0.04  & 0.13 & 0.54,0.02 & 0.13,0.03  \\
 $P/k$ [$f^{-1/2}$] (10$^5$ K cm$^{-3}$) & 1.8  & 2.2   & 5.5  & 28        & 4.2        \\
 $M_x$ [$f^{+1/2}$] (10$^5$ \msun)      &  0.6  & 0.8   &  3.1 & 1.3       & 3.5        \\
 $E_{th}$ [$f^{+1/2}$] (10$^{53}$ erg)  &  0.7  & 0.9   &  2.3 & 2.7       & 3.4        \\
 $t_c$ [$f^{+1/2}$] (Myr)               &  40   & 21    &  2.5 & 90        & 6.2         \\
\tableline
 \multicolumn{5}{l}{$^a$ $K=(10^{-9}/4\pi D^2) \int n_e n_p dV$} \\
\multicolumn{5}{l}{$^b$volume filling factor scaling in [ ] in Column 1.}\\
\end{tabular}
\end{center}
\end{table*}

\begin{figure}
\begin{center}

\includegraphics[angle=-90,width=\columnwidth]{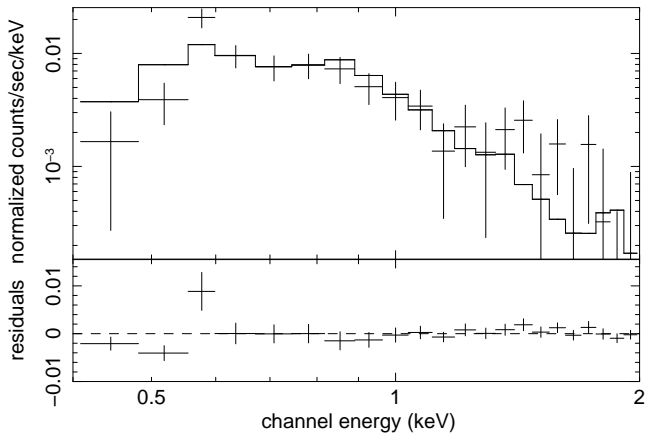}
\vspace{10pt}
\figcaption{Co-added spectrum of the 6 young star-forming regions.  
The best-fitting model, an absorbed two-temperature {\tt apec}
 model with the hydrogen column density allowed as a free parameter,
 and the fit residual are also shown.  
The spectrum has been re-binned to have a width of 73 eV for display purposes.
\label{f:x.young}}
\end{center}
\end{figure}

\subsection{Comparison to X-ray Properties}\label{s:blobx}

\begin{figure}
\begin{center}

\includegraphics[angle=-90,width=\columnwidth]{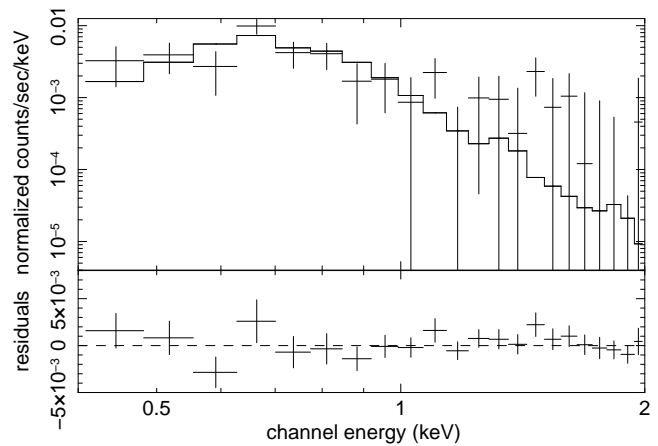}
\vspace{10pt}
\figcaption{Co-added spectrum of the 11 intermediate-age star-forming regions. 
The best-fitting model, an absorbed one-temperature {\tt apec}
 model with the hydrogen column density allowed as a free parameter,
 and the fit residual are also shown.  
The spectrum has been re-binned to have a  width of 73 eV for display purposes.
\label{f:x.intmed}}
\end{center}
\end{figure}

In most cases, 
 the X-ray emission from these individual star-forming regions is too
 faint to be detected as a discrete source using our standard source-detection
 algorithm. 
Therefore, we define the X-ray source radius 
 for each region to be
 the largest of the values determined from fitting the surface brightness
 distribution in the UV, mid-IR, and \ha\ bands.
Furthermore,
we combine (stack) the spectra of the young and of the intermediate age
 regions to accumulate sufficient counts for fitting of the 
spectra of these two age groups.
For this purpose,
we used the first observation only, in which all 
regions are imaged on the S3 CCD.  
We used the same background as was used for analysis of the residual emission 
(\S\ref{s:residual}) and
apply the same one- and two-temperature thermal emission models described in \S2 to the
 stacked spectra.  
Due to low counts, only the one-temperature models are applied 
 to the intermediate age region stacked spectrum. 
We calculated mass-weighted $A_V$ from the SED fitting (\S\ref{s:seds})
for both groups and calculated a corresponding $N_{\rm H}$, 
 namely 1.4$\times$10$^{21}$~cm$^{-2}$ for the young and 
 1.1$\times$10$^{21}$~cm$^{-2}$ for the intermediate regions.
The fitting results are shown in Tables~\ref{tb:xyoung} and \ref{tb:xint} for the young and intermediate 
age star-forming regions, respectively.   The 
spectra with best-fitting models are shown in Figure~\ref{f:x.young} for the young regions 
and Figure~\ref{f:x.intmed} for the intermediate regions.

Not surprisingly, the star-forming regions have X-ray temperatures and densities
similar to the \hii\ regions selected by their \ha\ brightness (Table~\ref{tb:xhiif} and \S\ref{s:hii}).
These values are somewhat lower than for isolated SNRs (Table~\ref{tb:xsnrf}) and the
densities are higher than deduced for the residual emission (Table~\ref{tb:xrf}).
The largest differences are found for the intermediate-age group: the 
derived electron density and pressure are a factor of two lower than that of 
the \hii\ regions. For models allowing the intervening column density to vary,
the resulting \nh\ are comparable to the Galactic value or about a factor of 
ten lower than for any of the other source populations considered including
the young star-forming region group.

\begin{table}
\begin{center}
\caption{Intermediate age regions X-ray fitting parameters \label{tb:xint}}

\begin{tabular}{lrrr}
%
\tableline \tableline 
 Fit Parameters & {\tt apec} & {\tt apec}  & {\tt apec} \\ 
\tableline 
%
$N_{\rm H}$ (10$^{21}$cm$^{-2}$)   & 0.4                   & 1.1                 & 0.4$^{+5.1}_{-0.0}$    \\
$T_{e_{1}}$ (MK)             & 3.1$^{+0.3}_{-0.4}$   & 3.0$^{+0.5}_{-0.5}$ & 3.1$^{+0.5}_{-1.5}$    \\
Normalization K$_{1}$ $^a$   & 0.6$^{+0.2}_{-0.2}$   & 1.0$^{+0.3}_{-0.2}$ & 0.6$^{+51}_{-0.1}$     \\
$L_{\rm X}$/10$^{37}$ \ergl\ & 1.0$^{+0.1}_{-0.2}$   & 1.0$^{+0.2}_{-0.2}$ & 1.0$^{+0.3}_{-0.6}$    \\
$L_{\rm X}^{int}$/10$^{37}$ \ergl\  & 1.2  & 1.8 &  1.2 \\ 
  C-stat/bin               &  98.5/108            &   98.7/108    & 98.5/108\\
\tableline
 \multicolumn{4}{l}{Derived Parameters$^b$}  \\ 
\tableline
 $V$    [$f$] ($10^{63}$ cm$^{3}$)      & 4.0   & 4.0   & 4.0  \\
 $n_e$ [$f^{-1/2}$] (cm$^{-3}$)         & 0.01  & 0.02  & 0.01 \\
 $P/k$ [$f^{-1/2}$] (10$^5$ K cm$^{-3}$) & 0.8  & 1.0   & 0.8  \\
 $M_x$ [$f^{+1/2}$] (10$^5$ \msun)      &  0.6  & 0.8   & 0.6  \\
 $E_{th}$ [$f^{+1/2}$] (10$^{53}$ erg)  &  0.7  & 0.9   & 0.7  \\
 $t_c$ [$f^{+1/2}$] (Myr)               &  60   & 50   & 60  \\
\tableline
 \multicolumn{4}{l}{$^a$ $K=(10^{-9}/4\pi D^2) \int n_e n_p dV$} \\
\multicolumn{4}{l}{$^b$volume filling factor scaling in [ ] in Column 1.}\\

\end{tabular}
\end{center}
\end{table}

These results indicate that the star clusters begin to show their age
 even after just 10$-$20~Myrs. Their hot gas densities and hence 
pressures begin to fall although their temperatures remain a moderate 
2$-$3~MK. These values of density and pressure are still higher than
the residual emission analyzed in \S\ref{s:residual}.
Note that most of the LyC-producing stars will have become SNe by 
the age of 10$-$20~Myrs so that the traditional current star-formation
tracer, \ha, will no longer be strong.

We can estimate the X-ray luminosities of the individual star-forming regions 
 by scaling by the ratio of counts in the stacked spectra (for each age group)
 to the number of counts detected in individual regions. 
The results are listed in Table~\ref{tb:19derived}. 
These luminosities have been corrected
for extinction local to each region using the values of $N_{\rm H}$ tabulated
in column~7 of Table~\ref{tb:19derived} (derived from their best-fitting $A_V$) 
and for Galactic extinction. 
The luminosities
were computed using the 
two temperature model with variable $N_{\rm H}$ for the young region group, and the one temperature model with variable $N_{\rm H}$ for the intermediate age regions.  

\begin{figure}
\begin{center}

\includegraphics[angle=-90,width=0.9\columnwidth]{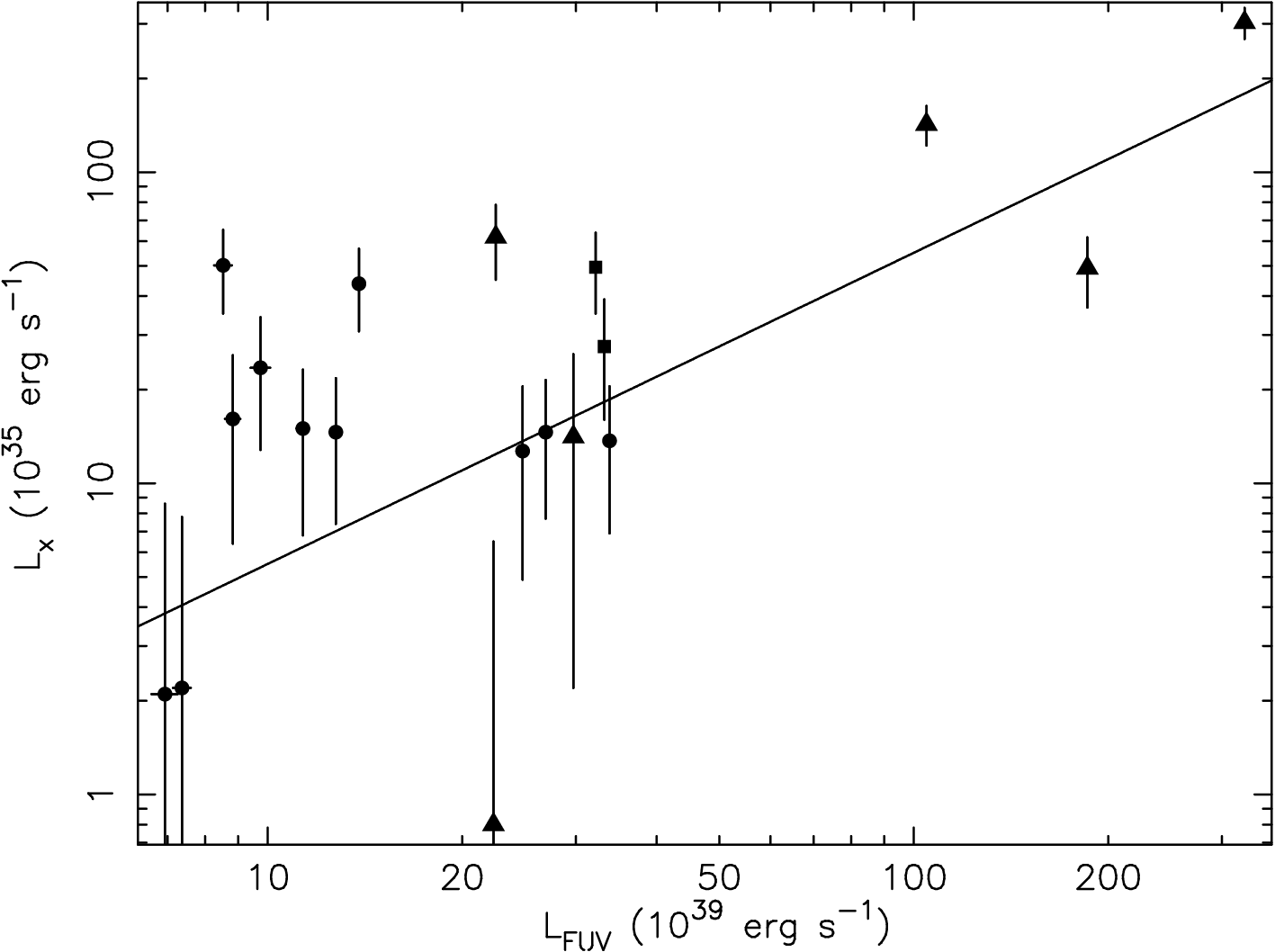}
\vspace{10pt}
\figcaption{$L_{\rm FUV}$ vs. $L_{\rm X}$ for the 19 star-forming regions.  
Luminosities are corrected for intrinsic and Galactic extinction.  
Triangles represent young (less than 2 Myrs), circles intermediate (2$-$20 Myes), and squares old (above 20 Myrs) star-forming regions. 
The straight line indicates the best-fitting linear function with a slope of 
 (5.5$\pm$0.8)$\times$10$^{-5}$. 
\label{f:x.vs.fuv}}
\end{center}

\begin{center}
\includegraphics[angle=-90,width=0.9\columnwidth]{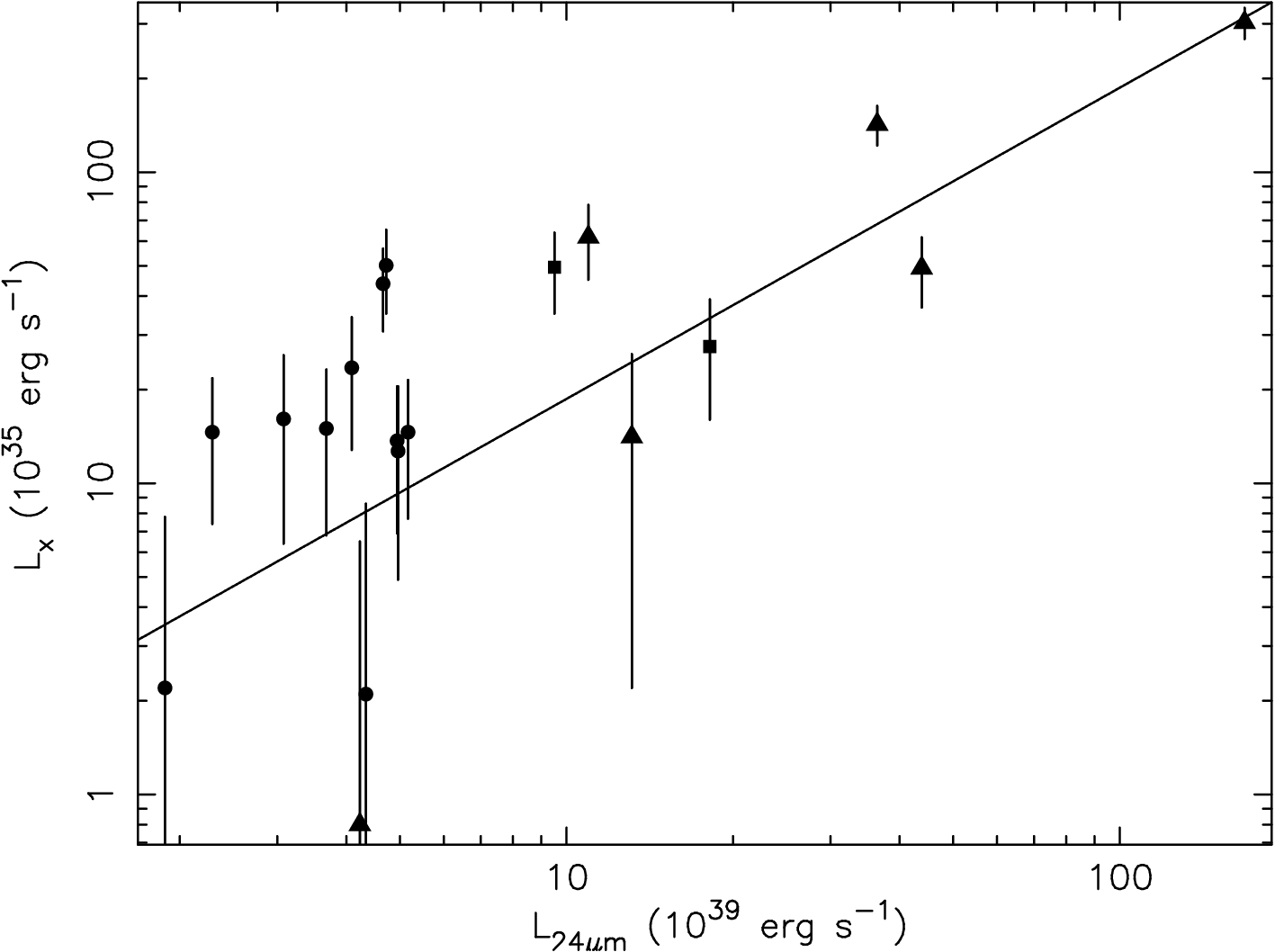}
\vspace{10pt}
\figcaption{$L_{24\mu\rm{m}}$ vs. $L_{\rm X}$ for the 19 star-forming regions.
Luminosities are corrected for intrinsic and Galactic extinction.  
Symbols are the same as in Figure~\ref{f:x.vs.fuv}.
The straight line indicates the best-fitting linear function with a slope of 
 (1.9$\pm$0.2)$\times$10$^{-4}$. 
\label{f:x.vs.24}}
\end{center}

\begin{center}
\includegraphics[angle=-90,width=0.9\columnwidth]{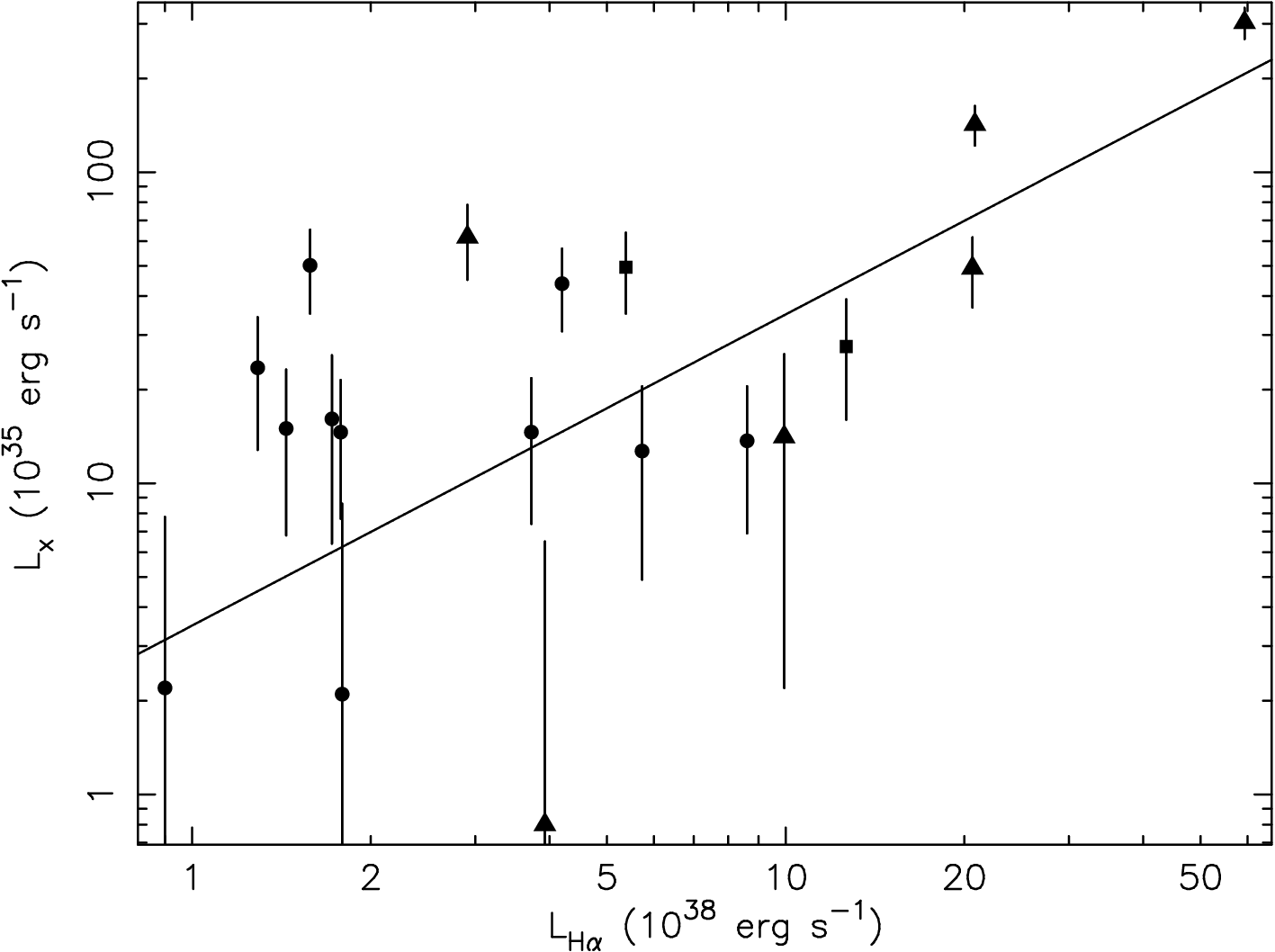}
\vspace{10pt}
\figcaption{$L_{\rm{H}\alpha}$ vs. $L_{\rm X}$ for the 19 star-forming regions.
Luminosities are corrected for intrinsic and Galactic extinction.  
Symbols are the same as in Figure~\ref{f:x.vs.fuv}.
The straight line indicates the best-fitting linear function with a slope of
 (3.5$\pm$0.5)$\times$10$^{-3}$.
\label{f:x.vs.ha}}
\end{center}
\end{figure}

Figures~\ref{f:x.vs.fuv} through~\ref{f:x.vs.ha} compare these
intrinsic X-ray luminosities
to the (extinction-corrected) FUV, 24\,\um,
and \ha\ luminosities, respectively. In all cases, there is only a weak
correlation between X-ray luminosity and these star-formation indicators.
This is due to the large scatter in the X-ray luminosities.

A simple calculation shows that in one or two regions a relatively high
 X-ray luminosity may be the result of a faint undetected X-ray point
 source in the region. 
From the XLF, \S\ref{s:pts}, we expect about 20 X-ray point sources with 
 0.5$-$2.0 keV luminosities in the range spanned by our sample of 
 star-forming regions and within the central 2\farcm5 radius disk of \ngc.
Scaling this to the total area occupied by the star-forming regions 
 results in an expected 1.7 X-ray point sources in this luminosity range 
 somewhere within our sample. 
(Of course, the brighter the point source, the lower the probability of
 coincidence since the XLF scales as $L^{-0.58}$.) 

\subsection{Comparison to Theoretical Expectations} \label{s:theory}

Stellar winds and supernovae within a young star cluster create a hot
low-density cavity in the interstellar medium (ISM) which persists 
beyond the $\sim$40~Myr lifetimes of the OB stars in the cluster.
The cumulative effect of winds and individual SNe is to gradually sweep the ISM
into a thin dense shell analogous to the shell around a stellar wind
bubble. The simple physical model of a stellar wind
bubble \citep{castor75,weaver77} can be taken over
directly to describe, qualitatively, these large \hii\ regions or 
superbubbles associated with OB associations \citep{mccray87,maclow88}. In this model,
massive stars inject kinetic energy into their surroundings creating
a freely expanding wind.  Hot shocked gas surrounds this free wind region and occupies most of the volume of the bubble.  Surrounding this hot gas is the dense cold swept-up shell.   Beyond the shell is the ambient interstellar medium.
Diffuse X-ray emission from star-forming regions arises from the tenuous shocked gas in the bubble interior and from higher density gas evaporated from the shell.
The dense evaporated component may dominate the X-ray emission which scales as the 
square of the density.

Although analytic versions of the original basic model have been 
 applied widely in the literature, 
 we note that numerous simplifying assumptions are made
 which may be poor approximations in reality.
Among these are spherical symmetry, constant energy injection rate, adiabaticity,
neglect of magnetic fields and of turbulence, and a homogeneous ambient medium.
For our present purposes, the most critical of these is the assumption
 of a constant energy injection rate or mechanical luminosity, $L_{\rm w}$.
Numerical simulations \citep[e.g.,][]{strickland99}
show that the expected 
X-ray luminosity is roughly proportional to  $L_{\rm w}$ and that $L_{\rm w}$ is 
far from constant.

The starburst synthesis models we have used here \citep{leitherer95}  
 show that $L_{\rm w}$ (injected by winds from stars of initial 
mass $>$15~\msun\ and by suitably time-averaged SN explosions)  
is a function of the age of the star-forming region,
 the mass of the star cluster, and the assumed IMF.
Specifically, 
only stellar winds from massive stars contributes to the mechanical
luminosity during the first 3~Myrs. Then supernova explosions begin to contribute
resulting in an increase in $L_{\rm w}$ by about a factor of 4 from an age of 3 to 6~Myrs. 
The luminosity decreases again as the contribution from stellar winds
declines due to the decreasing number of O stars. From $\sim$10 Myrs
and up to the last supernova, which is around 40 Myrs, $L_{\rm w}$ 
is almost constant then finally falls quickly to very low values.
The young and intermediate star-forming regions studied here
 have ages in the range where $L_{\rm w}$ changes significantly on short timescales
so that our estimates of $L_{\rm w}$ (Table~\ref{tb:19derived}) are reliable only to 
about a factor of four.

\begin{figure}
\begin{center}
\includegraphics[angle=-90,width=0.9\columnwidth]{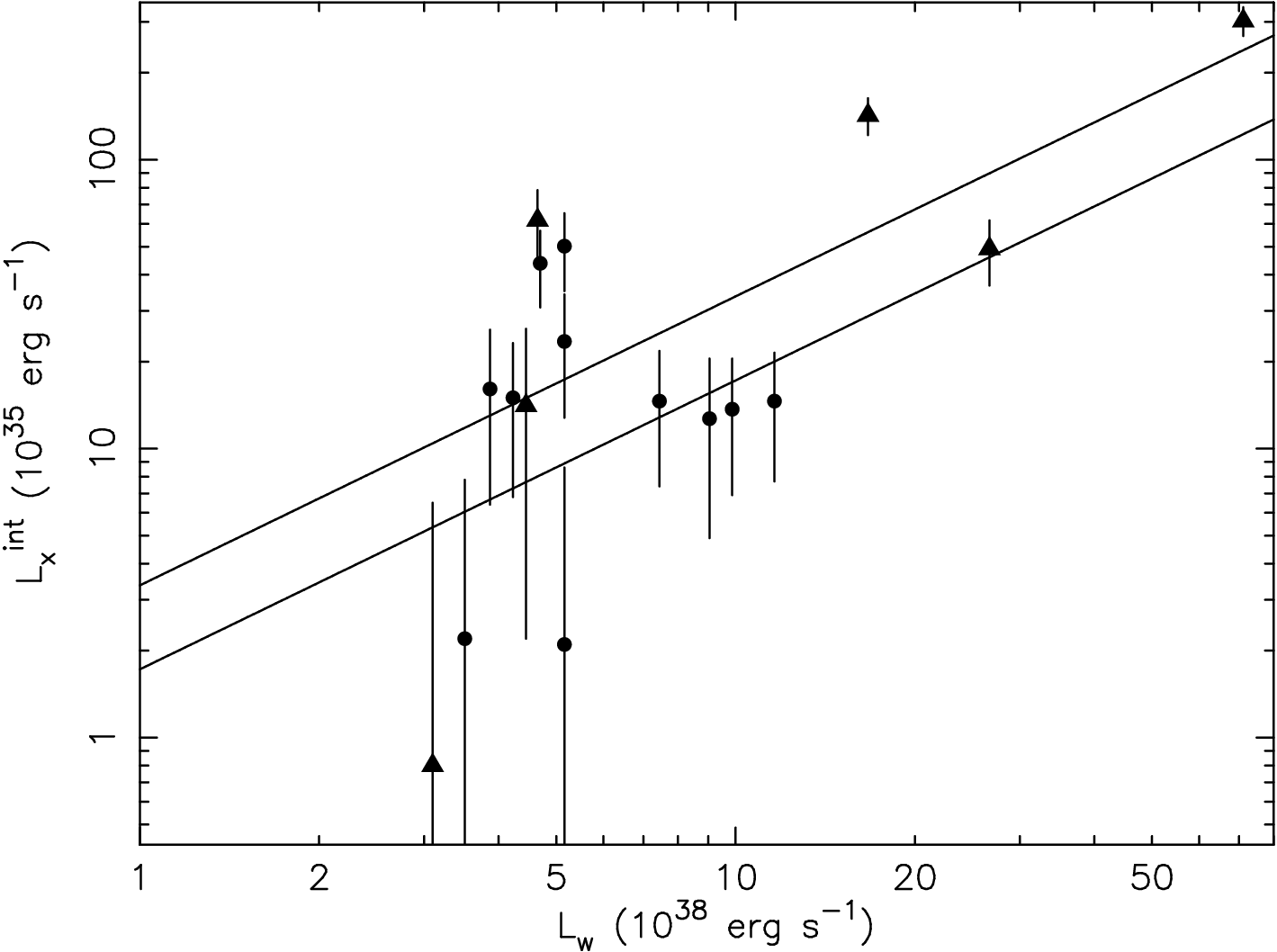}
\vspace{10pt}
\figcaption{Mechanical luminosity, $L_{\rm w}$ calculated from Starburst99 scaled by 
 the mass and age of each region is shown against the intrinsic
 X-ray luminosity.
Symbols are the same as in Figure~\ref{f:x.vs.fuv} 
 (the two regions older than 20 Myr are omitted because they have 
 very low mechanical luminosities according to Starburst99).
The lines indicate best-fitting slopes of 
 (3.4$\pm$0.5)$\times$10$^{-3}$ and (1.7$\pm$0.5)$\times$10$^{-3}$ 
 fitted to the young and intermediate-age regions, respectively. 
\label{f:XLw}}
\end{center}

\begin{center}
\includegraphics[angle=-90,width=0.9\columnwidth]{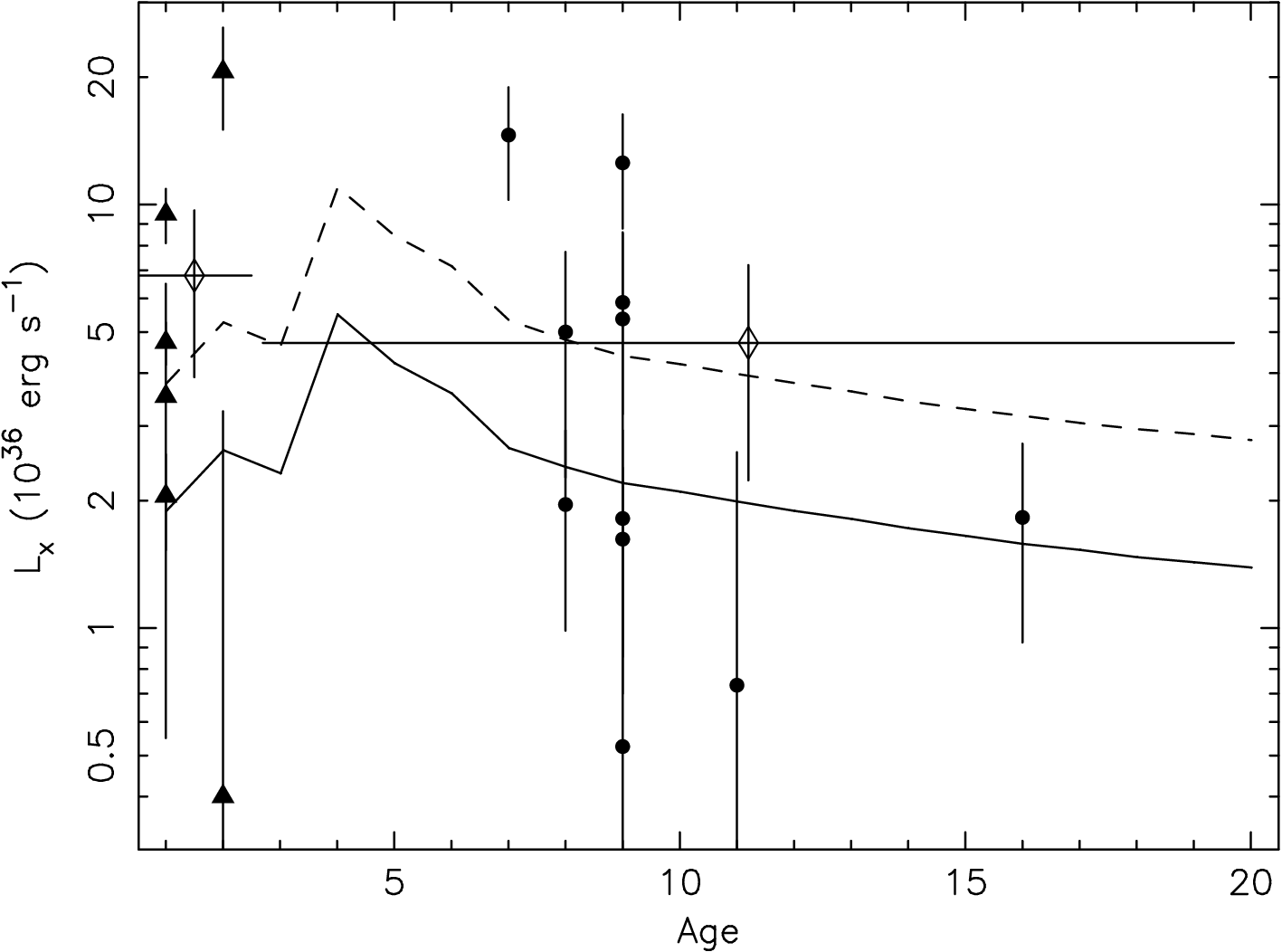}
\vspace{10pt}
\figcaption{Intrinsic X-ray luminosity is shown against age of star forming 
 regions.
Luminosites are scaled to an initial mass of 10$^{5}$ \msun.   
Symbols are the same as in Figure~\ref{f:x.vs.fuv}.  
Open diamonds depict the average for young and intermediate regions.  
The solid curve indicates the X-ray luminosity scaled from the mechanical luminosity, $L_{\rm w}$, assuming 
 $L_{\rm X}/L_{\rm w} = 0.0017$. 
The dashed curve indicates the X-ray luminosity scaled from $L_{\rm w}$ assuming 
 $L_{\rm X}/L_{\rm w} = 0.0034$.   
\label{f:Xtime}}
\end{center}
\end{figure}

Figure~\ref{f:XLw} displays the 
(extinction-corrected) X-ray luminosity estimated in \S\ref{s:blobx}
against the mechanical luminosity computed from the 
 \citep{leitherer95} models using the age and mass estimates of 
the star-forming regions deduced in  \S\ref{s:seds} (Table~\ref{tb:19derived}).
Filled triangles depict young regions, and filled circles represent intermediate-age regions.  We fitted with linear function for each group, and obtained 
the slope $L_{\rm X}/L_{\rm w}$.  
The resulting slopes are 0.0033$\pm$0.0005  for the young regions, 
and 0.0017$\pm$0.0005 for the intermediate-age regions.  
This is consistent with the \cxo\ result of the second brightest \hii\ 
region in the local group, NGC 604, whose $L_{\rm X}/L_{\rm w}$ is estimated as
0.002 \citep{tullmann08}. 

 Numerical simulations \citep{strickland99} found that $L_{\rm X}/L_{\rm w}$ for a thin galactic disk model is 0.5$-$2\%, using an $L_{\rm X}$ appropriate for the {\sl ROSAT} band.  
This must be adjusted for the different sensitivity of \cxo\ compared to  {\sl ROSAT}
for the observed spectral shape.
Here, the characteristic electron temperature is 2$-$3~MK so that
the fraction of the luminosity for \cxo\ band (0.5$-$2.0 keV) to {\sl ROSAT} band (0.1$-$2.5 keV) is 0.5  to 0.6. This translates $L_{\rm X}/L_{\rm w}$ to 0.25
 to 1\% for the \cxo\ band 
which is consistent with our results. 

Figure~\ref{f:Xtime} displays the (extinction-corrected)
X-ray emission against the estimated cluster age. 
Here, the X-ray luminosities are scaled to a fixed cluster mass of 10$^5$~\msun\
for easy comparison to $L_{\rm w}(t)$. The solid curve represents the best-fitting
scaling, $L_{\rm X}=0.0017L_{\rm w}$, from Figure~\ref{f:XLw} for the intermediate-age
 clusters and the dashed line 
is the factor-of-two higher scaling appropriate for the younger clusters. 
Note the roughly factor-of-four change in $L_{\rm w}$ at 
$t\sim4$~Myr as described above. 
Unfortunately, there is a considerable amount of scatter among the individual 
regions. For this reason, the $L_{\rm X}/L_{\rm w}$ ratio is not a sensitive 
independent measure of the age of a cluster.

\section{Discussion of Individual Star-Forming Regions} \label{s:6x6}

We have examined 19 star-forming regions in the central part of \ngc.
Their basic properties are listed in Tables~\ref{tb:19obsdata} 
 and~\ref{tb:19derived}.
They range in age from 1~to 16~Myr with two much older regions at 50 and 60~Myr.
These latter ages are likely poor estimates as these two regions are 
 also strong \ha\ sources which argues against an old age. 
The remaining star-forming regions fall into distinct age groups;
 young (1-3 Myr) and intermediate (7 to 16 Myr).
The bulk X-ray properties of the hot gas within these two age 
 groups, from model fits to their stacked X-ray spectra, 
 are summarized in Tables~\ref{tb:xyoung} and~\ref{tb:xint},
 respectively.

There is very little change in the X-ray temperatures 
and very little change in the X-ray luminosity per unit star cluster mass
with age (Figure~\ref{f:Xtime}).
There is also no significant evolution in the ratio of the
X-ray luminosity to the mechanical luminosity
(Figure~\ref{f:XLw}).
Thus, the basic X-ray observables, temperature and luminosity,
 do not trace evolutionary changes
 for this sample of young star-forming regions. 

However, the density of the hot gas derived from the spectral fitting
and hence the gas pressure in the young star clusters 
 is about a factor of 2 higher than in the intermediate age clusters. 
This is consistent with the trends deduced previously (\S\ref{s:X}) 
 where the youngest sources, the SNRs, have the highest densities and 
 pressures followed by the (\ha-selected) \hii\ regions with the 
 residual X-ray emission having the lowest density and pressure. 

There are three plausible scenarios for the nature of the hot 
residual gas consistent with these observed trends.
(1) If the decrease in density is temporal, then
 the residual emission comes from gas initially heated at times 
 considerably 
 more than $\sim$20~Myr in the past (the age of the intermediate group
 of star-forming regions); i.e., it is a relic of past star-formation
 activity.
This is consistent with the absence of a spatial correlation between
 tracers of recent star formation
  and this 
 residual emission.
(2) The hot residual gas may be escaping from active star-forming 
 regions into lower-density voids in the disk ISM.
These low-density voids may be, for instance,
 localized remnants of past star-forming activity. 
If this were the case, then we would expect some of this 
 hot gas to be surrounding the star-forming regions analyzed here.
We checked this possibility by taking successively
 larger X-ray source sizes
 for the 19 star-forming regions to estimate the true source extent.
We found that for sizes up to  twice the values adopted above
 (which were the maximum radii estimated from the UV and mid-IR images)
 there is a clear net X-ray excess (above the background).
Nevertheless, on larger scales, the morphology of the 
 residual hot gas does not correlate well with the star-formation
 tracers. 
(3) The hot residual gas may have moved out of the plane of the 
 galaxy down the density gradient into the halo.
However, again, as there is no evidence from the distribution of the
 residual gas that strongly correlates it with star-forming regions,
 it must be a relic of past activity.
In fact, there are no regions of active star formation that have
 the signature of blowout from the disk. 
Blowout requires that the bubble size be roughly a few times the 
 density scale height. 
The largest star-forming regions in the 
 central regions of \ngc\ are $\sim$200~pc compared to a 
 canonical scale height for late-type spirals of 250$-$500~pc
 (see also the discussion in \citeauthor{strickland04} 2004).

We note that the age of the hot residual gas inferred from all of these 
scenarios is roughly the same as the timescale for 
 gravitational instabilities to cause shells 
 surrounding OB associations to fragment \citep{mccray87}
at which point hot gas can vent out into the surroundings. 
It is also an age at which strong ionizing 
continua from massive stars has ended though SNe may still be active.
The current cooling time for the residual gas
exceeds 20~Myr and may be as high as 340~Myrs. Thus, regardless of the actual location
of the residual gas (disk or halo) and the time elapsed since it 
formed ($<$10$-$20~Myr if escaping from active star-forming regions
or perhaps factors of a few longer if remnants of past activity),
this gas is likely to continue to radiate at low levels for a time longer
than the characteristic timescales for localized star formation.

\section{Summary}\label{s:dis}

We have revisited the study of the 
 current and recent star formation activity in
 the central regions of normal late-type spiral galaxy 
 \ngc\ by re-evaluating the existing X-ray data
and including supporting observations at other key wavebands.
These include \sst\ images that trace dust heated by massive stars,
 \galex\ images showing the young stars, and \ha\ observations
that reveal nebulae photoionized by massive stars and SNe. 
Through analysis of this multi-wavelength data, we have obtained
 estimates of the mass, age, and line-of-sight extinction towards
 numerous young star clusters.
With few exceptions, these clusters and their environs are weak X-ray
 sources. 
Even the most powerful, a giant \hii\ region comparable to 30~Dor in the LMC and
 NGC~604 in M33, radiates at only $\sim$3$\times$10$^{37}$~\ergl\
 in the \cxo\ X-ray band which is about 0.5\% of its estimated mechanical
 luminosity.

We have also shown that, after carefully accounting for the point-source 
 (X-ray binary) population, SNRs and bright \hii\
 regions, there remains a residual X-ray component 
pervading the central few kpc of \ngc.
This component cannot be accounted for by faint sources 
  below our
 detection limit.
It is likely diffuse hot gas but not strongly correlated with 
current star-formation activity. It is likely a relic of star formation activity occurring some 20~Myrs (the age of the intermediate age regions) or older.

 The geometry of the residual hot gas is 
not well determined, which leads to additional 
uncertainties in its physical properties.
It is not clear whether this gas is confined to the disk
 of the galaxy or resides in the halo.  Nonetheless, it seems to be more 
 centrally-located than the star-forming regions.
Since the amount of hot gas represented by this residual emission 
 is substantial, it implies that star-formation activity was much higher
in the past. Combined with its central location,
 this conclusion is consistent with the suggestion by \citet{davidge02} 
 who suggested that an earlier episode of star formation 
 occurred in the central region of \ngc\ that
 has now propagated outward in the disk to its present radius of about a kpc
 from the galactic center.

\acknowledgements
 We gratefully acknowledge the anonymous referee
for careful reading and insightful comments that improved the paper. 
Support for this research was provided in part by
NASA through an Astrophysics Data Analysis Program grant NNX08AJ49G and through the
NASA/\cxo\ Award Number GO5-6089A issued by the \cxo\ X-ray Observatory 
Center, which is operated by the Smithsonian Astrophysical Observatory 
for and on behalf of NASA under contract NAS8-03060.
This work made use of observations made with the \cxo; 
of observations made with the Spitzer Space Telescope, which is operated by the 
 Jet Propulsion Laboratory, California Institute of Technology, under a 
 contract with NASA; 
of observations made with the Galaxy Evolution Explorer, 
 a NASA mission managed by the Jet Propulsion Laboratory; 
and of ground-based observations obtained as part of the 
 Spitzer Legacy Science project SINGS \citep{kennicutt03} 
 to which we are greatly indebted. 

%



\begin{center}
\includegraphics[angle=-90,width=0.3\columnwidth]{f8el.eps}
\includegraphics[angle=-90,width=0.3\columnwidth]{f8em.eps}
\includegraphics[angle=-90,width=0.3\columnwidth]{f8er.eps}

\includegraphics[angle=-90,width=0.3\columnwidth]{f8fl.eps}
\includegraphics[angle=-90,width=0.3\columnwidth]{f8fm.eps}
\includegraphics[angle=-90,width=0.3\columnwidth]{f8fr.eps}

\includegraphics[angle=-90,width=0.3\columnwidth]{f8gl.eps} 
\includegraphics[angle=-90,width=0.3\columnwidth]{f8gm.eps} 
\includegraphics[angle=-90,width=0.3\columnwidth]{f8gr.eps} 

\includegraphics[angle=-90,width=0.3\columnwidth]{f8hl.eps}
\includegraphics[angle=-90,width=0.3\columnwidth]{f8hm.eps}
\includegraphics[angle=-90,width=0.3\columnwidth]{f8hr.eps}
\newpage
\clearpage
\includegraphics[angle=-90,width=0.3\columnwidth]{f8il.eps}
\includegraphics[angle=-90,width=0.3\columnwidth]{f8im.eps}
\includegraphics[angle=-90,width=0.3\columnwidth]{f8ir.eps}

\includegraphics[angle=-90,width=0.3\columnwidth]{f8jl.eps}
\includegraphics[angle=-90,width=0.3\columnwidth]{f8jm.eps}
\includegraphics[angle=-90,width=0.3\columnwidth]{f8jr.eps}

\includegraphics[angle=-90,width=0.3\columnwidth]{f8kl.eps} 
\includegraphics[angle=-90,width=0.3\columnwidth]{f8km.eps} 
\includegraphics[angle=-90,width=0.3\columnwidth]{f8kr.eps} 

\includegraphics[angle=-90,width=0.3\columnwidth]{f8ll.eps}
\includegraphics[angle=-90,width=0.3\columnwidth]{f8lm.eps}
\includegraphics[angle=-90,width=0.3\columnwidth]{f8lr.eps}

\newpage

\includegraphics[angle=-90,width=0.3\columnwidth]{f8ml.eps}
\includegraphics[angle=-90,width=0.3\columnwidth]{f8mm.eps}
\includegraphics[angle=-90,width=0.3\columnwidth]{f8mr.eps}

\includegraphics[angle=-90,width=0.3\columnwidth]{f8nl.eps}
\includegraphics[angle=-90,width=0.3\columnwidth]{f8nm.eps}
\includegraphics[angle=-90,width=0.3\columnwidth]{f8nr.eps}

\includegraphics[angle=-90,width=0.3\columnwidth]{f8ol.eps} 
\includegraphics[angle=-90,width=0.3\columnwidth]{f8om.eps} 
\includegraphics[angle=-90,width=0.3\columnwidth]{f8or.eps} 

\includegraphics[angle=-90,width=0.3\columnwidth]{f8pl.eps}
\includegraphics[angle=-90,width=0.3\columnwidth]{f8pm.eps}
\includegraphics[angle=-90,width=0.3\columnwidth]{f8pr.eps}
\end{center}
\newpage

\begin{figure}
\figurenum{8 Cont.}

\includegraphics[angle=-90,width=0.3\columnwidth]{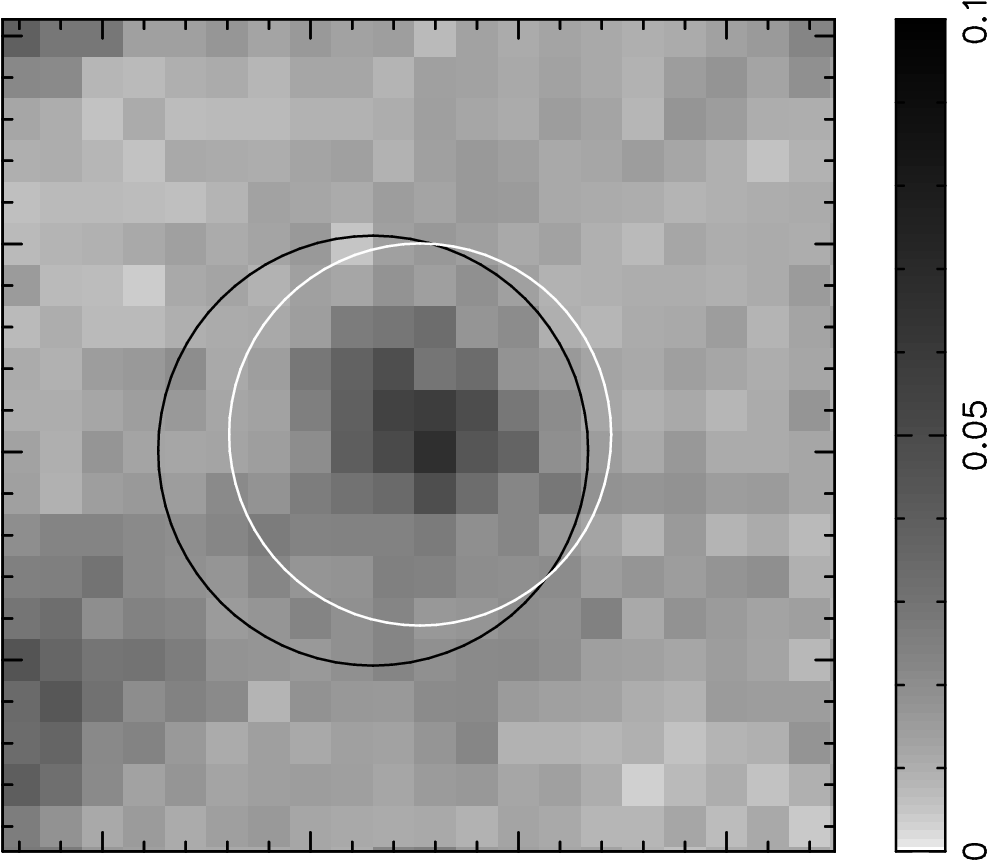}
\includegraphics[angle=-90,width=0.3\columnwidth]{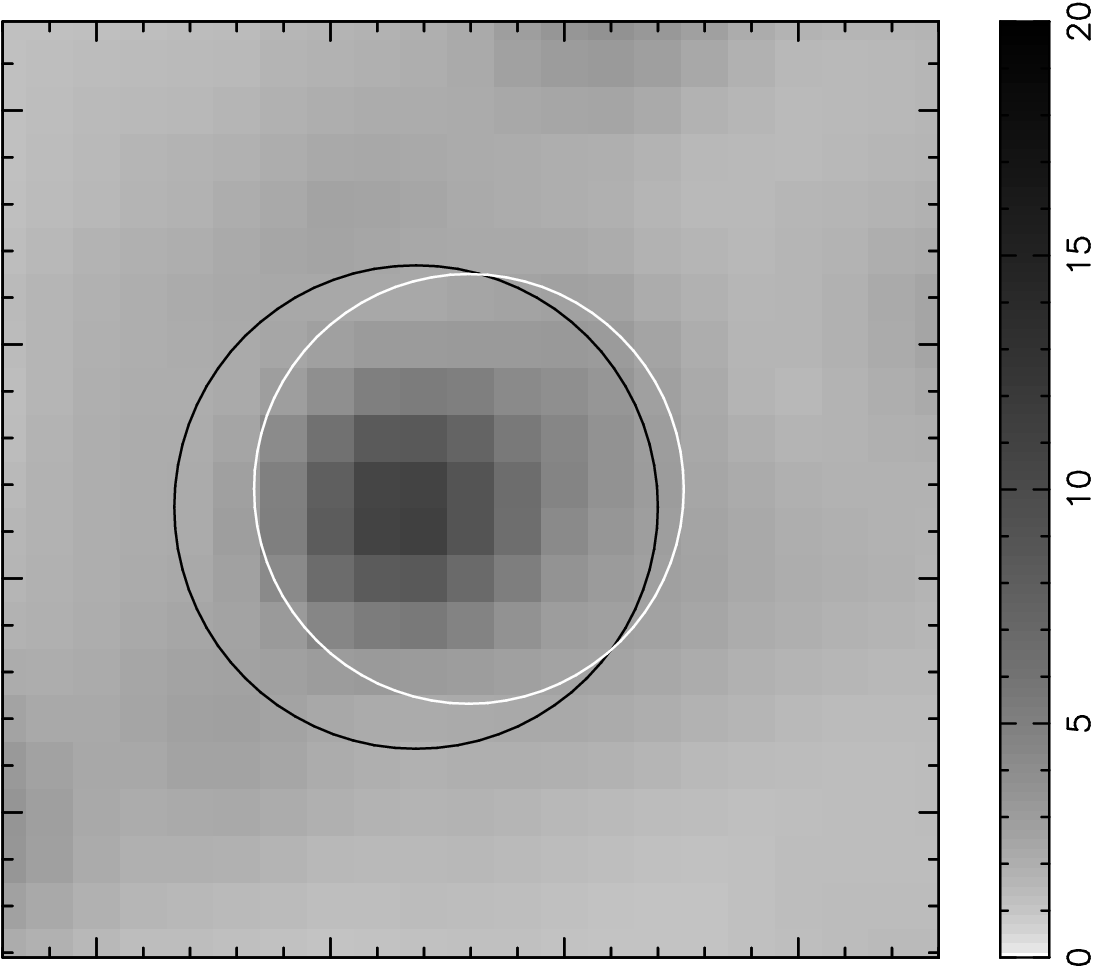}
\includegraphics[angle=-90,width=0.3\columnwidth]{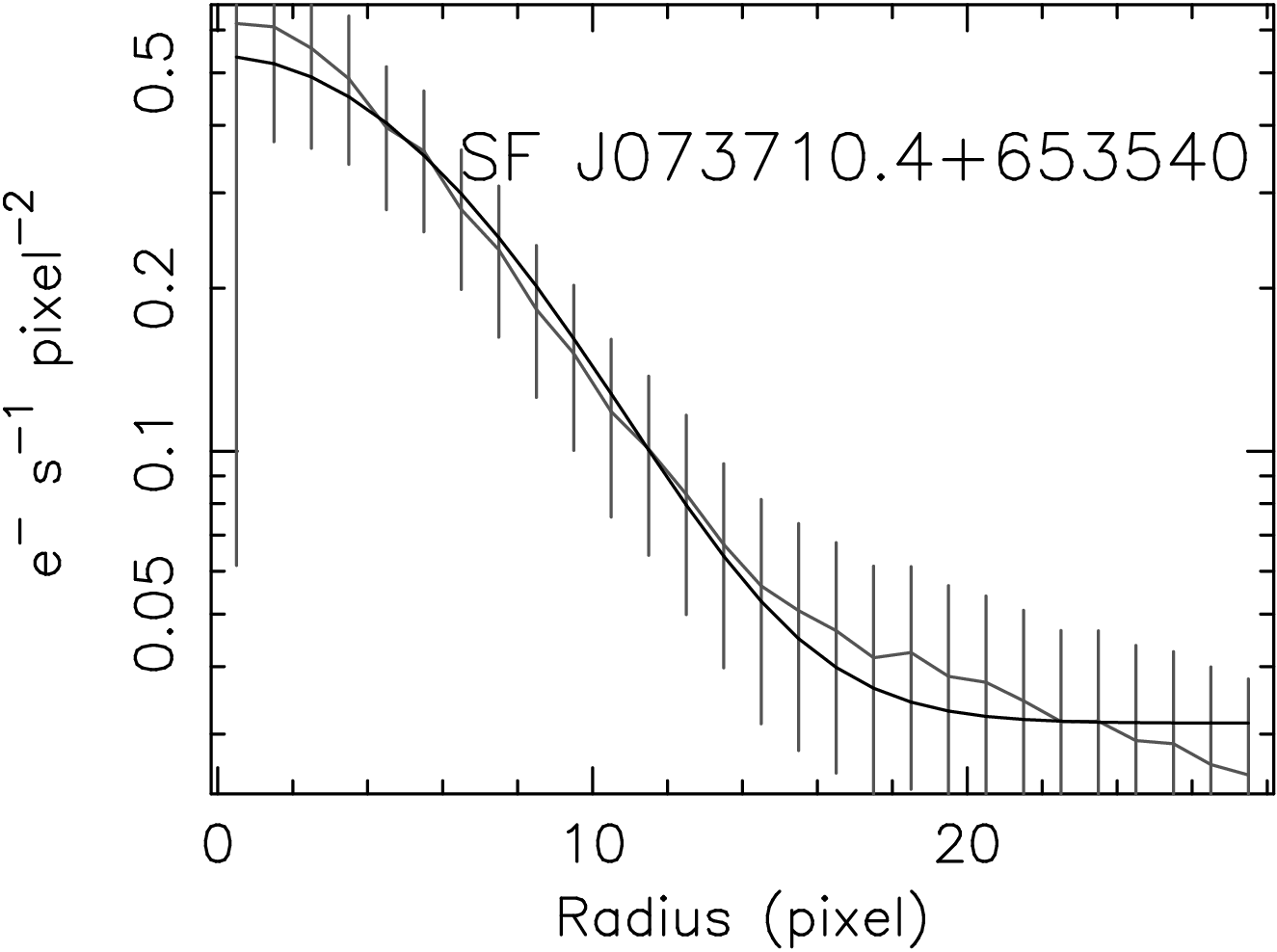}

\includegraphics[angle=-90,width=0.3\columnwidth]{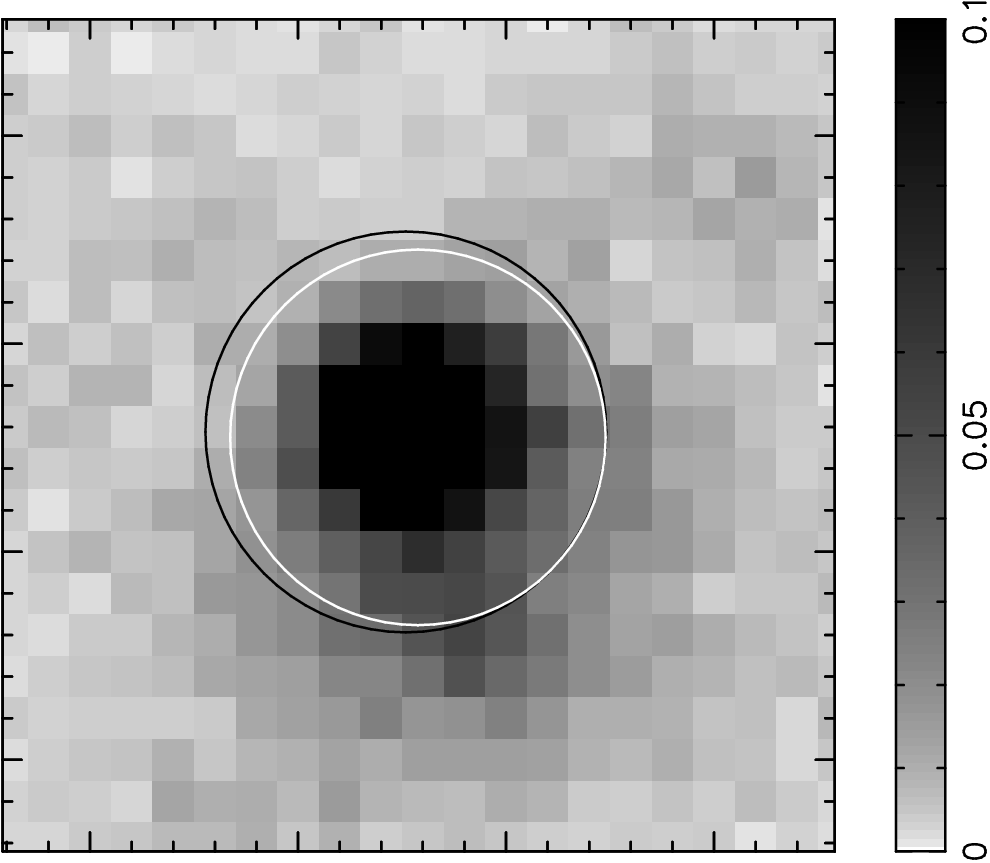}
\includegraphics[angle=-90,width=0.3\columnwidth]{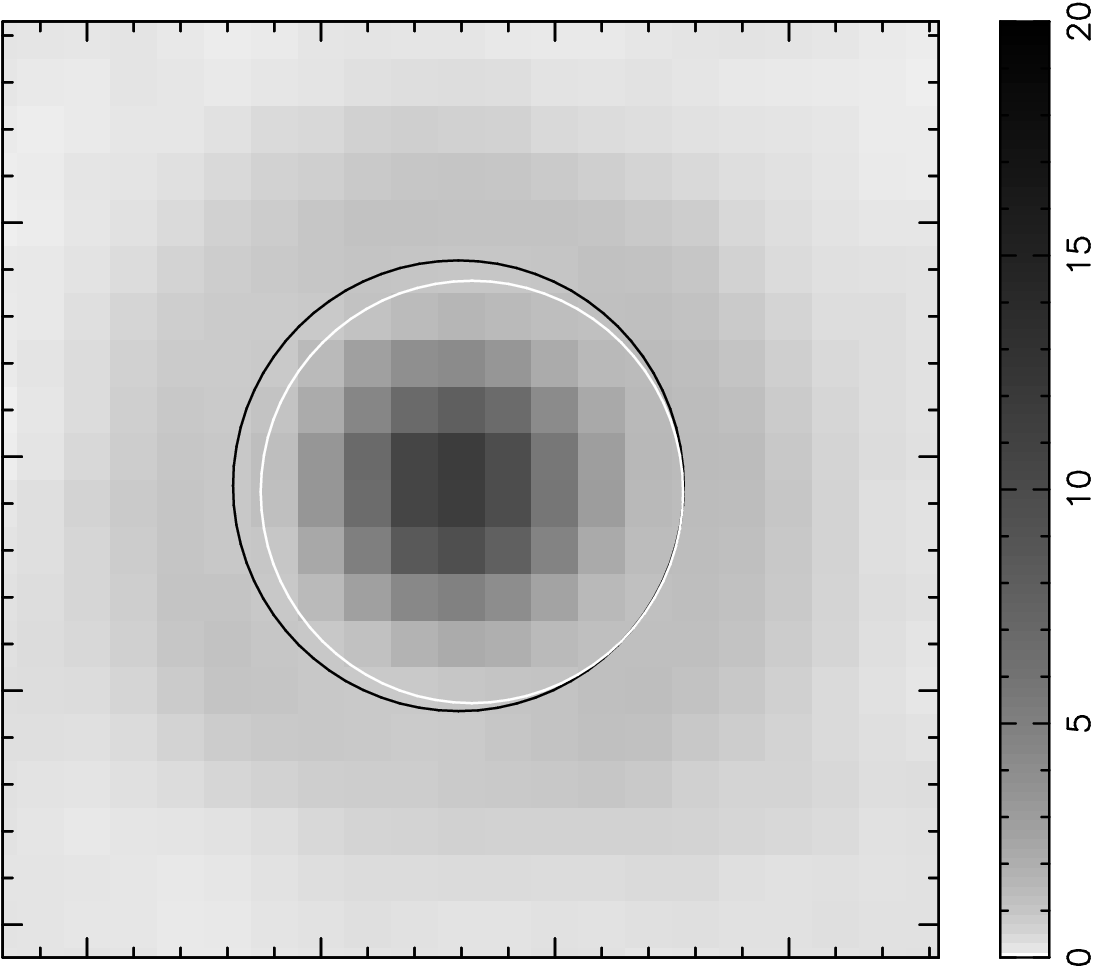}
\includegraphics[angle=-90,width=0.3\columnwidth]{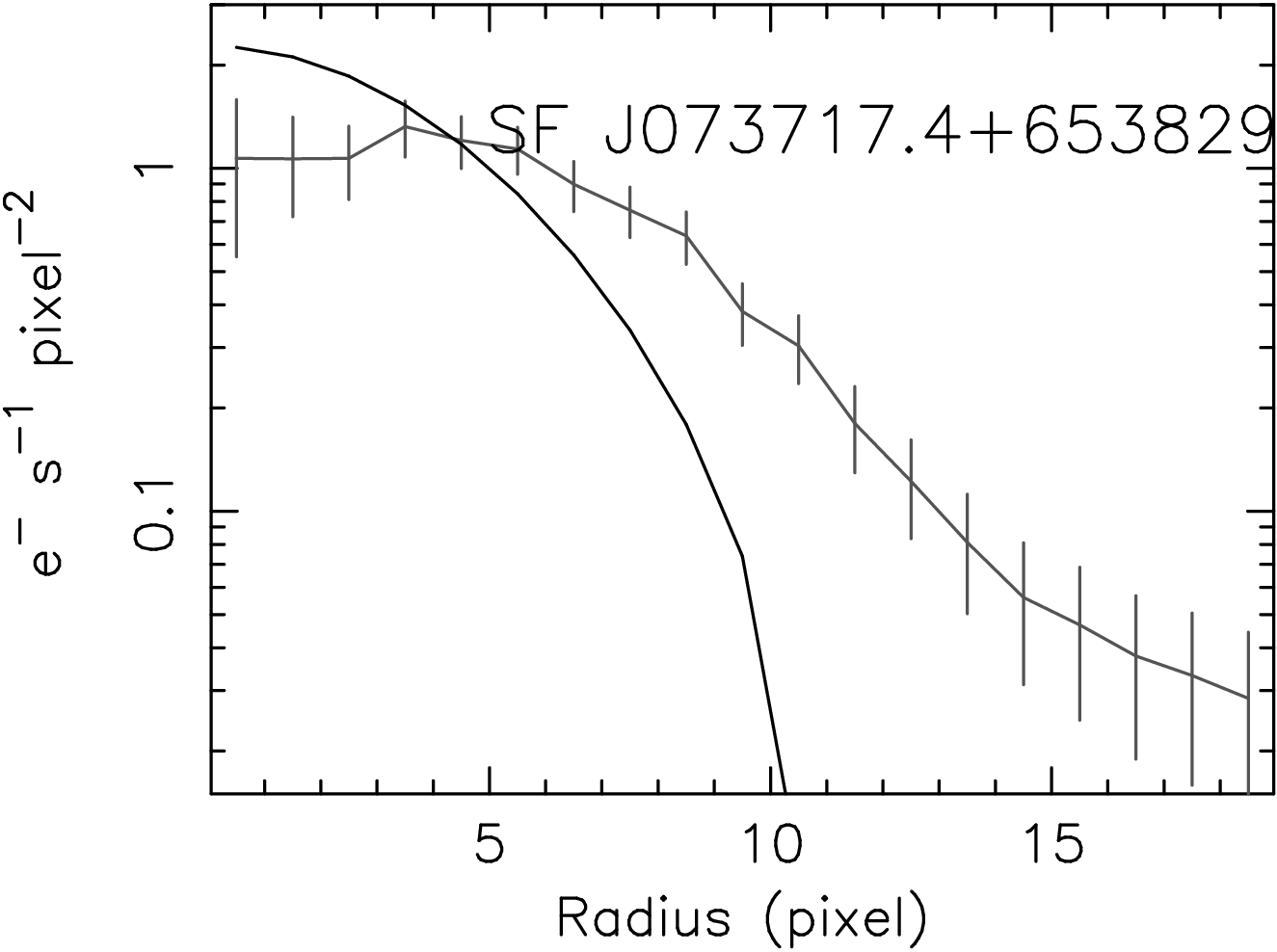}

\includegraphics[angle=-90,width=0.3\columnwidth]{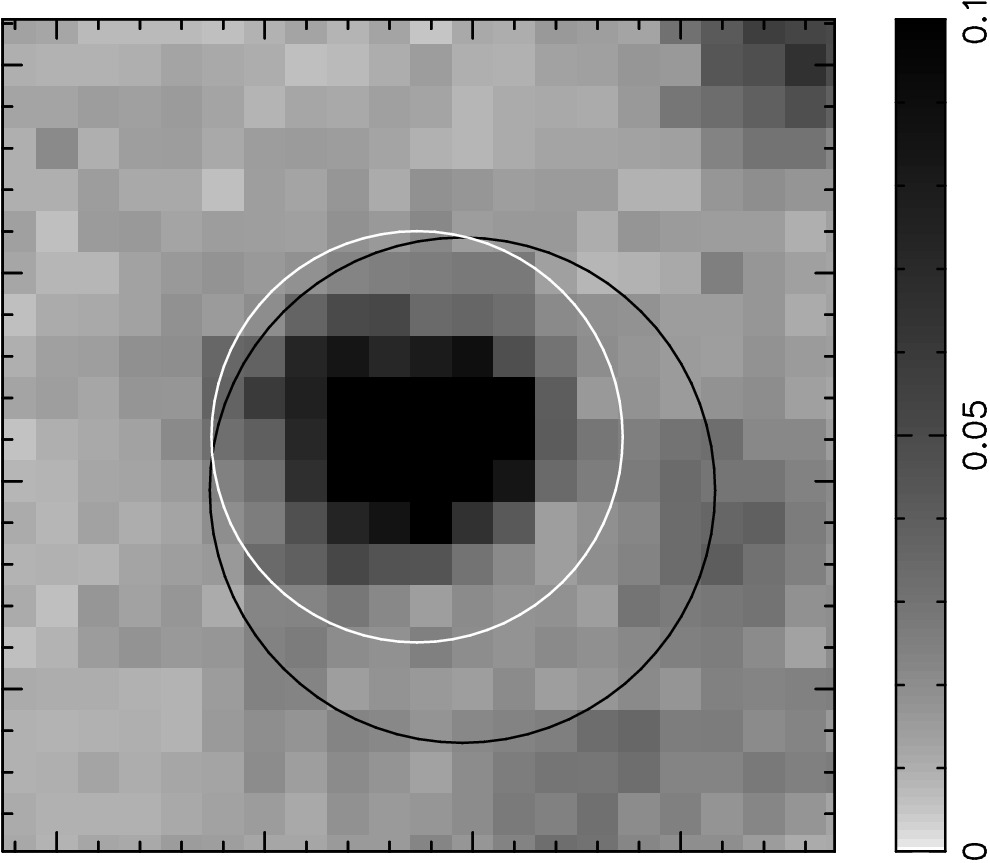} 
\includegraphics[angle=-90,width=0.3\columnwidth]{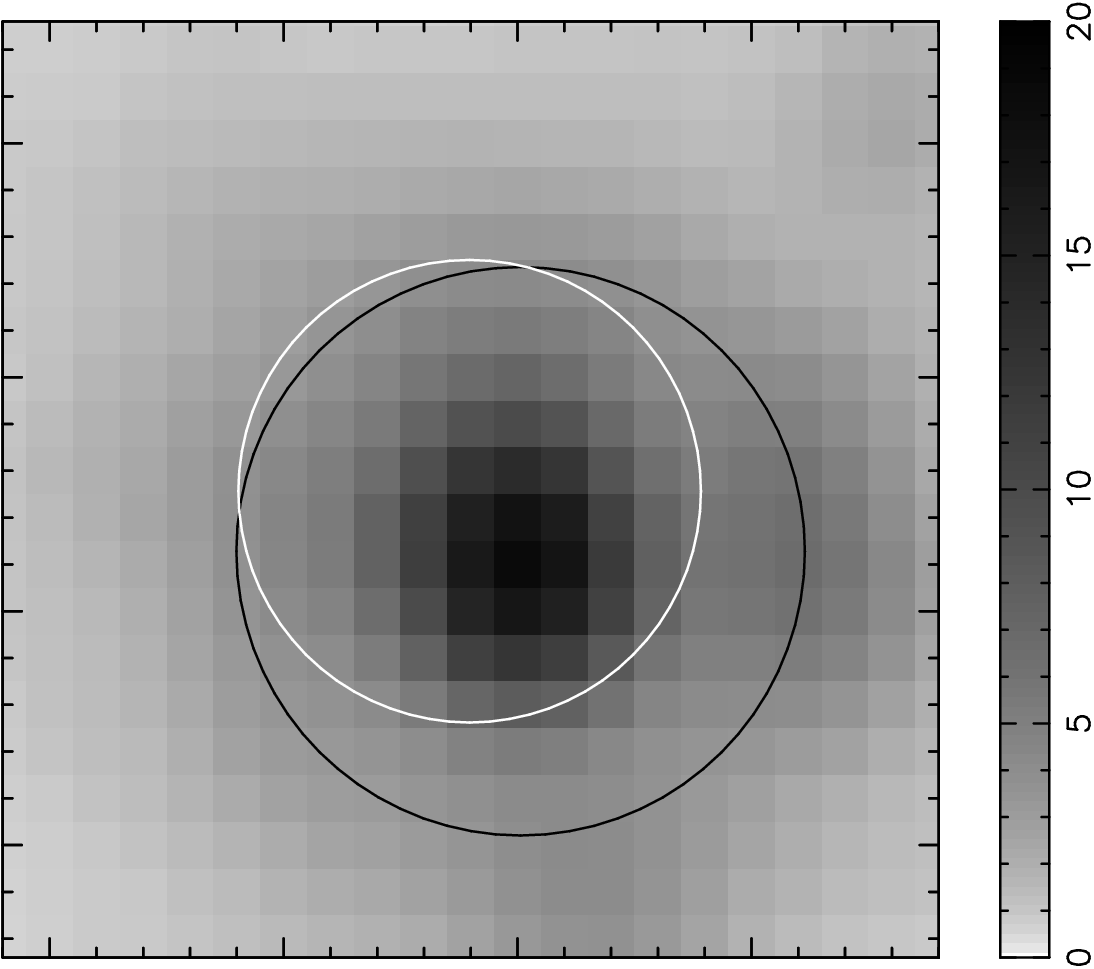} 
\includegraphics[angle=-90,width=0.3\columnwidth]{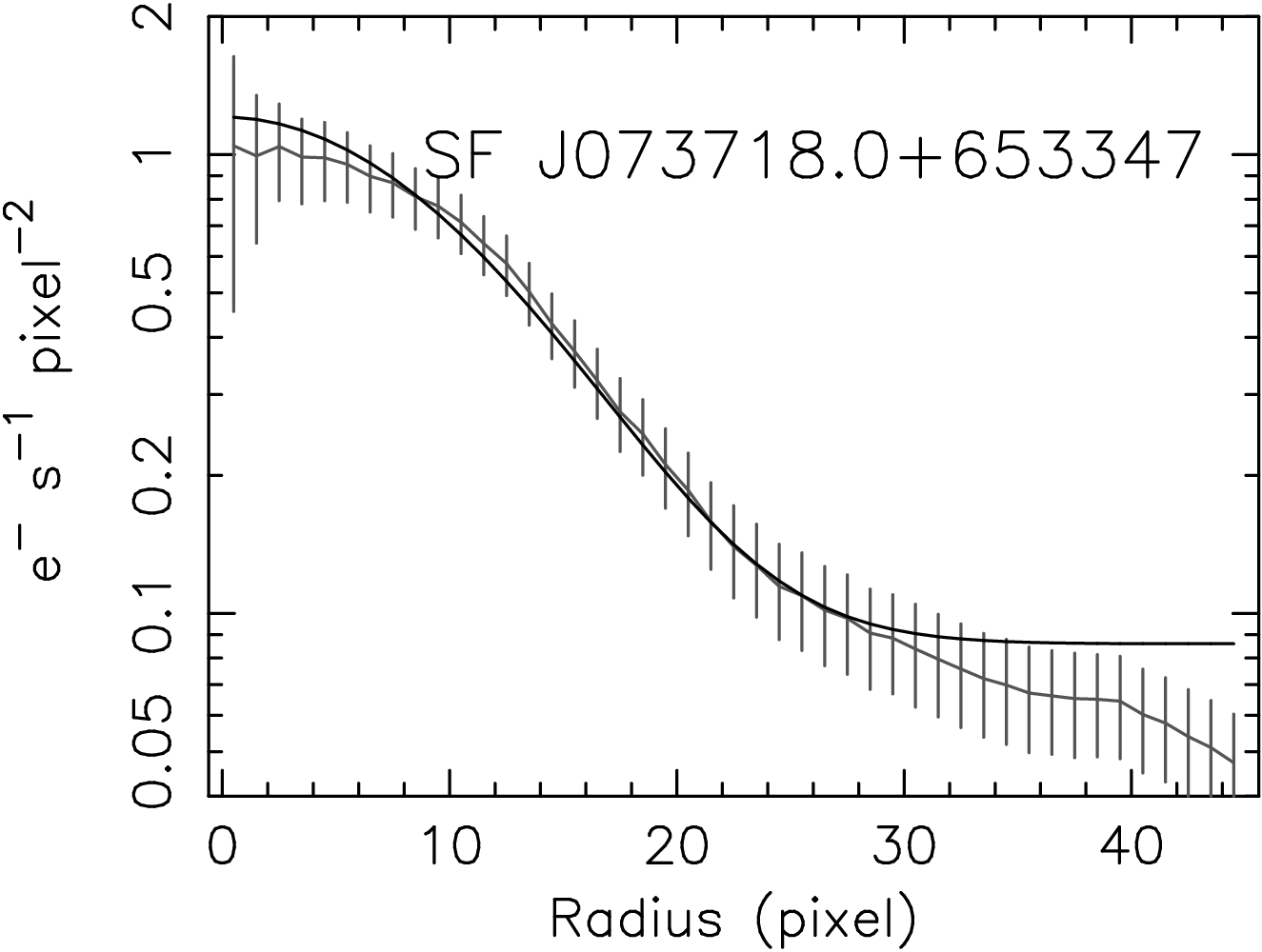} 

\figcaption{
30\arcsec$\times$30\arcsec\ close-up views of the rest 15 of 19 star-forming regions. 
FUV images are on the left and 24\,\um\ images are in the middle.
Circle locations, sizes, colors, and gray scales are the same as in Figure~\ref{f:allblobs}.
On the right are higher-resolution \ha\ radial profiles 
 shown with best-fitting Gaussian model curves.  
Each radial profile extends to 5 Gaussian widths.  One pixel corresponds to 0.\arcsec304. 
}
\end{figure}

\end{document}